\documentclass[journal,a4paper,10pt]{IEEEtran}

\usepackage{citesort,amsmath,upref,amssymb,color,bm}
\usepackage{epsfig,graphicx,rotating,subfigure,footmisc} 

\let\norm=\enVert

\newcommand{\E}[1]{{\rm E} {}}

\newcommand{\ist}{\hspace*{.3mm}}
\newcommand{\rmv}{\hspace*{-.3mm}}
\newcommand{\remark}[1]{}
\newcommand{\isum}[1]{\sum_{#1 = -\infty}^{\infty}} 
\newcommand{\esum}[2]{\sum_{#1}^{#2}} 

\def\C{\mathbb{C}}
\def\Z{\mathbb{Z}}

\def\NN{\mathbb{N}}

\def\M{M} 

\definecolor{grau}{gray}{0.6}

\newtheorem{theorem}{Theorem}

\newcommand{\eq}{\,=\,}
\renewcommand{\jmath}{j}

\newcommand{\be}{\begin{equation}}
\newcommand{\ee}{\end{equation}}

\newcommand{\supp}{\operatorname{supp}}

\newcommand{\vecr}{\operatorname{{rvec}}\hspace{-.5mm}}

\def\CM{\C^\M}          
\def\Q{Q}               
\def\CQ{\C^\Q}          
\def\CQM{\C^{\Q \times \M}}     
\def\blnum{B}               
\def\blidx{b}              
\def\Block{\mathcal{B}}    
\def\xx{\mathbf{x}}     
\def\yy{\mathbf{y}}     
\def\zz{\mathbf{z}}     
\def\AA{\mathbf{A}}
\def\UU{\mathbf{U}}
\def\VV{\mathbf{V}}
\def\bPhi{\mathbf{\Phi}}    
\def\taup{\tau_p}       
\def\nup{\nu_p}         
\def\Ts{T_{\rmv{\text{s}}}}    
\def\NLam{N_{\hspace{-.15mm}\Lambda}}   
\def\bijG{\mathsf S}         
\def\MCSidx{\theta}             
\def\MCSset{\Theta}             
\def\MCSsett{\theta \rmv\in\rmv \MCSset}
\def\MCSsettt{\theta \!\in\! \MCSset}
\def\MCSsettsub{\theta \in \MCSset}
\def\MCSnum{|\MCSset|}          
\def\Part{\mathcal{J}}          
\def\Group{\mathcal{I}}         
\def\gridx{b}                   
\def\grnum{B}                   
\def\cent{\zeta}                
\def\dT{d_\textnormal{T}}              
\def\dR{d_\textnormal{R}}              
\def\sb{\mathbf{s}}
\def\Sh{\mathbf{S}_h}
\def\rb{\mathbf{r}}
\def\HH{\mathbf{H}}
\def\aa{\mathbf{a}}
\def\Hlk{\HH_{l,k}}
\def\xxhat{\mathbf{\hat{x}}}
\def\Fmi{\mathbf{F}_{\rmv m,i}}
\def\Fmith{F_{\rmv m,i}^{(\tth)}}

\def\pp{\mathbf{p}}
\def\ps{\pp^{(\ss)}}
\def\PP{\mathbf{P}}
\def\Pils{\mathcal{P}^{(\ss)}}
\def\idx{j}
\def\idxx{\ell}

\newcommand{\Spset}[2][]{\Sigma^{#1}_{#2}}  

\newcommand{\GGmi}{\mathbf{G}_{m,i}}     
\newcommand{\GGmith}{G_{m,i}^{(\tth)}}
\newcommand{\pptra}{\pp^{(\tra)}}    
\newcommand{\tra}{s}             
\newcommand{\rec}{r}             
\newcommand{\ant}{\theta}        

\newcommand{\Pilsettra}{\mathcal{P}^{(\tra)}}   
\newcommand{\Grid}{\mathcal{G}}     
\newcommand{\Pil}{Q}                
\newcommand{\absb}[1]{\big|#1\big|} 

\newcommand{\Gmiant}{G_{m,i}^{(\ant)}}   
\newcommand{\Gmiantti}{\tilde{G}_{m,i}^{(\ant)}}   
\newcommand{\bPhitra}{\mathbf{\Phi}^{(\tra)}} 
\newcommand{\JD}{J\rmv D}
\newcommand{\UUtra}{\UU^{(\tra)}} 
\newcommand{\xxant}{\xx^{(\ant)}} 
\newcommand{\zzant}{\zz^{(\ant)}} 
\newcommand{\rectra}{(\rec,\tra)} 
\newcommand{\Ant}{\Theta}        
\newcommand{\xxanthat}{\hat{\xx}^{(\ant)}} 
\newcommand{\deff}{\triangleq}
\newcommand{\FFmi}{\mathbf{F}_{\rmv m,i}}     
\newcommand{\irange}{i\in\{-J/2,\dots,J/2 \mi 1\}}
\newcommand{\mrange}{m\in\{0,\dots,D \mi 1\}}
\newcommand{\mi}{\!-\!}
\newcommand{\HHlkhat}{\hat{\HH}_{l,k}}       
\def\bcalSi{\mathcal{S}}
\newcommand{\GG}{\mathbf{G}}     
\newcommand{\LK}{K \rmv L}
\newcommand{\KL}{K \rmv L}
\newcommand{\taupant}{\tau_p^{(\ant)}}  
\newcommand{\nupant}{\nu_p^{(\ant)}}    
\newcommand{\wwTp}{\ww_{\textnormal{T},p}^{(\tra)}}   
\newcommand{\wwRp}{\ww_{\textnormal{R},p}^{(\rec)}}   
\newcommand{\ww}{\mathbf{w}}     
\newcommand{\wTp}{w_{\textnormal{T},p}^{(\tra)}}   
\newcommand{\wRp}{w_{\textnormal{R},p}^{(\rec)}}   
\newcommand{\vvTp}{\vv_{\textnormal{T},p}}   
\newcommand{\vvRp}{\vv_{\textnormal{R},p}}   
\newcommand{\vTp}{v_{\textnormal{T},p}}      
\newcommand{\vRp}{v_{\textnormal{R},p}}      
\newcommand{\vv}{\mathbf{v}}     
\newcommand{\fc}{f_{\textnormal{0}}}         
\newcommand{\GGmiti}{\tilde{\GG}_{m,i}}     
\newcommand{\GGmitihat}{\hat{\tilde{\GG}}_{m,i}}     
\newcommand{\ggantti}{\tilde{\mathbf{g}}^{(\ant)}} 
\newcommand{\Partti}{\tilde{\mathcal{J}}}     

\newcommand{\NT}{N_{\text{T}}}
\newcommand{\NR}{N_{\text{R}}}
\newcommand{\rr}{r}
\renewcommand{\ss}{s}

\newcommand{\tth}{\theta}
\newcommand{\Tth}{\Theta}
\newcommand{\ttau}{\bm{\tau}}
\newcommand{\nnu}{\bm{\nu}}

\newcommand{\QQ}{Q}

\newcommand{\Lr}{L_{\textnormal{r}}}
\newcommand{\vecidx}{j}
\newcommand{\midx}{b'}

\allowdisplaybreaks

\begin{document}

\title{Multichannel Group Sparsity Methods for\\[-.5mm]
Compressive Channel Estimation in Doubly\\[-.5mm]
Selective Multicarrier MIMO Systems\\[-.5mm]
(Extended Version)\thanks{D.\ Eiwen and H.\ G.\ Feichtinger are with NuHAG, Faculty of Mathematics, University of Vienna, Austria
(\{daniel.eiwen,$\ist$hans.feichtinger\}@univie.ac.at).
G.\ Taub\"{o}ck is with the Acoustics Research Institute of the Austrian Academy of Sciences, Vienna, Austria (georg.tauboeck@oeaw.ac.at).
F.\ Hlawatsch is with the Institute of Telecommunications, TU Wien, Vienna, Austria (franz.hlawatsch@tuwien.ac.at).
This work was supported by the WWTF under grant MA 07-004 (SPORTS) and by the FWF under grants S10602-N13, S10603-N13,
and P27370-N30.
Parts of this work were previously presented at IEEE ICASSP 2010, Dallas, TX, Mar.\ 2010 and at
IEEE SPAWC 2010, Marrakech, Morocco, June 2010.}
\thanks{Extended version of a manuscript submitted to the IEEE Transactions on Signal Processing, \today.}\vspace*{1.3mm}}

\author{Daniel Eiwen, Georg Taub\"{o}ck, Franz Hlawatsch, and Hans Georg Feichtinger\vspace*{-3mm}}

\maketitle

\addtolength{\abovedisplayskip}{1.5mm}
\addtolength{\belowdisplayskip}{1.5mm}

\begin{abstract}
We consider
channel estimation within pulse-shaping multicarrier multiple-input multiple-output (MIMO) systems
transmitting over doubly selective MIMO channels. This setup includes MIMO orthogonal frequency-division multiplexing (MIMO-OFDM) systems
as a special case. We show
that the component channels tend to exhibit an approximate \emph{joint group sparsity} structure
in the delay-Doppler domain.
We then develop a compressive channel estimator that
exploits this
structure for improved performance.
The proposed channel estimator uses the methodology of multichannel group sparse compressed sensing,
which combines the methodologies of group
sparse compressed sensing and multichannel compressed sensing.
We derive an upper bound on the channel estimation error and analyze the estimator's computational complexity.
The performance of the estimator is
further improved by introducing a
basis expansion yielding enhanced
joint group sparsity, along with a
basis optimization algorithm that is able to utilize prior statistical information if available.
Simulations using a geometry-based channel simulator demonstrate the performance gains due to
leveraging
the 
joint group sparsity
and optimizing the basis.
\vspace{-.5mm}
\end{abstract}

\begin{keywords}
Channel estimation,
doubly selective channel,
group sparse compressed sensing,
MIMO-OFDM,
multicarrier modulation,
multichannel compressed sensing,
multiple-input multiple-output (MIMO) communications,
orthogonal frequency-division multiplexing (OFDM),
sparse reconstruction.
\vspace{-.5mm}
\end{keywords}


\section{Introduction}\label{sec:Intr}

Multiple-input multiple-output (MIMO) systems are a key methodology for meeting the growing
demand for higher data rates in wireless communications \cite{bigcalgold10}.
Here,
we consider the 
estimation of doubly selective
MIMO channels
based on
compressed sensing (CS) methods \cite{Can06,Don06,rauhutCSbook}.
We focus on multicarrier (MC) MIMO systems, which include orthogonal
frequency-division multiplexing (MIMO-OFDM) systems as a special case \cite{JiangHanzo07}.
MIMO-OFDM is used in
several important wireless standards
\cite{ieee802.11,ieee802.16,umts_lte,MIMO-OFDM_4G}.

Coherent detection in MIMO wireless communication systems requires channel state information
at the receiver. A common approach
is to embed pilot symbols into the transmit signal and to perform least-squares
or minimum mean-square error
channel estimation \cite{MIMO-OFDM_LS-MMSE}. More advanced pilot-based channel estimation methods include 
\cite{li98,Li1999,li_ieeewc02,Lu07,leus_acad11,rugini_acad11,GT_jstsp10,Bajwa10}.
In particular, \emph{compressive channel estimation}
\cite{GT_jstsp10,Bajwa10} uses CS
techniques \cite{rauhutCSbook,Can06,Don06}
to exploit an inherent sparsity of the channel that is related to the fact that doubly selective 
channels tend to be dominated by a relatively small number of clusters of significant propagation paths \cite{Raghavan07}.
While compressive
channel estimation within single-input single-output 
systems is
well explored
\cite{GT_jstsp10,Cotter02,WLi07,sharp-scaglione_icassp08,Bajwa10,DE_GT_icassp11,chen_2011,schniter1_2011,Schniter2_2012,ped12,prasad1,DE_GT_spawc10},
fewer works have addressed the MIMO case.
%
%
%
Existing
methods for MIMO channels either
exploit sparsity in the delay-Doppler-angle
domain \cite{Bajwa08c,Bajwa10} or
\emph{joint sparsity} of the component channels in the delay domain \cite{prasad2} or in the delay-Doppler
\cite{DE_GT_icassp10} domain.
%
%
%

The effective delay-Doppler (joint) sparsity 
is limited by leakage effects \cite{GT_jstsp10}, which 
correspond to a \emph{CS off the grid} scenario
\cite{chischarf11,tangrecht13}.
To reduce leakage effects, a basis optimization (dictionary learning) method
that aims at
maximizing
sparsity or joint sparsity has been proposed in \cite{GT_jstsp10,DE_GT_icassp10}.
Furthermore, methods that exploit the delay-Doppler structure of leakage---i.e., the similarity between the different delay-Doppler sparsity patterns---have
been proposed in \cite{DE_GT_spawc10}; these method rely on the concept of \emph{group sparsity}
\cite{vdBerg_Group}, which is closely related to \emph{block sparsity} \cite{Eldar_GBPDN,Stojnic_GBP,Eldar_BOMP} and \emph{model-based CS}
\cite{Bar_model08}.


Here,
we show that, in typical scenarios, there is a strong
similarity not only between the delay-Doppler sparsity patterns
but also between the delay-Doppler \emph{group} sparsity patterns of the MIMO component channels.
We exploit
this extended similarity by using \emph{multichannel group sparse CS} (MGCS),
which
combines
multichannel CS (MCS) \cite{gisttr06,tropp_MBP,dusabawaba05} 
and group sparse CS (GCS) \cite{vdBerg_Group,Eldar_GBPDN,Stojnic_GBP,Eldar_BOMP,Bar_model08}.
We thus propose an MGCS-based MIMO channel estimator that leverages 
joint group sparsity.
In contrast to previous approaches, including those in our conference publications
\cite{DE_GT_icassp10,DE_GT_spawc10}, our estimator
\emph{simultaneously} leverages group and joint sparsity in the delay-Doppler domain.
We also provide analytical
performance guarantees for the proposed estimator and analyze its computational complexity.
In addition, to reduce
leakage effects, we propose
a basis expansion that
maximizes joint group sparsity.
The optimum basis is computed by an algorithm that extends our previous
optimization procedures \cite{GT_jstsp10,DE_GT_icassp10,DE_GT_spawc10} to the
case of joint group sparsity. 
We demonstrate experimentally that the proposed MGCS-based channel estimator
significantly outperforms conventional compressive channel estimators for MIMO-OFDM systems, even if they exploit joint sparsity.

The rest of this paper is organized as follows.
The MC-MIMO system model is described in Section \ref{sec:Sys_Mod}.
In Section \ref{sec:CS}, we review
GCS, MCS, and
MGCS. In Section
\ref{sec:Del_Dop_Sp}, we analyze the multichannel (i.e., joint) group delay-Doppler
sparsity of doubly selective MC-MIMO channels.
In Section \ref{sec:Cha_Est}, we present the proposed MGCS channel estimator and a performance bound,
and we study the estimator's computational complexity.
Section \ref{sec:Bas_Opt} develops a basis optimization algorithm leading to
enhanced multichannel
group sparsity.
Finally, simulation results are presented in Section \ref{sec:Sim_Res}.

\vspace{.5mm}

\section{Multicarrier MIMO System}\label{sec:Sys_Mod} 

\vspace{.5mm}

We consider a pulse-shaping MC-MIMO system for the sake of generality
and because of its advantages over conventional cyclic-prefix (CP)
MIMO-OFDM \cite{kozmol98,Matz-Charlypaper07}. However, CP MIMO-OFDM is included as a special case.
The complex baseband  is considered throughout.



Let $\NT$ and $\NR$ denote the number of transmit and receive antennas, respectively.
The modulator generates a discrete-time transmit signal vector $\sb[n] \rmv\in \C^{\NT}\rmv$
\vspace{-.5mm}
according to \cite{kozmol98}
\begin{equation}
\label{trans-symbol}
 \sb[n] \,=\ist \esum{l=0}{L-1}\esum{k=0}{K-1} \rmv \aa_{l,k}\,g_{l,k}[n] \,.
\vspace{-.5mm}
\end{equation}
Here, $L$ and $K$ are
the number of
MC-MIMO symbols and the number of subcarriers, respectively;
$\aa_{l,k} \rmv\triangleq \big(a^{(1)}_{l,k} \cdots\ist a^{(\NT)}_{l,k} \big)^T\!\rmv\in \C^{\NT}\rmv$
denotes the
data symbol vectors;
and $g_{l,k}[n] \triangleq g[n - lN] \ist e^{\jmath 2 \pi \frac{k}{K} (n-lN)}$ is
a time-frequency shift of a transmit pulse $g[n]$
($N\!\geq\!K$ is the symbol duration).
Subsequently, $\sb[n]$ is converted into the continuous-time transmit
signal vector
\begin{equation*}
\sb(t) \,=\rmv \sum_{n=-\infty}^{\infty} \!\sb[n] \ist f_1(t \rmv-\rmv nT_{\rmv{\rm s}}) \,,
\vspace{-.8mm}
\end{equation*}
where $f_1(t)$ is the impulse response of an interpolation filter and $T_{\rmv{\text{s}}}$ is the sampling period.
Each transmit antenna $\ss\in\{1,\dots,\NT\}$ and receive antenna $\rr\in\{1,\dots,\NR\}$
are linked
by a
doubly selective channel with time-varying impulse response
$h^{(\rr,\ss)}(t,\tau)$.
This gives the MIMO channel output
\cite{matz_acad11}
\begin{equation}
\label{rec-cont}
\rb(t) \,= \int_{-\infty}^\infty \!\HH(t,\tau) \ist \sb(t \rmv-\rmv \tau) \ist d\tau \ist+\ist \zz(t) \,,
\vspace{-.5mm}
\end{equation}
where $\HH(t,\tau)$ is the $\NR \!\times\! \NT$ matrix with entries $h^{(\rr,\ss)}(t,\tau)$ and
$\zz(t)$ is a noise vector.
At the receiver, $\rb(t)$ is converted into the discrete-time signal vector $\rb[n] \rmv\in \C^{\NR}\rmv$ according to
\begin{equation*}
\rb[n] \,= \int_{-\infty}^\infty \!\rb(t) \ist f_2(n T_{\rmv{\text{s}}} \rmv-\rmv t) \ist dt \,,
\vspace{-.3mm}
\end{equation*}
where $f_2(t)$ is the impulse response of an anti-aliasing filter.
Subsequently, the demodulator
\vspace{-.5mm}
computes
\begin{equation}
\label{rec-symbol}
  \yy_{l,k}
  \,=\rmv \isum{n} \! \rb[n] \ist \gamma^*_{l,k}[n] \,,
\vspace{-.8mm}
\end{equation}
for $l \rmv\in\rmv \{ 0,\dots,L\!-\!1\}$ and $k \rmv\in\rmv \{ 0 \dots,K\!-\!1\}$,
where
$\gamma_{l,k}[n] \rmv\triangleq$\linebreak 
$\gamma[n - lN] \ist e^{\jmath 2\pi \frac{k}{K}(n-lN)}$
is a time-frequency
shift of a receive pulse $\gamma[n]$. Combining appropriate equations above,
we
\vspace{-.5mm}
obtain
\be
\label{io_channel_impresp}
\rb[n] \eq \!\rmv \sum_{m=-\infty}^{\infty}\!\! \HH[n,m] \ist \sb[n\!-\!m] \ist+\ist \zz[n] \,,
\vspace{-.5mm}
\ee
where $\HH[n,m] \triangleq \int_{-\infty}^{\infty}\int_{-\infty}^{\infty} \HH(t+n\Ts,\tau) \ist f_1(t-\tau+m\Ts) \ist f_2(-t) \ist\ist dt \ist d\tau$
and $\zz[n] \triangleq \int_{-\infty}^{\infty}\zz(t)f_2(n\Ts-t) \ist\ist dt$.
If $g[n]$ and 
$\gamma[n]$ 
are $1$ on $[0 ,N\!-\!1]$
and on $[N\!-\!K, N\!-\!1]$, respectively
and $0$ otherwise, we obtain a conventional CP MIMO-OFDM system with CP length $N\!-\!K \geq 0$ \cite{JiangHanzo07,MIMO-OFDM_4G}.



Neglecting intersymbol and intercarrier
interference, which is justified if the channel dispersion is not too strong
and if relevant system parameters are chosen appropriately
\cite{kozmol98,Matz-Charlypaper07}, equations \eqref{rec-symbol},
\eqref{io_channel_impresp}, and \eqref{trans-symbol}
\vspace{-.7mm}
yield
\begin{equation}
\label{gen.io.rel}
    \yy_{l,k} =\ist \Hlk\ist \aa_{l,k} \ist+\ist \zz_{l,k} \ist,
\;\; l \rmv\in\rmv \{0,\ldots,L\!-\!1\}, \; k \rmv\in\rmv \{0,\ldots,K\!-\!1\} ,
\vspace{-.1mm}
\end{equation}
where the channel coefficient
matrices $\Hlk \rmv\in \C^{\NR \times \NT}\rmv$ are given by
$\Hlk \triangleq\sum_{n = -\infty}^{\infty} \sum_{m=-\infty}^{\infty} \HH[n,m] \ist g_{l,k}[n-m] \ist \gamma^*_{l,k}[n] $;
furthermore, $\zz_{l,k} \triangleq \sum_{n=-\infty}^{\infty} \zz[n] \ist \gamma^*_{l,k}[n]$ \cite{kozmol98}.
Let $\gamma[n] \!=\! 0$
outside an interval $[0,L_{\gamma}]$. 
To calculate the $\yy_{l,k}$ in \eqref{rec-symbol},
$\rb[n]$ has to be known for $n\in\{0,\dots,\Lr-1\}$, where $\Lr \triangleq (L\!-\!1)N+L_\gamma+1$. 
For these $n$,
\eqref{io_channel_impresp} can be rewritten
\vspace{-.5mm}
as
\begin{equation}
\label{io_channel_spreading}
    \rb[n] \,=\!  \sum_{m=-\infty}^{\infty} \!\rmv \sum_{i=0}^{\Lr-1} \!\Sh[m,i] \ist \sb[n\!-\!m] \ist
    e^{\jmath 2 \pi \rmv \frac{ni}{\Lr}} +\,\zz[n] \,,
\vspace{-.5mm}
\end{equation}
with the \emph{discrete-delay-Doppler spreading function matrix}
\cite{Bel63,matz_acad11}
\vspace{-2.5mm}
\begin{equation}
\label{Spread_fct}
\Sh[m,i] \,\triangleq\ist
\frac{1}{\Lr} \! \sum_{n=0}^{\Lr-1} \!\rmv \HH[n,m] \ist e^{-\jmath 2 \pi\rmv\frac{in}{\Lr}} \ist, \;\;\; m,i \in \Z \,.
\vspace{-.5mm}
\end{equation}
Let us assume that the channel is causal with maximum discrete delay at most $K\!-\!1$,
i.e., $\HH[n,m]=\mathbf{0}$ for $m \notin \{0,\dots,K\!-\!1\}$. Then,
using \eqref{rec-symbol}, \eqref{io_channel_spreading}, \eqref{trans-symbol},
and the approximation $\Lr\!\approx\!LN$ (which is exact for CP MIMO-OFDM),
the channel coefficient matrices $\Hlk$ can be expressed
\vspace{-.5mm}
as
\begin{equation}
\label{System_Channel_Det}
  \Hlk  \ist=\sum_{m=0}^{K-1} \sum_{i=-L/2}^{L/2-1} \!\!
    \Fmi \, e^{-\jmath 2 \pi (\frac{km}{K}-\frac{li}{L} )} \ist.
\vspace{-.5mm}
\end{equation}
Here, $L$ is assumed
even for mathematical convenience and
\vspace{-.5mm}
\begin{equation}
\label{F_S_A}
\Fmi \,\triangleq \sum_{q=0}^{N-1} \rmv \Sh[m,i+qL] \, A^{*}_{\gamma,g}\rmv \bigg( \rmv m \ist , \frac{i+qL}{\Lr} \bigg) \ist,
\vspace{-1.5mm}
\end{equation}
where $A_{\gamma,g}(m,\xi) \triangleq \sum_{n=0}^{L_\gamma} \gamma[n]\,g^*[n\!-\!m] \ist e^{-\jmath 2\pi \xi n}$ is the
cross-ambiguity function of $\gamma[n]$ and $g[n]$ \cite{fla-book2}. The matrices $\Fmi$
represent the channel coefficient matrices $\Hlk$ in terms of a
discrete-delay variable $m$ and a discrete-Doppler variable $i$.


\section{CS
of Jointly Group-Sparse Signals}\label{sec:CS}

We will briefly review GCS
\cite{vdBerg_Group,Eldar_GBPDN,Stojnic_GBP,Eldar_BOMP,Bar_model08} and MCS
\cite{tropp_MBP,gisttr06,dusabawaba05}
and then discuss
their relationship with MGCS, which underlies the 
MIMO channel estimator presented in Section \ref{sec:Cha_Est}. 

\vspace{-.5mm}

\subsection{Group Sparse CS}\label{sub-sec:GCS}

We recall that a vector $\xx \rmv\in\rmv \CM$
is called (approximately) \emph{$S$-sparse} if at most $S$ of its entries are
(approximately) nonzero. To define group sparsity, let
$\Part={\{\Group_\gridx\}}_{\gridx=1}^{\grnum}$ be a partition of the index
set $\{1,\dots,M\}$ into ``groups'' $\Group_\gridx$, i.e.,
$\bigcup_{\gridx=1}^{\grnum}\Group_\gridx = \{1,\dots,M\}$ and $\Group_\gridx
\cap \Group_{\gridx'} = \emptyset$ for $\gridx \not= \gridx'$.
We do not require the groups $\Group_\gridx$ to consist of contiguous indices, which would be the special
case of \emph{block} sparsity.
For a vector $\xx \rmv\in\rmv \CM\rmv$, let $\xx[\gridx] \rmv\in \C^{|\Group_\gridx|}$ denote the subvector of $\xx$
comprising the entries ${[\xx]}_{\vecidx}$ of $\xx$ with $\vecidx \rmv\in\rmv \Group_\gridx$.
Then $\xx$ is called \emph{group $S$-sparse with respect to
$\Part$} if at most $S$ subvectors $\xx[\gridx]$ are not identically zero \cite{vdBerg_Group}. The set of all such vectors $\xx$ will be denoted by $\Spset{S|\Part}$.
We consider a linear model
\vspace{-1mm}
(or \emph{measurement equation})
\begin{equation}
\label{meas-eq-GCS}
\yy \ist=\ist \bPhi \xx+\zz \,,
\vspace{-1mm}
\end{equation}
where $\yy \rmv\in\rmv \CQ$ is an observed
(measured) vector,
$\bPhi \rmv\in\rmv \CQM$ is a known 
matrix, $\xx \rmv\in\rmv \CM$ is 
unknown 
but known to be (approximately) group $S$-sparse with respect to a given partition
$\Part\rmv$, i.e., $\xx \rmv\in\rmv \Spset{S|\Part}$, and $\zz \rmv\in\rmv \CQ$ is an unknown noise vector.
The
indices $\gridx$ for which 
$\xx[\gridx] \rmv\not=\rmv {\bf 0}$ 
are unknown.
Typically, the number of measurements is much smaller than the length of $\xx$, i.e., $Q \rmv\ll\rmv M$.
The goal is to reconstruct $\xx$ from $\yy$.

A trivial GCS reconstruction strategy is to use conventional CS methods
like basis pursuit denoising (BPDN) \cite{Troppjustrelax,Can06a,Don06a},
orthogonal matching pursuit (OMP) 
\cite{rauhutCSbook,Davis97OMP,Tropp04},
or compressive sampling matching pursuit (CoSaMP) \cite{netr09}, since a
group $S$-sparse vector is also $S'$-sparse, where $S'$ is the sum of the cardinalities of the $S$ groups with largest cardinalities. 
However, this strategy does not leverage the group structure of $\xx$. Therefore, some
CS recovery methods have been adapted to the group sparse case, as reviewed in what follows.

Let $\xx \rmv\in\rmv \CM\!$, not necessarily sparse or group sparse. 
For a
partition $\Part=\{\Group_\gridx\}_{\gridx=1}^{\grnum}$ of $\{1,\dots,M\}$, let
${\|\mathbf{x}\|}_{2|\Part} \triangleq \sum_{\gridx=1}^{\grnum} {\|\xx[\gridx]\|}_2$. The convex program
\be
\min_{\xx'\in\ist \CM} {\|\xx'\|}_{2|\Part} \quad \text{subject to} \; {\|\bPhi\xx' \!-\rmv \yy\|}_2\leq \epsilon
\label{G-BPDN}
\ee
is called \emph{group BPDN} (G-BPDN) \cite{Eldar_GBPDN,Stojnic_GBP}. The accuracy of
G-BPDN
depends on the measurement matrix $\bPhi$ \cite{Eldar_GBPDN}.
In particular, $\bPhi$
is said to satisfy the \emph{group restricted isometry property of order $S$ with respect to $\Part$}
if there is a constant $\delta \rmv\in\rmv (0,1)$
such that
\begin{align}
(1 \!-\rmv \delta) \ist {\|\tilde{\xx}\|}_2^2 \,\leq\, {\|\bPhi \tilde{\xx}\|}_2^2 \,\leq\ist (1 \!+\rmv \delta) \ist {\|\tilde{\xx}\|}_2^2 \quad
\text{for all} \;\ist \tilde{\xx} \rmv\in\rmv \Spset{S|\Part}\ist. \nonumber \\[-.5mm]
\label{eq:RIP} \\[-9.5mm]
\nonumber
\end{align}
The smallest such $\delta$ is denoted
$\delta_{S|\Part}$ and
called the \emph{group restricted isometry constant of order $S$ with respect to $\Part$}
(abbreviated as G-RIC); a small $\delta_{S|\Part}$ is desirable \cite{Eldar_GBPDN}.
\emph{Group OMP} (G-OMP; usually called block OMP)
\cite{Eldar_BOMP,Zvika_BOMP} 
is a greedy GCS reconstruction algorithm that iteratively identifies the support of the unknown vector.
Another greedy GCS method is obtained by specializing the model-based CoSaMP
algorithm \cite{Bar_model08} to the group sparse setting. This method, which we abbreviate as G-CoSaMP,
differs from the classical CoSaMP algorithm
in that, in each iteration, the support estimate is adapted in terms of entire
groups of $\Part$ instead of single indices.

A matrix $\bPhi$ satisfies the conventional restricted isometry property of order $S'$ with restricted isometry constant (RIC)
$\delta_{S'}$ if the 
double bound in \eqref{eq:RIP}
is satisfied for every $S'$-sparse vector $\tilde{\xx}$ 
\cite{Can06a,Candes_BPDN} 
(and $\delta_{S'}$ is the smallest $\delta$ in \eqref{eq:RIP}).
Since
a group $S$-sparse vector
is also $S'$-sparse,
where $S'$ is the sum of the cardinalities of the $S$ groups with largest cardinalities, the G-RIC of $\bPhi$ satisfies $\delta_{S|\Part} \leq \delta_{S'}$.
The following result has
been shown in \cite{rauhwar13}, cf.\ also
\cite{CandesTaoIT06,rudelversh08,rauhutCSbook}. Let $\bPhi$ be a $Q 
\rmv\times\rmv M$ matrix that is constructed by choosing uniformly at random
$Q$ rows from a unitary $M \rmv\times\rmv M$ matrix $\UU$ and properly
scaling the resulting matrix. Then for any prescribed $\gamma \rmv\in\rmv
(0,1)$ and $\eta \rmv\in\rmv (0,1)$, $\bPhi$ will, with probability at least
$1\rmv\rmv-\rmv \eta$, satisfy the restricted isometry property of order $S'$
with RIC $\delta_{S'} \rmv<\rmv \gamma$  if
\begin{equation}\label{RIP_unitary}
Q\,\geq\, C \, \frac{\mu_{\UU}^2\ist S' \max\{ \log^3(S') \log(M), \log(1/\eta)  \} }{\gamma^{2}} \,,
\end{equation}
where $C$ is a constant and $\mu_{\UU} \triangleq \sqrt{M} \max_{i,j} |{[\UU]}_{i,j}|$. Clearly, if $\delta_{S'} \rmv<\rmv \gamma$, also $\delta_{S|\Part} \rmv<\rmv \gamma$.
Unfortunately, there are so far no results
that improve on the above result by
exploiting the available group structure.

\vspace{-.5mm}

\subsection{Multichannel CS} \label{sub-sec:MCS}

A collection of vectors $\xx^{(\MCSidx)} \rmv\rmv\in\rmv
\CM\!$, $\MCSsett \triangleq \{\MCSidx_1,\MCSidx_2,\ldots,$\linebreak 
$\MCSidx_{\MCSnum}\}$  is called \emph{jointly $S$-sparse} if the
$\xx^{(\MCSidx)}\rmv$ share a common $S$-sparse support, i.e.,
$\big|\bigcup_{\MCSsettsub}\supp(\xx^{(\MCSidx)})\big| \leq S$ with
$\supp(\xx^{(\MCSidx)}) \triangleq \big\{\vecidx \in \{1,\dots,M\} \, \big|
\, {[\xx^{(\MCSidx)}]}_{\vecidx} \!\neq\rmv 0 \big\}$
\cite{HolgerAverageCaseAnalysis}. We consider the \emph{simultaneous sparse
reconstruction problem}, where the unknown, (approximately) jointly
$S$-sparse vectors $\xx^{(\MCSidx)} \rmv\rmv\in \CM\rmv$, $\MCSsett$ are
to be reconstructed simultaneously from
measurements vectors $\yy^{(\MCSidx)} \!\in\rmv \CQ$ given
\vspace{-.5mm}
by
\begin{equation}
\label{meas-eq-MCS}
\yy^{(\MCSidx)}=\ist \bPhi^{(\MCSidx)} \xx^{(\MCSidx)}+\zz^{(\MCSidx)} \,, \quad \MCSsett \,.
\vspace{-.5mm}
\end{equation}
Here, the $\bPhi^{(\MCSidx)} \rmv\rmv\in \CQM$ are known 
and the $\zz^{(\MCSidx)} \rmv\rmv\in \CQ$ are unknown.
The
supports $\supp(\xx^{(\MCSidx)})$ are unknown, and typically $Q \rmv\ll\rmv M$. 
Note that the conventional sparse reconstruction problem is
reobtained for $\MCSnum \rmv=\rmv 1$.

Because each vector $\xx^{(\MCSidx)}$ is itself (approximately) $S$-sparse, any conventional CS method,
such as BPDN, OMP, or CoSaMP, can be used to reconstruct each vector individually. 
True MCS methods that leverage the common structure of the
vectors 
include \emph{distributed compressed sensing --- simultaneous OMP} (DCS-SOMP) \cite{dusabawaba05} 
and \emph{CoSOMP} \cite{duarte_COSOMP}.
For the special case where
$\bPhi^{(\MCSidx)} \rmv=\rmv \bPhi$ for all
$\MCSsett$, \emph{multichannel BPDN} (M-BPDN) 
\cite{tropp_MBP,cotter05MMV}
and \emph{simultaneous OMP} (SOMP)
\cite{gisttr06}
are
popular MCS methods.

\vspace{-.5mm}

\subsection{Multichannel Group Sparse CS} \label{sub-sec:MGCS}

We now combine
the notions of group sparsity and joint sparsity. We call a collection of vectors $\xx^{(\MCSidx)} \rmv\!\in\rmv
\CM\!$, $\MCSsettt$ \emph{jointly group S-sparse with respect to the
partition} $\Part \!= {\{\Group_\gridx\}}_{\gridx=1}^{\grnum}$ if the
vectors $\big( \norm{\xx^{(\MCSidx)}[1]}_{2} \,\cdots$
$\norm{\xx^{(\MCSidx)}[\grnum]}_{2}\big)^T\!$,
$\MCSsettt$ are jointly $S$-sparse. Furthermore, we consider the \emph{simultaneous group sparse reconstruction problem}, i.e.,
reconstructing vectors $\xx^{(\MCSidx)}\!$, $\MCSsett$ that are (approximately)
jointly group $S$-sparse with respect to a given partition $\Part$ from
the observations $\yy^{(\MCSidx)}\!$, $\MCSsett$ given by \eqref{meas-eq-MCS}. Once again, we could use
a conventional CS method for each
$\xx^{(\MCSidx)}$ individually, since each $\xx^{(\MCSidx)}$ is (approximately)
$S'$-sparse
as mentioned in Section \ref{sub-sec:GCS}.
Alternatively, we could use a GCS method for each $\xx^{(\MCSidx)}\!$, since each $\xx^{(\MCSidx)}$ is itself (approximately)
group $S$-sparse with respect to $\Part$. Finally, we could use an MCS
technique since the $\xx^{(\MCSidx)}$ are (approximately) jointly $S'$-sparse. However, none of these trivial MGCS approaches fully leverages the
combined group
and joint sparsity.


To overcome these limitations, we use
the well-known fact that
the simultaneous sparse reconstruction problem can be recast as a group sparse
reconstruction problem \cite{duarte_COSOMP,Eldar_GBPDN}.
We apply this principle to
simultaneous \emph{group} sparse reconstruction
as follows.
Let the vectors $\xx^{(\MCSidx)} \rmv\!\in\rmv \CM\!$, $\MCSsett$ be jointly group $S$-sparse with respect to the partition $\Part= {\{\Group_\gridx\}}_{\gridx=1}^{\grnum}$ of $\{1,\dots,M\}$.
Then, we consider the associated index set $\{1,\dots,M \MCSnum\}$,
and we define an associated partition
$\tilde{\Part} \triangleq {\{\tilde{\Group}_\gridx\}}_{\gridx=1}^{\grnum}$
of $\{1,\dots,M \MCSnum\}$
with groups $\tilde{\Group}_\gridx$ of size $|\tilde{\Group}_\gridx|=|\Group_\gridx|\MCSnum$
given by
\vspace{-.5mm}
\be
\label{Group_joint}
\tilde{\Group}_\gridx \,\triangleq\, \big\{ \vecidx+ (\xi \!-\! 1) M \ist\big| \ist\ist \vecidx \rmv\in\rmv \Group_\gridx \ist ,\ist \xi \rmv\in\rmv \{1,\dots,\MCSnum\} \big\} \,.
\vspace{-1mm}
\ee
Furthermore, we define the stacked 
vectors $\xx \triangleq \big( \xx^{(\MCSidx_1)\ist T} \cdots$\linebreak 
$\xx^{(\MCSidx_{\MCSnum})\ist T}\big)^T\!$
of length $M\MCSnum$ and 
$\yy \triangleq \big( \yy^{(\MCSidx_1)\ist T} \cdots\ist \yy^{(\MCSidx_{\MCSnum})\ist T}\big)^T\!$ of length $Q\MCSnum$, 
and the block-diagonal matrix of size $Q\MCSnum \times M\MCSnum$ given by
\vspace{-2.5mm}
\begin{equation}
\label{Phi_MCS_GCS}
\bPhi \,\triangleq\ist \left( \begin{array}{ccc}
\!\!\bPhi^{(\MCSidx_1)} &\!\!\!\!\!\! & \!\mathbf{0}\!\! \\[-2.5mm]
&  \!\!\!\!\!\!\!\!\ddots\!\!\! & \!\!\\[-1.5mm]
\!\!\mathbf{0} & \!\!\!\!\!\! & \!\bPhi^{(\MCSidx_{\MCSnum})}\!\!
\end{array} \right) .
\vspace{-1mm}
\end{equation}
Then, the equations \eqref{meas-eq-MCS}
can be written in the form of \eqref{meas-eq-GCS},
i.e., $\yy = \bPhi \xx+\zz$,
with $\zz \triangleq \big( \zz^{(\MCSidx_1)\ist T} \cdots\ist \zz^{(\MCSidx_{\MCSnum})\ist T}\big)^T\!$.
It is now easily verified that
\begin{equation}
\label{x_b_theta_vector}
\hspace*{-2mm}\xx[b] \ist=\ist \big( \big(\xx^{(\MCSidx_1)}[b]\big)^T \!\cdots \big(\xx^{(\MCSidx_{\MCSnum})}[b]\big)^T \big)^T \!, \;\; b \rmv\in\rmv \{1,\ldots,B\} \ist. \!\!
\end{equation}
(Note that the $b$ on the left-hand side refers to the partition
$\tilde{\Part}\rmv$ whereas the $b$ on the right-hand side refers to the
partition $\Part\rmv$.) Therefore, if the
$\xx^{(\MCSidx)}$ are jointly group $S$-sparse with respect to $\Part\rmv$, the stacked vector
$\xx$ is group $S$-sparse with respect to $\tilde{\Part}\rmv$. Hence, by
applying a GCS reconstruction method---such as G-BPDN, G-OMP, or
G-CoSaMP---to the measurement equation $\yy = \bPhi \xx+\zz$,
we can fully exploit the structure given by the simultaneous
group
and joint sparsity of the $\xx^{(\MCSidx)}\!$. We will then say that the respective GCS reconstruction method ``operates in MGCS mode.''
It is furthermore easy to show that the G-RIC of $\bPhi$ with respect to $\tilde{\Part}$
satisfies 
\be
\delta_{S|\tilde{\Part}} \ist\ist=\ist\ist \max_{\tth \in \Tth} \, \delta^{(\tth)}_{S|\Part} \,,
\label{eq-delta-max}
\vspace{-.5mm}
\ee
where $\delta^{(\tth)}_{S|\Part}$ is the G-RIC of $\bPhi^{(\tth)}$ with respect to $\Part$.

An alternative MGCS method,
referred to as G-DCS-SOMP, extends DCS-SOMP (see Section \ref{sub-sec:MCS}) to incorporate group sparsity. This method
adds entire groups to the joint support in each iteration, rather than adding only single indices.




\section{Joint Group Sparsity in the Delay-Doppler\\[-.5mm]
Domain}\label{sec:Del_Dop_Sp}

Next, we demonstrate that the
matrices $\Fmi$ (see \eqref{System_Channel_Det} and \eqref{F_S_A}) exhibit a
joint group sparsity structure. This structure will be exploited by our channel estimator.

\vspace{-.5mm}

\subsection{Delay-Doppler Spreading Model} \label{sub-sec:dd-model}

Let
$\tth\!\triangleq\!(\rr,\ss)$ index the channel between transmit antenna
$\ss$ and receive antenna $\rr$, and let
$\Tth\triangleq \big\{\tth \!=\! (\rr,\ss) \,|\, \rr\rmv\in\rmv\{1,\dots,$\linebreak 
$\NR\}, \, \ss\rmv\in\rmv\{1,\dots,\NT\} \big\}$.
For real-world (underspread \cite{matz_acad11})
wireless
channels and practical transmit and receive pulses, the entries $\Fmith
\rmv\triangleq\rmv {[\Fmi]}_{r,s} $ of the
$\Fmi$ in \eqref{F_S_A} are effectively supported in some small rectangular region
$[0,D\!-\!1]\times$\linebreak 
$[-J/2,J/2\!-\!1]$
about the origin of the discrete delay-Doppler ($(m,i)$) plane.
Here, $D$ and $J$ are chosen such that
$\Delta K\!\triangleq\!K/D$ and $\Delta L \!\triangleq\! L/J$ are integers $\geq \! 1$, and $J$ (and therefore also $L$) is
even.
Hence, the summation intervals in \eqref{System_Channel_Det} can be replaced by $m \rmv\in\rmv \{0,\dots,D\!-\!1\}$ and $i \rmv\in \{-J/2,\dots,J/2\!-\!1\}$.
In what follows, let $\vecr_{m,i}\big\{ \Fmith \big\} \triangleq$\linebreak 
$\big( F_{0,-J/2}^{(\theta)} \; F_{0,-J/2+1}^{(\theta)} \cdots F_{1,-J/2}^{(\theta)} \cdots F_{D-1,J/2-1}^{(\theta)} \big)^T\!$, $\MCSsett$ denote the result of a rowwise stacking with respect to $m,i$ of the $D \!\times\! J$
``matrices'' $\Fmith$ into $J\rmv D$-dimensional vectors.
To assess the joint group sparsity of the 
$\Fmith\ist$, $\tth \rmv\in\rmv \Tth$, we will show
that the vectors $\vecr_{m,i}\big\{ \Fmith \big\}$ are approximately
jointly group sparse with respect to some partition $\Part$, to be
specified in Section \ref{sub-sec:Jo_Gr_Sp}. As a preparation,
we first discuss the special cases of group sparsity and joint sparsity in
Sections \ref{sub-sec:Gr_Sp} and \ref{sub-sec:Jo_Sp}, respectively. For this
discussion, we omit the matrix-to-vector stacking operations
and thus deal directly with two-dimensional (2D) functions.

Because of \eqref{F_S_A}, analyzing the joint group sparsity of the $\Fmith$,
$\tth \rmv\in\rmv \Tth$ basically amounts to studying
the spreading functions $S^{(\tth)}_h[m,i]$. Indeed, \eqref{F_S_A} written entrywise
\vspace{-.5mm}
reads
\be \label{F_S_A entry}
\Fmith \,= \sum_{q=0}^{N-1} \rmv S_h^{(\tth)}[m,i+qL] \, A^{*}_{\gamma,g}\rmv \bigg(
\rmv m \ist , \frac{i+qL}{\Lr} \bigg) \ist,
\vspace{-1.5mm}
\ee
and neither the multiplication
by $A^{*}_{\gamma,g}\big( m \ist , \frac{i+qL}{\Lr} \big)$ nor the summation
with respect to $q$
can create any ``new'' nonzeros.
Let us assume that each channel comprises $P$
propagation paths (multipath components) corresponding to the same set of $P$
specular scatterers with channel-dependent delays $\tau_p^{(\tth)}\rmv$ and
Doppler frequency shifts $\nu_p^{(\tth)}\!$, for $p \rmv\in\rmv \{1,\dots,P\}$.
\vspace{-.5mm}
Thus,
\begin{equation}
\label{impresp_cont}
h^{(\tth)}(t,\tau) \eq\rmv \sum_{p=1}^P\eta_p^{(\tth)} \ist\delta\big(\tau \rmv-\tau_p^{(\tth)} \big) \,e^{\jmath 2 \pi \nu_p^{(\tth)} t} \ist, \;\;\; \tth \rmv\in\rmv \Tth \ist,
\vspace{-1.5mm}
\end{equation}
where the $\eta_p^{(\tth)}$ are complex path gains. This model is often a
good approximation of real mobile radio channels
\cite{molisch_book,chmodel_vehicular_09,chmodel_cellular_09}. We emphasize
that we use it only for analyzing the sparsity
of the $\Fmith$ and for motivating the basis optimization in Section \ref{sec:Bas_Opt}; it is not
required for the proposed channel estimator.
Inserting \eqref{impresp_cont} into \eqref{Spread_fct}, we
\vspace{-.5mm}
obtain
\begin{equation}
\label{Spread_fct_mod}
S_h^{(\tth)}[m,i] \eq\rmv \sum_{p=1}^P \eta_p^{(\tth)} \, e^{\jmath \pi \big(\nu_p^{(\tth)} \Ts - \frac{i}{\Lr} \big) (\Lr-1)} \, \Lambda_p^{(\tth)}[m,i] \,,
\vspace{-.5mm}
\end{equation}
with the \emph{shifted leakage \vspace{-1mm} kernels}
\begin{equation}
\label{leakage_kernel}
\Lambda_p^{(\tth)}[m,i] \,\triangleq\, \phi^{(\nu_p^{(\tth)})} \bigg( m - \frac{\tau_p^{(\tth)}}{\Ts} \bigg) \ist \psi \big( i - \nu_p^{(\tth)} \Ts \Lr \big) \,,
\vspace{-.5mm}
\end{equation}
where 
\vspace*{-1mm}
\[
\phi^{(\nu)} (x) \ist\triangleq \int_{-\infty}^\infty \! f_1(\Ts x - t) \ist f_2(t) \, e^{-\jmath 2 \pi \nu t} dt
\vspace{-2mm}
\]
and
\[
\psi(x) \ist\triangleq\ist \frac{\sin (\pi x)}{\Lr \sin (\pi x / \Lr)} \,.
\]
As shown in \cite{GT_jstsp10},
each $\Lambda_p^{(\tth)}[m,i]$ is effectively supported in a rectangular
region of some delay length $\Delta \widetilde m \!\in\! \NN$ and Doppler
length $\Delta \tilde i \!\in\!\NN$, centered about the delay-Doppler point
$\cent_p^{(\tth)}\!\triangleq\! \big(\tau_p^{(\tth)} \! / \ist \Ts \ist,
\nu_p^{(\tth)}\Ts \Lr\big)$. Therefore, each $\Lambda_p^{(\tth)}[m,i]$ is
approximately $\Delta \widetilde m \ist \Delta \tilde i$-sparse. Here,
$\Delta \widetilde m$ and $\Delta \tilde i$ can be chosen to achieve a prescribed approximation quality.
Typically, $\Delta \widetilde m$ can be chosen quite small because the functions $\phi^{(\nu)}(x)$ decay
rather rapidly, whereas $\Delta \tilde i$ has to be
larger because $\psi(x)$ decays more slowly.

\begin{figure}[t]
\vspace*{2mm} \hspace*{73mm} \centering \hspace*{-90mm}
\includegraphics[height=7cm]{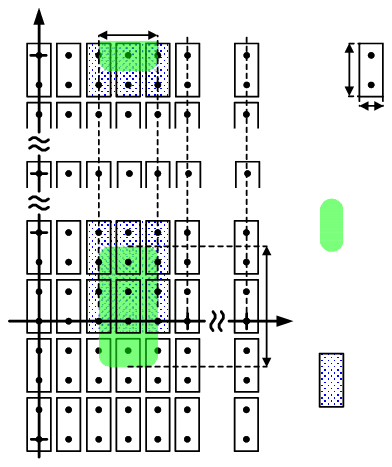} \small
\put(-16,145){$\Delta m'$} \put(-32,167){$\Delta i'$}
\put(-122,189.5){$\Delta \widetilde m$} \put(-51,67){$\Delta \tilde i$}
\put(-46,51){$m$} \put(-161,193){$i$} \put(-181,174.5){$\Lr\!-\!1$}
\put(-188,122.5){$L/2\!-\!1$} \put(-187,97){$J/2\!-\!1$} \put(-174,58){$0$}
\put(-182.5,7){$-J/2$} \put(-99.5,188){$D\!-\!1$}
\put(-73,188){$K\rmv\!-\!1$} \put(4,168){2D block $\Block_\blidx$}
\put(-13,107.6){effective support} \put(-13,94){of $\Lambda_p^{(\tth)}[m,i]$}
\put(-13,51.5){$\widetilde{N}_\Lambda$ blocks contain-} 
\put(-13,41){ing the support of} 
\put(-13,28.5){$\Lambda_p^{(\tth)}[m,i]$ within}
\put(-13,16.7){$[0,K\!\!-\!\rmv 1] \rmv\times\rmv [0,\Lr\!-\!\rmv 1]$} 
\normalsize
\vspace*{-1mm}
\caption{Illustration of the 2D block tiling $\{\Block_\blidx\}$, the effective support of a shifted leakage kernel $\Lambda_p^{(\tth)}[m,i]$,
and the $\widetilde{N}_\Lambda$ blocks containing this effective support. In this example, $\Delta \widetilde{m}=2$, $\Delta \tilde{i}=4$, $\Delta m'=1$, $\Delta i'=2$,
and $\widetilde{N}_\Lambda=9$.}
\label{fig.blocks}
\end{figure}

\vspace{-.5mm}

\subsection{Group Sparsity} \label{sub-sec:Gr_Sp}

We first analyze the group sparsity of $\Fmith$ for a single
$\tth$.
Consider a tiling of $\Z \rmv\times\rmv \Z$ into 2D blocks $\Block_\gridx$ of
equal size $\Delta m' \!\times\rmv \Delta i'$, where $\Delta m'$ divides $D$,
$\Delta i'$ divides $J/2$, and
$[0,\Delta m'\!-\!1]\times [0,\Delta i'\!-\!1]$ is one of
these blocks. As visualized in Fig.\ \ref{fig.blocks},
the effective support of each shifted leakage kernel $\Lambda_p^{(\tth)}[m,i]$
within $[0,K\!-\!1] \times [0,\Lr\rmv-\!1]$ is contained in at
most $\widetilde{N}_\Lambda$ blocks $\Block_\gridx$, where
\vspace{-1mm}
\be
\label{NLam_tilde} \widetilde{N}_\Lambda \,\triangleq\ist \bigg( \bigg\lceil
\frac{\Delta \widetilde m}{\Delta m'} \bigg\rceil + 1 \rmv\bigg)
  \bigg( \bigg\lceil \frac{\Delta \tilde i}{\Delta i'} \bigg\rceil +1 \rmv\bigg) \ist.
\vspace{-1.5mm}
\ee
Thus, by
\eqref{Spread_fct_mod}, the support of $S_h^{(\tth)}[m,i]$ is contained in at most $P \widetilde{N}_\Lambda$ blocks.
Since $\Delta i'$ divides $L$ (because $\Delta i'$ divides $J/2$ and $J$
divides $L$), the summation in \eqref{F_S_A entry}
only adds up whole blocks. Hence,
the nonzeros contained in a single
block are not
spread over several blocks
and thus no ``new'' nonzero blocks within the fundamental region
$[0,D\rmv-\rmv 1] \times [-J/2,J/2\rmv-\rmv 1]$ are
created. Also, since $\Delta
m'$ divides $D$ and $\Delta i'$ divides $J/2$, the support restriction of
$\Fmith$ to $[0,D\!-\!1] \times [-J/2,J/2\!-\!1]$ is
compatible with the block boundaries.
Thus, the effective support of $\Fmith$ within $[0,D\rmv-\rmv 1] \times [-J/2,J/2\rmv-\rmv 1]$
is contained in at most $P \widetilde{N}_\Lambda$ blocks. Therefore, $\Fmith$ is
approximately group $P \widetilde{N}_\Lambda$-sparse with respect to the tiling $\{\Block_\blidx\}$.

\vspace{-.5mm}

\subsection{Joint Sparsity} \label{sub-sec:Jo_Sp}

Next, we analyze the joint sparsity of the $\Fmith$, $\tth \rmv\in\rmv \Tth$.
We first consider the shifted leakage kernels $\Lambda_p^{(\tth_1)}[m,i]$ and
$\Lambda_p^{(\tth_2)}[m,i]$ of two different channels $\tth_1=(\rr_1,\ss_1)$
and $\tth_2=(\rr_2,\ss_2)$, corresponding to the same scatterer $p$. As
mentioned in Section \ref{sub-sec:dd-model}, these leakage kernels are
effectively supported in rectangular regions of equal size $\Delta \widetilde
m \rmv\times\rmv \Delta \tilde i$ that are centered about
$\cent_p^{(\tth_1)}\!=\! \big(\tau_p^{(\tth_1)} \! / \ist \Ts \ist,
\nu_p^{(\tth_1)}\Ts \Lr\big)$ and $\cent_p^{(\tth_2)}\!=\!
\big(\tau_p^{(\tth_2)} \! / \ist \Ts \ist, \nu_p^{(\tth_2)}\Ts \Lr\big)\rmv$.
We will show that, typically, these center points are very close to each
other,
and therefore the supports of the two leakage kernels strongly
overlap.

\begin{figure}[t]
\vspace*{2mm}
\centering
\includegraphics[width=7.0cm]{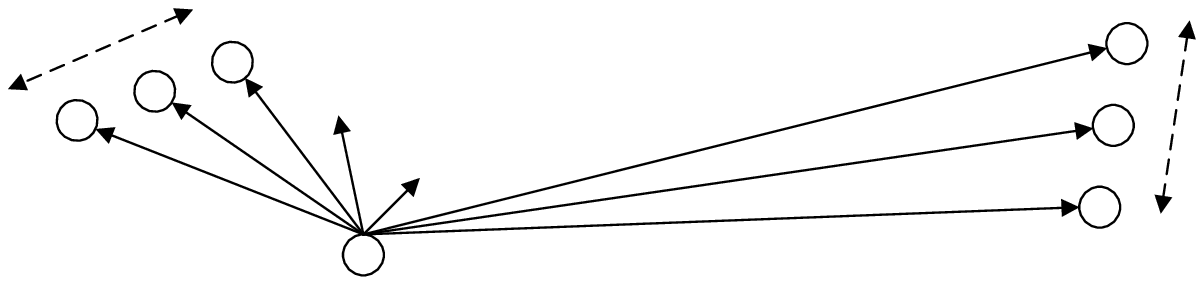}
\small
\put(-190,42){$\dT$}
\put(-1,27){$\dR$}
\put(-177,4){$\wwTp$}
\put(-85,32){$\wwRp$}
\put(-133,20){$\vvTp$}
\put(-152,32){$\vvRp$}
\put(-158,-10){scatterer $p$}
\put(-210,53){transmit antennas}
\put(-45,50){receive antennas}
\normalsize
\vspace*{1mm}
\caption{Illustration of the propagation paths from the transmit antennas
to the receive antennas 
via a scatterer $p$ in a $3 \!\times\! 3$ MIMO system.}
\label{fig.paths}
\vspace*{-1mm}
\end{figure}


\subsubsection{Time delay}
Consider the transmit antennas, the receive antennas, and
some
scatterer $p$, as shown
in Fig.\ \ref{fig.paths}.
Let $\wwTp$ and $\wwRp$ denote the vectors connecting scatterer $p$ with
transmit antenna $\tra$ and receive antenna $\rec$, respectively, and let
$\wTp \deff \big\| \wwTp \big\|_{2}$ and $\wRp \deff \big\| \wwRp
\big\|_{2}$.
The time delay $\taupant$ for scatterer $p$ and antenna pair $\ant=\rectra$
is then obtained 
\vspace{-1mm}
as
\be
\taupant =\ist
\frac{\wTp+\wRp}{c}\,,
\ee
where $c$ denotes the speed of light. We can bound
the difference between the time delays of two
channels $\ant_1$ and $\ant_2$, $\Delta \tau_p^{(\tth_1,\tth_2)} \triangleq \big|\tau_p^{(\tth_1)}
\!-\rmv \tau_p^{(\tth_2)} \big|$,
\vspace{-.5mm}
as
\begin{align}
\Delta \taup^{(\ant_1,\ant_2)}
&=\ist \frac{1}{c} \ist \big| w_{\textnormal{T},p}^{(\tra_1)} \rmv+
w_{\textnormal{R},p}^{(\rec_1)} \rmv- w_{\textnormal{T},p}^{(\tra_2)} \rmv-
w_{\textnormal{R},p}^{(\rec_2)} \big| \nonumber \\[1mm]
&\leq\ist \frac{1}{c} \ist \big( \big|
w_{\textnormal{T},p}^{(\tra_1)} \rmv- w_{\textnormal{T},p}^{(\tra_2)} \big|
  + \big| w_{\textnormal{R},p}^{(\rec_1)} \rmv- w_{\textnormal{R},p}^{(\rec_2)} \big| \big)\,.
\label{Deltataup12} \\[-6.5mm]
\nonumber
\end{align}
From
geometric considerations, the difference between the transmitter-scatterer path lengths, $\big|
w_{\textnormal{T},p}^{(\tra_1)} \rmv- w_{\textnormal{T},p}^{(\tra_2)} \big|$,
cannot be larger than the distance between the two transmit antennas $\tra_1$
and $\tra_2$. This distance, in turn, is bounded by the maximum distance
between any two transmit antennas, denoted by $\dT$. Thus,
$\big| w_{\textnormal{T},p}^{(\tra_1)} \rmv- w_{\textnormal{T},p}^{(\tra_2)}
\big| \le \dT$. Using the same argument for the scatterer-receiver path, we
obtain $\big| w_{\textnormal{R},p}^{(\rec_1)} \rmv-
w_{\textnormal{R},p}^{(\rec_2)} \big| \le \dR$, where $\dR$ denotes the
maximum distance between any two receive antennas. Inserting these bounds
into \eqref{Deltataup12} gives
\be
\label{delay bound}
\Delta\tau_p^{(\tth_1,\tth_2)} \leq \tau_{\text{B}}
\triangleq \frac{\dT+\dR}{c} \,.
\ee


\subsubsection{Doppler frequency shift}
Next, we consider the Doppler frequency
shift $\nupant$ for scatterer $p$ and antenna pair $\ant=\rectra$. If the
source of a sinusoidal
wave with frequency $\fc$ moves towards an observer with relative velocity $v$, at an angle $\alpha$ relative
to the observer-source direction, the Doppler frequency shift
is approximately
$\nu= \fc\frac{v}{c} \cos{\alpha}$ \cite{molisch05}. In our case, because
transmitter, receiver, and scatterers are moving, the Doppler effect occurs
twice. Let $\vvTp$ and $\vvRp$ denote the velocity vectors of scatterer $p$
relative to transmitter and receiver, respectively, and let
$\vTp\deff\norm{\vvTp}_{2}$ and $\vRp\deff\norm{\vvRp}_{2}$. First, we
consider the transmission from antenna $s$ to scatterer $p$, with carrier
(center) frequency $\fc$. Let $\alpha$ denote the angle between $\vvTp$ and $\wwTp$, and note that
$\cos{\alpha} = \frac{\vvTp^T \wwTp}{\vTp\wTp}$.
The carrier frequency observed at scatterer $p$ is approximately
$f_1=\fc+\nu_{\textnormal{T},p}^{(\tra)}\ist$,
\vspace{-2mm}
with
\be
\label{Doppler T}
\nu_{\textnormal{T},p}^{(\tra)} \ist=\ist \fc \ist
\frac{\vTp}{c} \ist \cos{\alpha} \,=\ist \fc \ist\frac{\vTp}{c} \ist
\frac{\vvTp^T\wwTp}{\vTp \wTp}
  \ist=\ist \frac{\fc}{c}\ist \frac{\vvTp^T\wwTp}{\wTp}\,.
\ee
After transmission from scatterer $p$ to receive antenna $r$,
the observed carrier frequency at
receive antenna $\rec$ is given by
$f_2=f_1+\nu_{\textnormal{R},p}^{(\rec)}\ist$, with the Doppler frequency
shift
\vspace{-1mm}
(cf. \eqref{Doppler T})
\be
\label{Doppler R}
\nu_{\textnormal{R},p}^{(\rec)} \ist=\ist \frac{f_1}{c} \ist
\frac{\vvRp^T\wwRp}{\wRp}\,.
\vspace{-1.5mm}
\ee
Inserting for $f_1$ yields $f_2=\fc +\nu_{\textnormal{T},p}^{(\tra)} +\nu_{\textnormal{R},p}^{(\rec)}$. Thus, the total Doppler frequency shift is
\vspace{-1mm}
\be
\nupant =\ist \nu_{\textnormal{T},p}^{(\tra)} + \nu_{\textnormal{R},p}^{(\rec)} \,.
\ee

We can now bound the difference between the Doppler frequency shifts of two
channels $\ant_1$ and $\ant_2$, $\Delta \nu_p^{(\tth_1,\tth_2)}
\triangleq \big|\nu_p^{(\tth_1)} \!-\rmv \nu_p^{(\tth_2)}\big|$, as
\begin{align}
\Delta
\nup^{(\ant_1,\ant_2)}
&=\ist \big| \nu_{\textnormal{T},p}^{(\tra_1)} \rmv+ \nu_{\textnormal{R},p}^{(\rec_1)} \rmv- \nu_{\textnormal{T},p}^{(\tra_2)}
\rmv- \nu_{\textnormal{R},p}^{(\rec_2)} \big| \nonumber \\[.5mm]
&\leq\ist \big|
\nu_{\textnormal{T},p}^{(\tra_1)} \rmv- \nu_{\textnormal{T},p}^{(\tra_2)}
\big|
  + \big| \nu_{\textnormal{R},p}^{(\rec_1)} \rmv- \nu_{\textnormal{R},p}^{(\rec_2)} \big| \,.
\label{Deltanup12}
\end{align}
For the transmitter-scatterer path, we obtain using \eqref{Doppler T}
\begin{align*}
\big| \nu_{\textnormal{T},p}^{(\tra_1)} \rmv- \nu_{\textnormal{T},p}^{(\tra_2)} \big|
&\eq \frac{\fc}{c} \left|\vvTp^T \!\left(\frac{\ww_{\textnormal{T},p}^{(\tra_1)}}{w_{\textnormal{T},p}^{(\tra_1)}}
  - \frac{ \ww_{\textnormal{T},p}^{(\tra_2)}} {w_{\textnormal{T},p}^{(\tra_2)}}\right)\right| \\[.8mm]
&\overset{(a)}{\leq}\ist \frac{\fc}{c} \, \vTp \, \Bigg\| \frac{\ww_{\textnormal{T},p}^{(\tra_1)}}{w_{\textnormal{T},p}^{(\tra_1)}}
  - \frac{ \ww_{\textnormal{T},p}^{(\tra_2)}} {w_{\textnormal{T},p}^{(\tra_2)}} \Bigg\|_2 \\[.8mm]
&\overset{(b)}{\leq}\ist \frac{\fc}{c} \, \vTp \ist \frac{ \big\| \ww_{\textnormal{T},p}^{(\tra_1)}
  \rmv- \ww_{\textnormal{T},p}^{(\tra_2)} \big\|_2}{w_{\textnormal{T},p,\textnormal{min}}^{(\tra_1,\tra_2)}} \\[-1mm]
&\overset{(c)}{\leq}\ist \frac{\fc}{c} \, \frac{\vTp \ist \dT }{w_{\textnormal{T},p,\textnormal{min}}^{(\tra_1,\tra_2)} }  \,,
\end{align*}
with $w_{\textnormal{T},p,\textnormal{min}}^{(\tra_1,\tra_2)} \deff \min
\rmv\big\{ w_{\textnormal{T},p}^{(\tra_1)},w_{\textnormal{T},p}^{(\tra_2)}
\big\}$. Here, $(a)$ is due to the Cauchy-Schwarz inequality, $(b)$ follows from the inequality
\be \label{norm_diffs_ineq}
\norm{\frac{ \mathbf{a} }{ {\|\mathbf{a}\|}_2 } - \frac{ \mathbf{b} }{ {\|\mathbf{b}\|}_2 }}_2 \ist\leq\,
\frac{ \norm{\mathbf{a} \rmv-\rmv \mathbf{b}}_2 }{ \min\{\norm{\mathbf{a}}_2,\norm{\mathbf{b}}_2\}} \;,
\ee
which is proven in Appendix A,
and $(c)$ is a consequence of $\big\| \ww_{\textnormal{T},p}^{(\tra_1)} \rmv-
\ww_{\textnormal{T},p}^{(\tra_2)} \big\|_2 \leq \dT$. 
A similar derivation for the scatterer-receiver path, using \eqref{Doppler R}, yields $\big|
\nu_{\textnormal{R},p}^{(\rec_1)} \rmv- \nu_{\textnormal{R},p}^{(\rec_2)}
\big| \le \frac{f_1}{c} \ist \frac{\vRp \ist \dR
}{w_{\textnormal{R},p,\textnormal{min}}^{(\rec_1,\rec_2)} }$, where
$w_{\textnormal{R},p,\textnormal{min}}^{(\rec_1,\rec_2)} \deff \min
\rmv\big\{ w_{\textnormal{R},p}^{(\rec_1)},w_{\textnormal{R},p}^{(\rec_2)}
\big\}$. Inserting these bounds into \eqref{Deltanup12}, we obtain
\be
 \label{Doppler bound}
\Delta \nu_p^{(\tth_1,\tth_2)} \leq\ist \nu_{\text{B},p}^{(\tth_1,\tth_2)}
\triangleq\ist \frac{1}{c} \! \left( \frac{\fc \ist \vTp \ist \dT}{w_{\textnormal{T},p,\textnormal{min}}^{(\tra_1,\tra_2)} }
    + \frac{f_1 \vRp \ist \dR }{w_{\textnormal{R},p,\textnormal{min}}^{(\rec_1,\rec_2)} } \right) \rmv.
\ee


\subsubsection{Joint sparsity of the
$\Fmith$}
From
\eqref{delay bound}
and
\eqref{Doppler bound},
it follows that the center points $\cent_p^{(\tth_1)}\!=\! \big(\tau_p^{(\tth_1)} \! /
\ist \Ts \ist, \nu_p^{(\tth_1)}\Ts \Lr\big)$ and $\cent_p^{(\tth_2)}\!=\!
\big(\tau_p^{(\tth_2)} \! / \ist \Ts \ist, \nu_p^{(\tth_2)}\Ts \Lr\big)\rmv$
of
$\Lambda_p^{(\tth_1)}[m,i]$ and $\Lambda_p^{(\tth_2)}[m,i]$ differ by at most $\tau_{\text B} / \Ts$ in the
$m$-direction and by at most $\nu_{\text{B},p}^{(\tth_1,\tth_2)} \ist \Ts
\Lr$ in the $i$-direction. Since this holds
for \emph{any} pair of channels
$\tth_1$ and $\tth_2$, we
conclude that all $\Lambda_p^{(\tth)}[m,i]$, $\MCSsett$ are (approximately) jointly
$\Delta m \ist \Delta i$-sparse, where
\[
\Delta m \,\triangleq\, \Delta \widetilde m + \lceil\tau_{\text B} / \Ts\rceil \quad\; \text{and} \quad\; \Delta i \,\triangleq\, \Delta \tilde i + \lceil\nu_{\text B} \ist \Ts \Lr \rceil \,,
\]
with $\nu_{\text{B}} \triangleq \max_{p,\ist\tth_1 \neq \tth_2} \!\big\{
\nu_{\text{B},p}^{(\tth_1,\tth_2)}\big\}$. (Here, we used the fact that the
effective supports of $\Lambda_p^{(\tth_1)}[m,i]$ and
$\Lambda_p^{(\tth_2)}[m,i]$ have size $\Delta \widetilde m \rmv\times\rmv
\Delta \tilde i$.)
With 
\eqref{Spread_fct_mod}, it then follows that the spreading functions
$S_h^{(\tth)}[m,i]$ are (approximately) jointly $P \Delta m \ist \Delta
i$-sparse. Finally, because of \eqref{F_S_A entry}, the same is true for the
$\Fmith$ (as discussed before).

Since the antenna spacings are typically much smaller than the path lengths,
i.e., $\dT \ll w_{\textnormal{T},p}^{(\tra)}$ and $\dR \ll
w_{\textnormal{R},p}^{(\rec)}\ist$,
and for practical
velocities $\vTp$ and $\vRp\ist$,
$\lceil \tau_{\text B} / \Ts \rceil$ and $\lceil \nu_{\text{B}} \ist \Ts \Lr
\rceil$ will be small compared to $\Delta \widetilde m$ and $\Delta \tilde
i$, respectively. Therefore, the \emph{joint} sparsity order $P \Delta m \ist
\Delta i = P \ist \big(\Delta \widetilde m + \lceil\tau_{\text B} / \Ts\rceil
\big) \ist \big(\Delta \tilde i + \lceil\nu_{\text B} \ist \Ts \Lr \rceil
\big)$ will not be much larger than the \emph{individual} sparsity orders $P
\Delta \widetilde m \ist \Delta \tilde i$ of the 
$\Fmith$.

\vspace{-.5mm}

\subsection{Joint Group Sparsity} \label{sub-sec:Jo_Gr_Sp}

We reconsider the tiling of $\Z \rmv\times\rmv \Z$ into the 2D blocks
$\Block_\gridx$ as previously considered in Section \ref{sub-sec:Gr_Sp},
restricting it to $[0,D\!-\!1] \times [-J/2,J/2\!-\!1]$ (see
Fig.\ \ref{fig.blocks}).
Within that region, we obtain a finite number $\blnum$ of blocks
$\Block_\gridx$, $b \rmv\in\rmv \{1,\dots,B\}$. We recall that the blocks
are of equal size $|\Block_\gridx| = \Delta m' \Delta i'$, and hence $\blnum
= \frac{J \rmv D}{\Delta m' \Delta i'}$.
Since the leakage kernels $\Lambda_p^{(\tth)}[m,i]$, $\MCSsett$ are
(approximately) jointly $\Delta m \ist \Delta i$-sparse, as we just showed in
Section \ref{sub-sec:Jo_Sp}, it follows by the same reasoning as in Section
\ref{sub-sec:Gr_Sp}
that their effective supports are jointly contained in at most $\NLam$ blocks
$\Block_\gridx$, where (cf. \eqref{NLam_tilde})
\[
\NLam \,\triangleq\ist \bigg( \bigg\lceil \frac{\Delta m}{\Delta m'} \bigg\rceil + 1 \rmv\bigg)
  \bigg( \bigg\lceil \frac{\Delta i}{\Delta i'} \bigg\rceil +1 \rmv\bigg) \ist.
\vspace{-1mm}
\]
Thus, again because of \eqref{Spread_fct_mod} and \eqref{F_S_A entry}, the $\Fmith\ist$,
$\MCSsett$ are (approximately) jointly group $P \NLam$-sparse with respect to
the tiling ${\{\Block_\blidx\}}_{\blidx=1}^{\blnum}$.

This joint group sparsity of the $\Fmith$ translates into a joint group
sparsity of the vectors $\vecr_{m,i}\big\{ \Fmith \big\}$, $\MCSsett$
defined in Section \ref{sub-sec:dd-model}.
The stacking operator $\vecr_{m,i}\{ \cdot \}$ corresponds to the one-to-one
2D\,$\to$\,1D index mapping
$\bijG \!: \{0,\dots,D\!-\!1\} \times \{-J/2,\dots,J/2\!-\!1\}  \rightarrow  \{1,\dots,J \rmv D\}$ given by
\be
\label{bij}
\bijG(m,i) \,\triangleq\, mJ+i+ \frac{J}{2} +1\ist.
\ee
Under this index mapping, the 2D blocks $\Block_\blidx$ are converted
into the 1D \emph{groups} $\Group_\blidx \triangleq \bijG(\Block_\blidx)
\subseteq \{1,\dots,J \rmv D\}$, $\blidx\in\{1,\dots,\blnum\}$, which are of
equal size $|\Group_\blidx| = |\Block_\blidx| = \Delta m' \Delta i'$.
Clearly, $\Part \deff {\{\Group_\blidx\}}_{\blidx=1}^{\blnum}$ constitutes a
partition of $\{1,\dots,JD\}$, as required by the definition of joint group
sparsity in Section \ref{sub-sec:MGCS}. Then, because
the effective supports of all the $F^{(\tth)}_{m,i}$ within
$\{0,\dots,D\!-\!1\}\times \{-J/2,\dots,J/2\!-\!1\}$ are jointly contained in
at most $P \NLam$ blocks $\Block_\blidx$, it follows that the
vectors $\vecr_{m,i}\big\{ \Fmith \big\}$ are (approximately) jointly group
$P \NLam$-sparse with respect to the 1D partition $\Part$.

\vspace{.5mm}

\section{Compressive MIMO Channel Estimation\\[-.3mm]
Exploiting Joint Group Sparsity}\label{sec:Cha_Est}

\vspace{1mm}

The MGCS-based MC-MIMO channel estimator presented in this section
exploits the joint group sparsity
of doubly selective MC-MIMO
channels studied in Section \ref{sec:Del_Dop_Sp}. It generalizes
the estimators previously presented in
\cite{GT_jstsp10,DE_GT_icassp10,DE_GT_spawc10}.

\vspace{-.5mm}

\subsection{Subsampled Time-Frequency Grid and Pilot Arrangement}\label{sub-sec:Pilots} 

As in Section \ref{sec:Del_Dop_Sp}, we assume that the support of $\Fmi$ is
contained in $[0,D\!-\!1] \times [-J/2,J/2\!-\!1]$.
By the 2D discrete Fourier
transform (DFT) relation
in \eqref{System_Channel_Det}, the channel coefficient matrices $\Hlk$ are then
uniquely determined by their values on the \emph{subsampled time-frequency
grid}
\begin{align*}
\mathcal{G} &\ist\triangleq\ist \big\{(l,k) = (\lambda \,\Delta L, \kappa \,\Delta K)\, \big|\, \lambda\in\{0,\dots,J\!-\!1\},\\[0mm] 
&\hspace{50mm} \kappa\in\{0,\dots,D\!-\!1\} \big\} \,. \\[-6mm]
\end{align*}
Due to \eqref{System_Channel_Det}, these subsampled values are given
\vspace{-.5mm}
by
\begin{equation}
\label{System_Channel_Det_subs}
  \HH_{\lambda\,\Delta L, \kappa\,\Delta K} \eq \rmv\sum_{m=0}^{D-1}\sum_{i=-J/2}^{J/2-1} \!\!\Fmi \,
    e^{-\jmath 2 \pi ( \frac{\kappa m}{D}-\frac{\lambda i}{J} )} ,
\vspace{-.5mm}
\end{equation}
with jointly group sparse coefficient matrices $\Fmi$ (cf.\ Section \ref{sec:Del_Dop_Sp}).
However, the sparsity
is impaired by leakage effects. To
reduce them
and, thereby, improve the joint group
sparsity of $\Fmi$, we generalize \eqref{System_Channel_Det_subs} to an
orthonormal 2D basis
\vspace{-.5mm}
expansion
\be
\HH_{\lambda\,\Delta L, \kappa\,\Delta K}
\eq \rmv\sum_{m=0}^{D-1}\sum_{i=-J/2}^{J/2-1} \!\!\GGmi \,
    u_{m,i}[\lambda,\kappa] \,,
\label{Gen_Basis_Exp}
\vspace{-.5mm}
\ee
with some
orthonormal 2D basis $\{u_{m,i}[\lambda,\kappa]\}$. The construction of a basis yielding
improved joint group sparsity of the
$\GGmith \rmv\triangleq\rmv {[\GGmi]}_{r,s}$ (recall that $\tth\!=\!(\rr,\ss)$) will be studied
in Section \ref{sec:Bas_Opt}. Clearly,
the 2D DFT \eqref{System_Channel_Det_subs} is a special case of
\eqref{Gen_Basis_Exp} with $\GGmi=\sqrt{J\rmv D} \,\Fmi$ and
$u_{m,i}[\lambda,\kappa]=\frac{1}{\sqrt{J\rmv D}} \, e^{-\jmath 2 \pi
(\frac{\kappa m}{D} - \frac{\lambda i}{J})} $.

\begin{figure}[t]
\vspace*{2mm}
\centering
\hspace*{-7.5mm}\includegraphics[width=6.5cm]{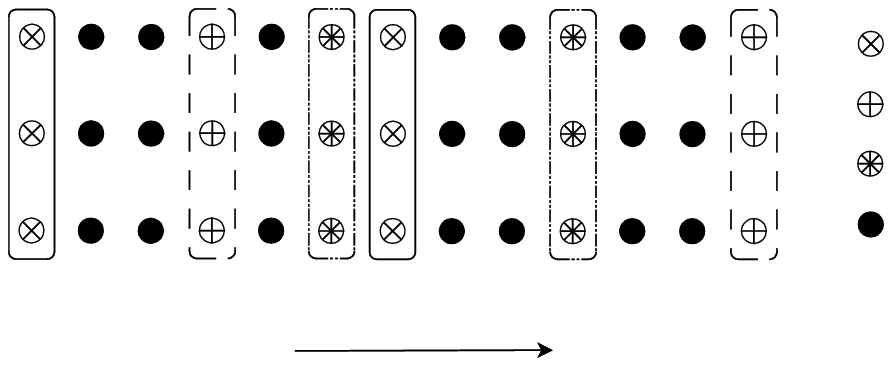}\scriptsize
\put(-205,66){$\ss\!=\!1$}
\put(-205,45.5){$\ss\!=\!2$}
\put(-205,25){$\ss\!=\!3$}
\put(4,64.9){$(l,k) \!\in\! \mathcal{P}^{(1)}$}
\put(4,52.1){$(l,k) \!\in\! \mathcal{P}^{(2)}$}
\put(4,39.2){$(l,k) \!\in\! \mathcal{P}^{(3)}$}
\put(4,26.5){data symbol}
\put(-139,0){$l,k$}
\put(-186,12){$\mathbf{p}^{(1)}$}
\put(-148,12){$\mathbf{p}^{(2)}$}
\put(-123.5,12){$\mathbf{p}^{(3)}$}
\put(-109.5,12){$\mathbf{p}^{(1)}$}
\put(-34,12){$\mathbf{p}^{(2)}$}
\put(-72.5,12){$\mathbf{p}^{(3)}$}
\normalsize
\vspace*{-1mm}
\caption{Example of a pilot arrangement
for
$\NT \rmv=\rmv 3$ transmit antennas. For each $\ss \in \{1,2,3\}$, the same length-$3$ pilot vector $\ps$ is transmitted at
all time-frequency positions $(l,k) \in \mathcal{P}^{(\ss)}\rmv$. Note that
$\Pil= |\Pilsettra| \rmv=\rmv 2$.}
\label{fig.pilots}
\vspace{-1mm}
\end{figure}

Let $\mu\rmv\triangleq\rmv (l,k)$ index the (nonsubsampled) time-frequency positions.
For pilot-aided channel estimation, we choose $\NT$ linearly independent pilot vectors $\ps\!\triangleq\rmv \big(p_1^{(\ss)} \rmv\cdots\ist p_{\NT}^{(\ss)}\big)^T\!$,
$\ss\in$\linebreak 
$\{1,\dots,\NT\}$ and $\NT$ pairwise disjoint sets of pilot time-frequency positions $\Pils\!$, $\ss\in\{1,\dots,\NT\}$.
The
$\Pils$ are \nolinebreak 
sub\-sets of the subsampled time-frequency grid $\mathcal{G}$, i.e., $\Pilsettra \!\rmv\subseteq\rmv \Grid$,
with
equal size $\Pil\triangleq | \Pilsettra |$.
Let $\mu_q^{(\ss)}\!$, $q\in\{1,\dots,\QQ\}$ denote the pilot time-frequency positions in $\Pils\!$.
For each $\ss \in \{1,$\linebreak 
$\dots,\NT\}$, the
pilot vector $\ps$ is transmitted at all $\QQ$ time-frequency positions $\mu_q^{(\ss)} \rmv\!\in\!\Pils\!$, i.e.,
$\aa_{\mu_q^{(\ss)}}\!=\!\ps$ for all $q \in\{1,$\linebreak 
$\dots,\QQ\}$ in \eqref{trans-symbol}.
Thus, $\NT \QQ$ pilot vectors,
or $\NT^2\QQ$ pilots symbols,
are transmitted in total. An example of such a pilot arrangement is shown in Fig.\ \ref{fig.pilots}.
Note that the individual entries of
$\aa_{\mu_q^{(\ss)}}\!=\!\ps$ correspond to pilots at the same time-frequency position, but transmitted from different antennas.

\vspace{-1.5mm}

\subsection{The Estimator}\label{sub-sec:Estimator} 

Let $y^{(\rr)}_{\mu} \rmv\!\triangleq\rmv {[\yy_{\mu}]}_{\rr}$ (cf.\ \eqref{rec-symbol})
denote the demodulated symbol at receive antenna $\rr$ and time-frequency position $\mu$, with
$z^{(\rr)}_{\mu} \rmv\!\triangleq {[\zz_{\mu}]}_{\rr}$ (cf.\ \eqref{gen.io.rel}) the associated noise component.
Furthermore let
\be
\Gmiantti \ist\triangleq\, \mathbf{g}^{(\rr)T}_{m,i} \pptra ,
\label{eq_G-k}
\vspace{-1mm}
\ee
where $\mathbf{g}^{(\rr)T}_{m,i}\!$ denotes the $\rr\ist$th row of $\GGmi$.
Then, writing $\mu_q^{(\ss)}\rmv\triangleq (\lambda_q^{(\ss)} \Delta L, \kappa_q^{(\ss)} \Delta K)$ for the pilot time-frequency positions
and using \eqref{gen.io.rel}, \eqref{Gen_Basis_Exp}, and \eqref{eq_G-k}, we obtain
\begin{align}
y^{(\rr)}_{\mu_q^{(\ss)}} &\,=\, {\big[\yy_{\!\mu_q^{(\ss)}}\big]}_{\rr} \nonumber\\
&\ist\overset{(\ref{gen.io.rel})}{=}\, {\big[\HH_{\mu_q^{(\ss)}} \pptra + \zz_{\mu_q^{(\ss)}}\big]}_{\rr}\nonumber\\
&\rmv\overset{(\ref{Gen_Basis_Exp})}{=} \Bigg[\Bigg(\sum_{m=0}^{D-1}\sum_{i=-J/2}^{J/2-1} \rmv\!\GG_{m,i}\,u_{m,i}\big[\lambda_q^{(\ss)}\rmv,\kappa_q^{(\ss)}\big] \rmv\Bigg) \ist\pptra
  + \zz_{\mu_q^{(\ss)}} \Bigg]_{\rr} \nonumber \\
&\,=\, \sum_{m=0}^{D-1} \sum_{i=-J/2}^{J/2-1} \!\mathbf{g}_{m,i}^{(\rr)T} \pptra \ist u_{m,i}\big[\lambda_q^{(\ss)}\rmv,\kappa_q^{(\ss)}\big] \ist+\ist z_{\mu_q^{(\ss)}}^{(\rr)}\nonumber\\
&\rmv\overset{(\ref{eq_G-k})}{=} \sum_{m=0}^{D-1} \sum_{i=-J/2}^{J/2-1} \!\rmv \Gmiantti \,u_{m,i}\big[\lambda_q^{(\ss)}\rmv,\kappa_q^{(\ss)}\big] \,+\, z^{(\rr)}_{\mu_q^{(\ss)}} \,,
  \label{System_Channel_Det_pil}\\[-6.5mm]
\nonumber
\end{align}
for all
$q\in\{1,\dots,\QQ\}$, $\ss\in\{1,\dots,\NT\}$, and $\rr\in\{1,\dots,\NR\}$.
To rewrite this relation in vector-matrix notation, we define $\UU$ to be the unitary $J\rmv D \times\! J\rmv D$
matrix whose $(mJ+ i \rmv+ J/2 + 1)\ist$th column 
is given by
$\vecr_{\lambda,\kappa}\big\{ u_{m,i}[\lambda,\kappa] \big\}
= \big( u_{m,i}[0,0]$\linebreak 
$u_{m,i}[0,1] \,\cdots\, u_{m,i}[1,0] \,\cdots\, u_{m,i}[J\!-\!1,D\!-\!1] \big)^T\!$
(this denotes the rowwise stacking with respect to $\lambda,\kappa$ of the $J
\!\times\! D$ ``matrix'' $u_{m,i}[\lambda,\kappa]$ into a $J\rmv
D$-dimensional vector). Furthermore, we set
\vspace*{-.5mm}
\be
\bPhitra \triangleq\ist
\sqrt{\frac{\JD}{\Pil}} \, \UUtra \in\ist \mathbb{C}^{\QQ\times J\rmv D} ,
\label{eq_Usdef}
\vspace{.5mm}
\ee
where $\UU^{(\ss)}\rmv$ denotes the $\QQ \!\times\!
J\rmv D$ submatrix of $\UU$ constituted by the $\QQ$ rows corresponding to
the pilot positions $\mu_q^{(\ss)}\rmv\!\in\rmv \Pils\rmv$,
$q\in\{1,\dots,\QQ\}$. We also define the vectors
\be
\xxant \,\triangleq\, \sqrt{\frac{\Pil}{\JD}} \,
\vecr_{m,i}\big\{\Gmiantti\big\} \in\ist \mathbb{C}^{J\rmv D},
\label{eq_D-x-G}
\ee
as well as the vectors $\yy^{(\tth)} \triangleq \big( y^{(\tth)}_{1} \rmv\cdots\ist y^{(\tth)}_{\QQ} \big)^T \!\!\in \mathbb{C}^{\QQ}$ with
$y^{(\tth)}_q \!\triangleq y^{(\rr)}_{\mu_q^{(\ss)}}$
and $\zz^{(\tth)} \!\triangleq \rmv \big( z^{(\tth)}_{1} \rmv\cdots\ist z^{(\tth)}_{\QQ} \big)^T \!\! \in \mathbb{C}^{\QQ}$ with $z^{(\tth)}_q \rmv\triangleq z^{(\rr)}_{\mu_q^{(\ss)}}$.
We can then rewrite (\ref{System_Channel_Det_pil})
\vspace{-.5mm}
as
\be
\label{MIMO ChaEst eq vec}
\yy^{(\tth)} =\ist \bPhi^{(\ss)} \xx^{(\tth)} \rmv+ \zz^{(\tth)},  \quad \tth = (\rr,\ss) \in \Tth \ist.
\vspace{-.5mm}
\ee
Thus, we
obtained measurement equations of the form
\eqref{meas-eq-MCS}, of dimension $\QQ \times \! J \rmv D$ (i.e., $M \rmv=\rmv J \rmv
D$). In practice, $\QQ \rmv\ll\! J\rmv D$. Since the coefficients $\GGmith$ are
(approximately) jointly group sparse,
the functions $\Gmiantti = \mathbf{g}^{(\rr)T}_{m,i} \pptra\rmv$ and,
consequently, the vectors $\xx^{(\tth)}\rmv$ are (approximately) jointly
group sparse as well.
Therefore, \eqref{MIMO ChaEst eq vec} is recognized as an instance of the MGCS problem introduced in Section
\ref{sub-sec:MGCS}, and thus any MGCS reconstruction method
can be used to reconstruct the $\xx^{(\tth)}\rmv$.

We can now state the overall channel estimation algorithm:
\begin{enumerate}

\vspace{.5mm}

\item Stack the demodulated symbols at the pilot positions, $y^{(\rr)}_{\mu_q^{(\ss)}}$, into 
vectors $\yy^{(\tth)}$ (see
above)
and obtain estimates $\hat\xx^{(\tth)}$ of the
$\xx^{(\tth)}$ via an MGCS reconstruction method based on the measurement matrices $\bPhi^{(\ss)}\!$.

\vspace{1mm}

\item Rescale these estimates $\hat\xx^{(\tth)}$ with
    $\sqrt{\JD/\Pil}$ to obtain estimates
    $\hat{\tilde{G}}_{m,i}^{(\ant)}$ of $\Gmiantti$, i.e., calculate
    $\vecr_{m,i}\{\hat{\tilde{G}}_{m,i}^{(\ant)}\} = \sqrt{\JD/\Pil} \, \hat{\xx}^{(\ant)}$ (cf.\ the definition of $\xx^{(\ant)}$ in \eqref{eq_D-x-G}).

\vspace{.5mm}

\item Calculate 
\vspace{-1mm}
(cf.\ \eqref{eq_G-k})
\[
\hat{\mathbf{g}}^{(\rr)}_{m,i} \ist=\ist \PP^{-T} \big(\hat{\tilde{G}}_{m,i}^{(\rec,1)} \cdots\, \hat{\tilde{G}}_{m,i}^{(\rec,\NT)}\big)^T
\vspace{-.5mm}
\] 
with
$\PP \triangleq \big( \pp^{(1)} \cdots\ist \pp^{(\NT)} \big)$.
Note that
$\PP$ is nonsingular as
the
$\ps\rmv$ were chosen linearly independent.

\vspace{1mm}

\item From $\hat{\mathbf{g}}^{(\rr)}_{m,i}$, calculate estimates $\hat{\HH}_{\lambda\,\Delta L, \kappa\,\Delta K}$ of the subsampled channel coefficient matrices
                   $\HH_{\lambda\,\Delta L, \kappa\,\Delta K}$ according to \eqref{Gen_Basis_Exp} with $\GG_{m,i}$ replaced by $\hat{\GG}_{m,i}$.

\vspace{1mm}

\item Calculate estimates of the 2D DFT coefficients $\Fmi$
according to the inversion of \eqref{System_Channel_Det_subs},
\vspace{-1.5mm}
i.e.,
\be
\label{step5}
\hat{\mathbf{F}}_{\rmv m,i} \ist=\ist \frac{1}{\JD} \sum_{\lambda=0}^{J-1} \sum_{\kappa=0}^{D-1} \hat{\HH}_{\lambda\,\Delta L, \kappa\,\Delta K}
  \, e^{-\jmath 2 \pi (\frac{i \lambda}{J} - \frac{m \kappa}{D})}
  \vspace{-1mm}
\ee
for $\mrange$ and $\irange$. Set $\hat{\mathbf{F}}_{\rmv m,i} \rmv=\rmv \mathbf{0}$ otherwise.

\vspace{1mm}

\item Calculate estimates $\HHlkhat$ of all
$\Hlk$ by using the 2D DFT expansion \eqref{System_Channel_Det} with $\FFmi$ replaced by $\hat{\mathbf{F}}_{m,i}$.

\vspace{1.5mm}
\end{enumerate}

In the special case where the 2D DFT basis is used,
steps 4 and 5 can be omitted because
$\FFmi \rmv=\rmv \GGmi/\sqrt{\JD}$.

According to their definition in \eqref{eq_Usdef},
the measurement matrices $\bPhi^{(\ss)}$ are constructed by selecting those $|\Pils| \!=\! \QQ$ rows of the scaled unitary matrix
$\sqrt{\JD/\Pil} \, \UU$
that correspond to the pilot positions $\mu_q^{(\ss)}\rmv\!\in\rmv \Pils\!$.
Therefore, motivated by
the construction of measurement matrices described in Section \ref{sub-sec:GCS},
we choose these rows---or, in other words, the pilot positions
$\mu_q^{(\ss)}\rmv$---uniformly at random from the subsampled grid $\mathcal{G}$.
More precisely, we first choose a subset of $\mathcal{G}$ of size $\NT \QQ$ uniformly at random,
and then we partition it into $\NT$ pairwise disjoint sets $\Pilsettra\!$, $\ss \rmv\in\rmv \{1,\dots,\NT\}$ of equal size $\QQ$.
This construction differs from the construction of measurement matrices explained in Section \ref{sub-sec:GCS}
in that here the pilot sets $\Pilsettra$ (i.e. the rows of the unitary matrix $\sqrt{\JD/\Pil} \,\UU$) have to be chosen pairwise disjoint, 
which contradicts the assumption underlying \eqref{RIP_unitary} that each row of $\sqrt{\JD/\Pil} \,\UU$ be chosen with equal probability.
Unfortunately, an analysis of the G-RIC for this exact scenario does not seem to exist.
Nevertheless, we can expect $\bPhitra$ to satisfy the group restricted isometry property with a small G-RIC with high probability (as explained in Section \ref{sub-sec:GCS}) if the pilot sets are chosen sufficiently large.
We also note that CS channel estimation methods have been shown to be robust
to the choice of the pilot pattern
\cite{Lampe13}.

The pilot positions are chosen
in a design phase before the start of data transmission
and then remain fixed.
Therefore, once pilot sets $\Pilsettra$ yielding matrices $\bPhitra$ with ``good'' MGCS reconstruction properties are
found, they can be
used for all future data transmissions.

\vspace{-.5mm}

\subsection{Performance Analysis} \label{sub-sec:Perf_Ana}

We next present upper bounds on the
estimation error of our
channel estimator.
Let $\Part \rmv=\rmv {\{\Group_\gridx\}}_{\gridx=1}^{\grnum}$ be the partition used by
the MGCS reconstruction method in step $1$, let
$\bcalSi \rmv\subseteq\rmv \{1,\ldots,$\linebreak 
$\grnum \}$ comprise
the indices $\gridx$ of those $S$
groups $\Group_\gridx$ that contain the effective joint support of the
vectors $\vecr_{m,i}\big\{ \GGmith \big\}$, and let $\big(
\rmv\vecr_{m,i}\big\{ \GGmith \big\}\big)[\gridx] \rmv\in
\C^{|\Group_\gridx|}$ be composed
of those entries of $\vecr_{m,i}\big\{ \GGmith
\big\}$ whose indices are
in $\Group_\gridx$. Then, the
leakage of the vectors $\vecr_{m,i}\big\{ \GGmith \big\}$ outside $\bcalSi$
and, thus, the error of a joint group $S$-sparsity assumption (i.e., the
error incurred when the entries of the vectors $\vecr_{m,i}\big\{ \GGmith
\big\}$ outside these $S$ groups are set to zero) can be quantified by
\be 
\label{C_GSJ} 
C_{G,\bcalSi,\Part} \ist\triangleq\, \sum_{\gridx \ist\notin \bcalSi} \rmv \bigg( \sum_{\tth \in \Tth} \big\| \big(
\rmv\vecr_{m,i}\big\{ \GGmith \big\}\big)[\gridx] \big\|_{2}^2 \bigg)^{\!\rmv 1/2} .
\ee
We also define 
the root mean square error (RMSE) of channel estimation
\vspace{-1mm}
\be 
\label{RMSE}
E \,\triangleq\ist \Bigg( \sum_{\tth \in \Tth} \sum_{l=0}^{L-1} \sum_{k=0}^{K-1} \big| \hat{H}^{(\tth)}_{l,k} \rmv-\rmv H^{(\tth)}_{l,k} \big|^2 \Bigg)^{\!\rmv 1/2} ,
\ee
with $H^{(\tth)}_{l,k} \rmv\triangleq {[\HH_{l,k}]}_{\rr,\ss}$ and $\hat{H}^{(\tth)}_{l,k} \rmv\triangleq {[\hat{\HH}_{l,k}]}_{\rr,\ss}$.
In the following theorem, we consider the MGCS-based channel estimator described in Section \ref{sub-sec:Estimator},
where the MGCS reconstruction method in step $1$ is G-BPDN
or G-CoSaMP (see Section \ref{sub-sec:GCS})
operating in the \emph{MGCS mode}
(see Section \ref{sub-sec:MGCS}), so that joint
and group sparsity are leveraged simultaneously. The proof of the theorem is given in 
\pagebreak 
Appendix B.

\begin{theorem}\label{Theorem_MIMO}
Assume that the noise $\zzant$ in \eqref{MIMO ChaEst eq vec}
satisfies\footnote{\label{chi_tail} For 
independent and identically distributed zero-mean circularly symmetric complex Gaussian
$z^{(\tth)}_q$ with variance $\sigma^2_{\text{p}}$ at the pilot positions, we have 
${\rm P} \big\{\big(\sum_{\tth \in \Tth} {\|\zzant\|}_2^2\big)^{1/2} \!\rmv\leq\!
\epsilon \big\} \rmv\geq\rmv 1 \rmv-\rmv (a^2 e^{1-a^2})^{\NT \NR \QQ}$ for any $\epsilon =a \, \sigma_{\text{p}} \sqrt{\NT \NR \QQ} $ with $a>1$, so that the condition
is satisfied with ``overwhelming''
probability.} 
$\big( \sum_{\tth \in \Tth} {\|\zzant\|}_2^2\big)^{1/2} \rmv\leq\rmv \epsilon$ for some $\epsilon > 0$.
First, let the MGCS reconstruction method in step 1 be G-BPDN operating in MGCS mode.
If all 
$\bPhitra$ in \eqref{MIMO ChaEst eq vec} satisfy the
group restricted isometry property of order $2S$ with respect to $\Part$
with G-RIC $\delta_{2S|\Part}^{(\tra)} \rmv\leq\rmv \sqrt{2}-1$,
then
\vspace{-2.5mm}
 \begin{equation}
\label{th MGCE  BPDN}
E  \ist\leq\, C_0'  \frac{C_{G,\bcalSi,\Part}}{\sqrt{S}} + C_1' \epsilon \,,
\vspace{-1.5mm}
\end{equation}
with the constants $C_0'\deff c_0 \sqrt{\LK/(\JD)} \, \|\PP\| \ist \|\PP^{-1}\|$ and $C_1'\rmv\deff c_1 \sqrt{\LK/\Pil}\ist \|\PP^{-1}\|$, where
$c_{0}=\frac{2 \ist (1 \ist-\ist \delta_{2S|\Part})}{1 \ist-\ist (1\ist+\ist \sqrt{2})\ist\delta_{2S|\Part}}$,
\vspace{-.7mm}
$c_{1}=\frac{4\ist\sqrt{1 \ist+\ist \delta_{2S|\Part}}}{1\ist-\ist (1\ist+\ist \sqrt{2})\ist\delta_{2S|\Part}}$,
and $\norm{\cdot}$ denotes the spectral norm 
\cite{Horn99}.
Alternatively, assume that 
step 1 uses G-CoSaMP with $n$ iterations operating in MGCS mode.
If all 
$\bPhitra$ satisfy the group restricted isometry property
of order $4S$ with respect to $\Part$ with G-RIC
$\delta_{4S|\Part}^{(\tra)} \rmv\leq\rmv 0.1$,
\vspace{-1.5mm}
then
\begin{equation}
\label{theorem_est_error}
 E \ist\leq\, C_0'' \Big(1+\frac{1}{\sqrt{S}}\Big) \ist C_{G,\bcalSi,\Part} +\ist C_1'' \epsilon \ist+\ist C_2''(n) \,,
\vspace{-1.5mm}
\end{equation}
with the constants $C_0'' \deff 20 \ist \sqrt{\LK/(\JD)} \, \|\PP\| \ist \|\PP^{-1}\|$,
$C_1'' \deff 20 \ist \sqrt{\LK/\Pil}\ist \|\PP^{-1}\|$, and
$C_2''(n) \rmv\deff\rmv 2^{-n} \ist \|\PP\| \ist \|\PP^{-1}\| \ist \big( \LK/(\JD)$\linebreak 
$\times \sum_{\tth \in \Tth} \sum_{m=0}^{D-1}\sum_{i=-J/2}^{J/2-1} \absb{\Gmiant}^2 \big)^{1/2}\rmv$.

\vspace{2mm}


\end{theorem}

Note that $C_2''(n)$ can be made arbitrarily small by increasing the number $n$ of G-CoSaMP iterations.
Furthermore, as mentioned at the end of Section
\ref{sub-sec:Estimator}, the
$\bPhitra$ can be expected to satisfy
the group restricted isometry property with sufficiently small G-RICs
$\delta_{2S|\Part}^{(\tra)}$ and $\delta_{4S|\Part}^{(\tra)}$ (with high
probability) if the size of the pilot sets $\Pilsettra$ is
sufficiently large. Finally, while
the bounds \eqref{th MGCE BPDN} and \eqref{theorem_est_error} are
pessimistic in
general, i.e., the actual
estimation accuracy
is typically much better, they
express the dependence on various system
parameters and can therefore provide valuable design guidelines.

\subsection{Computational Complexity}\label{sub-sec:compl}

To analyze the complexity of the proposed method, we consider each step individually.
The complexity of step 1 depends on the MGCS algorithm used and will be denoted as $\mathcal{O}(\text{MGCS})$.
The rescaling performed in step 2 requires $\mathcal{O}\big(\NT\NR J\rmv D\big)$ operations.
The complexity of step 3 is $\mathcal{O}\big( \NT^2 \NR J\rmv D \big)$, because $\PP^{-T}$ is of size $\NT \times \NT$,
and thus each of the $\NR$ matrix-vector products has complexity $\mathcal{O}\big(\NT^2\big)$
(note that $\PP$ has to be inverted only once before the start of data transmission).
Evaluating \eqref{Gen_Basis_Exp} in step 4 in an entrywise (i.e., channel-by-channel) fashion
requires $\mathcal{O}\big(\NT\NR (J\rmv D)^2\big)$ operations; a more
efficient computation of \eqref{Gen_Basis_Exp} may be possible if the matrix
$\UU$ with columns $\vecr_{\lambda,\kappa}\big\{ u_{m,i}[\lambda,\kappa]
\big\}$ has a suitable structure. In step 5, again proceeding entrywise, the
calculation of \eqref{step5} can be performed efficiently in
$\mathcal{O}\big(\NT\NR J\rmv D \log(J\rmv D)\big)$ operations by using the
FFT. By the same reasoning, step 6 has complexity $\mathcal{O}\big(\NT\NR
K\rmv L \log(K\rmv L)\big)$. Therefore, the overall complexity of the
proposed channel estimator is obtained \vspace{-1mm} as \be
\label{MCE complexity} 
\mathcal{O}(\text{MGCS}) +\ist \mathcal{O}\big(\NT\NR \ist (\JD)^2\big) +\ist \mathcal{O}\big(\NT\NR\KL\log(\KL)\big) \,,
\vspace{-1mm}
\ee
since typically $\NT \rmv\ll\rmv \JD$ in practice.

Usually, the term $\mathcal{O}(\textnormal{MGCS})$ will dominate the overall complexity.
The complexity of the various MGCS algorithms depends on the implementation.
For G-CoSaMP (see Section \ref{sub-sec:GCS}), we have 
$\mathcal{O}(\textnormal{MGCS})= \mathcal{O}(\textnormal{G-CoSaMP})=n_{\textnormal{G-CoSaMP}} \ist\ist \mathcal{O}(\bPhi)$,
where $n_{\textnormal{G-CoSaMP}}$ is the number of G-CoSaMP iterations and $\mathcal{O}(\bPhi)$ denotes the complexity of
multiplying $\bPhi$ or $\bPhi^H\rmv$ by a vector of appropriate length. Taking advantage of the block-diagonal structure of $\bPhi$
(see \eqref{Phi_MCS_GCS} and \eqref{MIMO ChaEst eq vec}), we have $\mathcal{O}(\bPhi) \leq \NR\sum_{\tra=1}^{\NT} \mathcal{O}(\bPhitra)$ and, hence,
$\mathcal{O}(\textnormal{G-CoSaMP}) \leq n_{\textnormal{G-CoSaMP}} \ist \NR\sum_{\tra=1}^{\NT} \mathcal{O}(\bPhitra)$.
For G-DCS-SOMP (see Section \ref{sub-sec:MGCS}),
following the implementation of OMP in \cite{Blumensath_OMP}, the complexity in the special setting \eqref{MIMO ChaEst eq vec},
where only $\NT$ different matrices are involved, is
$\mathcal{O}(\textnormal{G-DCS-SOMP})=$\linebreak 
$n_{\textnormal{G-DCS-SOMP}} \ist \NR\sum_{\tra=1}^{\NT} \mathcal{O}(\bPhitra)
+ \mathcal{O}\big(\NT \JD \ist (n_{\textnormal{G-DCS-SOMP}}')^2 \big)$. Here,
$n_{\textnormal{G-DCS-SOMP}}$ denotes the number of G-DCS-SOMP iterations and
$n_{\textnormal{G-DCS-SOMP}}'$ denotes the sum of the cardinalities of the chosen groups.
Finally, we note that a complexity analysis of G-BPDN does not seem to be available.

As mentioned in Section \ref{sub-sec:Estimator}, if the 2D DFT  basis is used, steps 4 and 5 can be omitted;
the second term in \eqref{MCE complexity} then is replaced by $\mathcal{O}(\NT^2\ist \NR\ist \JD)$ (which is due to step 3). More importantly, also
the complexity of the MGCS algorithms is typically reduced, because
the vector-matrix products can be calculated using FFT methods.

\section{Basis Optimization}\label{sec:Bas_Opt}

We now consider the design
of
the
basis $\{u_{m,i}[\lambda,\kappa]\}$ with the
goal of maximizing the joint group sparsity of the coefficients $\GGmith$ in
the expansion \eqref{Gen_Basis_Exp}.
The proposed basis optimization methodology extends the methodology presented for 
single-channel, nonstructured sparsity
in \cite{GT_jstsp10}
to the 
case of joint group sparsity.
We note that separate extensions to group sparsity and joint sparsity individually were presented
in \cite{DE_GT_spawc10} and \cite{DE_GT_icassp10}, respectively.

\vspace{-1mm}

\subsection{Basis Optimization Framework}\label{sec:Bas_Opt_frame}

Following \cite{GT_jstsp10}, we
\vspace{-1mm}
set
\be
\label{eq:u_v}
u_{m,i}[\lambda,\kappa] \,\triangleq\, \frac{1}{\sqrt{D}} \,
v_{m,i}[\lambda] \, e^{-\jmath 2 \pi \frac{\kappa m}{D}} ,
\vspace{-1.5mm}
\ee
where ${\{ v_{m,i}[\lambda] \}}_{i=-J/2}^{J/2\rmv-1}$ is an orthonormal 1D basis for each $m \in \{0,\ldots,D\rmv-\!1\}$.
Note that with respect to its dependence on $\kappa$, $\{u_{m,i}[\lambda,\kappa]\}$ conforms to the 2D Fourier basis underlying \eqref{System_Channel_Det_subs};
this is motivated by the fact that the leakage effects
in the $m$ direction are relatively weak
(as noted below \eqref{leakage_kernel}), and thus little improvement of the joint group sparsity can be achieved by
optimizing the $\kappa$ dependence.
However, with respect to $\lambda$,  $\{u_{m,i}[\lambda,\kappa]\}$ uses optimized 1D basis functions $v_{m,i}[\lambda]$;
this accounts for
the fact that the leakage effects in the $i$ direction are relatively strong
(again as noted below \eqref{leakage_kernel}).

Our development is motivated by the channel model \eqref{impresp_cont} but does not require
knowledge of the
parameters $P$, $\eta_p^{(\tth)}\!$, $\tau_p^{(\tth)}\!$, and $\nu_p^{(\tth)}\!$. More specifically, 
we consider elementary single-scatterer channels
$h^{(\tth)}(t,\tau)= \delta(\tau-\tau_1^{(\tth)}) \, e^{\jmath 2 \pi \nu_1^{(\tth)} t}\rmv$, $\MCSsett$, where
$\tau_1^{(\tth)}$ and $\nu_1^{(\tth)}$ are 
modeled as random variables with
some
probability density function (pdf) $p(\ttau_1,\nnu_1) \rmv\triangleq\rmv
p \big(\tau_1^{(\tth_1)}\!,$\linebreak 
$\ldots,\tau_1^{(\tth_{\NT\NR})}\!,\nu_1^{(\tth_1)}\!,\dots,\nu_1^{(\tth_{\NT\NR})} \big)$
representing \emph{a priori} knowledge about the distribution of the delays and Doppler frequency shifts.
If such knowledge is unavailable, an
uninformative pdf is used, e.g., a uniform distribution on some feasible delay-Doppler
region.
We emphasize that the optimized basis is not restricted to a specular channel model of the form \eqref{impresp_cont}
but can be used for general doubly selective MIMO channels as defined in \eqref{rec-cont}.

To optimize the 1D basis functions $v_{m,i}[\lambda]$,
we consider a given 2D tiling ${\{\Block_\blidx\}}_{\blidx=1}^{\blnum}$ with
corresponding 1D partition $\Part = {\{\Group_\blidx\}}_{\blidx=1}^{\blnum}$
of $\{1,\dots,J\rmv D\}$ defined by the groups $\Group_\blidx =
\bijG(\Block_\blidx)$ (see Section \ref{sub-sec:Jo_Gr_Sp}).
We wish to find $\{v_{m,i}[\lambda]\}_{i=-J/2}^{J/2-1}$, $\mrange$ such that, for the
random single-scatterer channels $h^{(\tth)}(t,\tau)$ discussed above,
the vectors $\vecr_{m,i}\big\{ \GGmith \big\}$, $\MCSsett$ are maximally
jointly group sparse with respect to $\Part$ \emph{on average}.
Let  
$\mathbf{G} \in \mathbb{C}^{J\rmv D \times \NT\NR}$ denote the matrix with columns
$\vecr_{m,i}\big\{ \GGmith \big\}$, 
\vspace{-1mm}
i.e., 
\be
\label{Gmith}
{[\mathbf{G}]}_{\bijG(m,i),\xi} \rmv=G_{m,i}^{(\tth_\xi)}\,, \quad \xi \in \{1,\dots,|\Theta|\} = \{1,\dots,\NT\NR\} \ist.
\ee
Motivated by (M)GCS
theory---see Sections \ref{sub-sec:GCS} and \ref{sub-sec:MGCS}---we measure
the joint group sparsity of $\vecr_{m,i}\big\{ \GGmith \big\}$, $\MCSsett$
with respect to $\Part$ 
\vspace{-2mm}
by
\be
\label{eq:norm_F_J}
{\| \mathbf{G} \|}_{\textnormal{F}|\Part} \,\triangleq\,
\sum_{\blidx=1}^{\blnum} \rmv \Bigg( \sum_{(m,i) \in \Block_\blidx} \ist \sum_{\theta \in \Theta} \big|\GGmith\big|^2 \Bigg)^{\!\!1/2}.
\vspace{-.5mm}
\ee
We note for later use that this norm can also be written 
\vspace{-.8mm}
as
\be
\label{eq:norm_F_J_b}
{\| \mathbf{G} \|}_{\textnormal{F}|\Part} \ist= \sum_{\gridx=1}^{\grnum} {\|\mathbf{G}[\gridx]\|}_{\textnormal{F}} \,\ist,
\vspace{-1.5mm}
\ee
where $\mathbf{G}[\gridx] \in \mathbb{C}^{|\Group_\gridx| \times |\MCSset|}$
denotes the matrix
that is constituted by the rows of $\mathbf{G}$
indexed by $\Group_\gridx$ and ${\|\rmv\cdot\rmv\|}_{\textnormal{F}}$ denotes the Frobenius norm.
Furthermore, 
${\| \mathbf{G} \|}_{\textnormal{F}|\Part}
={\|\rmv\mathbf{g}\rmv\|}_{2,\tilde\Part}$, where
$\mathbf{g}\triangleq$\linebreak 
$\big(\mathbf{g}^{(\tth_1)T} \cdots\ist \mathbf{g}^{(\tth_{|\Tth|})T}\big)^T$ with
$\mathbf{g}^{(\tth)}\triangleq\vecr_{m,i}\big\{\GGmith\big\}$ corresponds
to the stacking explained in Section \ref{sub-sec:GCS} and the associated
partition $\tilde\Part$ of $\{1,\ldots,J\rmv D \NT\NR \}$ defined in
Section \ref{sub-sec:MGCS} is used.

We 
aim to minimize $\mathbb{E}\big\{ {\| \mathbf{G} \|}_{\textnormal{F}|\Part} \big\}$ (expectation 
with respect to $(\ttau_1,\nnu_1)$)
with respect to 
${\{ v_{m,i}[\lambda] \}}_{i=-J/2}^{J/2\rmv-1}\ist$, $m \in \{0,\ldots,D\rmv-\!1\}$.
This minimization can be rephrased as follows. Let $\VV \triangleq \textnormal{diag} \{\VV_0,\dots,\VV_{D-1}\} \in \C^{J\rmv D \times J\rmv D}$
be the unitary block diagonal matrix with 
$\VV_m \in \mathbb{C}^{J \!\times\! J}$ given by ${[\VV_m]}_{i+J/2+1,\,\lambda+1} \rmv\triangleq v^*_{m,i}[\lambda]$,
$i \in \{-J/2,\dots,J/2-1\}$, $\lambda \in \{0,\dots,J-1\}$.
Furthermore,
\vspace{-.7mm}
let
\begin{align}
&\hspace{-2mm}C^{(\nu)}[m,\lambda] \nonumber\\[.5mm]
&\hspace{-2mm}\;\triangleq \sum_{i=-J/2}^{J/2-1} \sum_{q=0}^{N-1} \psi^{(\nu)}(i+qL) \, A^*_{\gamma,g} \bigg(m,\frac{i+qL}{\Lr}\bigg) \, e^{\jmath 2 \pi \frac{\lambda i}{J}} \rmv, \!
\label{BasOpt_C}\\[-6.5mm]
\nonumber
\end{align}
with $\psi^{(\nu)}(i) \rmv\triangleq\rmv e^{\jmath \pi (\nu \Ts - \frac{i}{\Lr})(\Lr-1)} \psi(i-\nu \Ts \Lr)$, 
\pagebreak 
and define 
\[
\mathbf{c}_m(\tau,\nu) \ist\triangleq \sqrt{D} \, \phi^{(\nu)} \rmv\bigg(m-\frac{\tau}{\Ts}\bigg) \big( C^{(\nu)}[m,0] \ist\cdots\ist C^{(\nu)}[m,J \!-\! 1] \big)^T
\]
and, in 
\vspace{-.5mm}
turn, 
\[
\mathbf{c}(\tau,\nu) \ist\triangleq\ist \big( \mathbf{c}_0^T(\tau,\nu) \ist\cdots\ist \mathbf{c}_{D-1}^T(\tau,\nu) \big)^T \rmv.
\]
We evaluate $\mathbf{c}(\tau,\nu)$ at 
$\tau \rmv=\rmv \tau_1^{(\tth)}$ and 
$\nu \rmv=\rmv \nu_1^{(\tth)}$ for all 
$\tth$
and arrange the resulting vectors $\mathbf{c}\big(\tau_1^{(\tth_\xi)},\nu_1^{(\tth_\xi)}\big)$, $\xi \in \{1,\dots,\NT\NR\}$ into the matrix
\begin{align*}
\mathbf{C}(\ttau_1,\nnu_1) &\ist\triangleq\ist \Big( \mathbf{c}\big( \tau_1^{(\tth_1)}\!,\nu_1^{(\tth_1)}\big) \;\, \mathbf{c}\big(\tau_1^{(\tth_2)}\!,\nu_1^{(\tth_2)}\big) \\[0mm]
&\hspace{13mm}  \ist\cdots\ist\ist \mathbf{c}\big(\tau_1^{(\tth_{\NT\NR})}\!,\nu_1^{(\tth_{\NT\NR})}\big) \Big) \in \mathbb{C}^{\JD \times \NT\NR}\,.\\[-7mm]
\end{align*}
Then, it is shown in Appendix C
\vspace*{-1mm}
that
\begin{equation}
\label{BasOpt_Calc}
\mathbf{G} = \VV \mathbf{C}(\ttau_1,\nnu_1)\,,
\vspace*{-1mm}
\end{equation}
and thus we can rephrase our minimization problem as
\be
\label{bas_opt_0}
\hat{\VV}_{\!\text{opt}} \ist=\, \underset{\VV \in\ist
\mathcal{U}^{\hspace*{0.2mm}\text{bl}}}{\arg \min}\ \mathbb{E}\big\{ {\| \VV
\mathbf{C}(\ttau_1,\nnu_1) \|}_{\textnormal{F}|\Part} \big\} \,,
\vspace{-1mm}
\ee
where $\mathcal{U}^{\hspace*{0.2mm} \text{bl}}$ denotes the set of unitary block
diagonal $J\rmv D \times J\rmv D$ matrices with blocks of equal size $J\rmv \times \rmv J$ on the diagonal. Finally, with a view towards a numerical
algorithm, we
use the following Monte-Carlo approximation of
\eqref{bas_opt_0}:
\be
\label{bas_opt}
\hat{\VV} \ist=\, \underset{\VV \in\ist
\mathcal{U}^{\hspace*{0.2mm} \text{bl}}}{\arg \min}\, \sum_{\rho} {\| \VV
\mathbf{C}((\ttau_1,\nnu_1)_{\rho}) \|}_{\textnormal{F}|\Part} \,,
\vspace{-1.5mm}
\ee
where the $(\ttau_1,\nnu_1)_{\rho}$ denote samples of the random vector
$(\ttau_1,\nnu_1)$ independently drawn from its pdf $p(\ttau_1,\nnu_1)$.

\vspace{-1mm}

\subsection{Basis Optimization Algorithm}\label{sec:Bas_Opt_algo}

Because the set $\mathcal{U}^{\hspace*{0.2mm} \text{bl}}$ is not convex, the
minimization problem \eqref{bas_opt} is not convex. An approximate solution
can be obtained by an algorithm that extends the basis optimization algorithm
presented for single-channel, nonstructured sparsity in \cite{GT_jstsp10} to
the
case of joint group sparsity. 
We first exploit the fact that \eqref{bas_opt}---which is a minimization
problem of dimension $DJ^2$---can be decomposed into
$D/\Delta m'$ separate minimization problems of dimension $\Delta m' J^2$ each.
To obtain this decomposition, we first partition the set
$\{0,\ldots,D \mi 1\}$ into the $D/\Delta m'$ pairwise disjoint subsets
$\mathcal{M}_{\midx} \triangleq\{(\midx \!-\!1)\Delta m',(\midx \!-\!1)\Delta
m' + 1,\dots,\midx \Delta m' -1\}$, for $\midx\in\{1,\dots,D/\Delta m'\}$,
and we consider the $D/\Delta m'$ sets $\mathcal{A}_{\midx}$,
$\midx\in\{1,\dots,D/\Delta m'\}$ that consist of the indices $b$ of all
those $J/\Delta i'$ blocks $\Block_\blidx$ that contain pairs $(m,i)$ with $m
\in \mathcal{M}_{\midx}$, i.e., $\mathcal{A}_{\midx} \triangleq \big\{\blidx
\in \{1,\dots,\blnum\}\,\big|\, \exists \,m \!\in\! \mathcal{M}_{\midx}
\;\textnormal{such that}\; (m,i) \rmv\in\rmv \Block_\blidx \big\}$.
(In Fig.\ \ref{fig.blocks},
$\mathcal{A}_{\midx}$ corresponds to all
the blocks placed on top of each other in the $\midx$th vertical column within the
fundamental domain.) Now recall (cf.\ \eqref{eq:norm_F_J_b}) that
\vspace{-2mm}
\be\label{recall_norms}
\|\VV
\mathbf{C}(\ttau_1,\nnu_1)\|_{\textnormal{F}|\Part} \ist=\ist
\sum_{\blidx=1}^{\blnum} \big\|\big(\VV
\mathbf{C}(\ttau_1,\nnu_1)\big)[\blidx] \big\|_{\textnormal{F}}\,, 
\vspace{-.7mm}
\ee 
where
$\big(\VV \mathbf{C}(\ttau_1,\nnu_1)\big)[\blidx] \in
\mathbb{C}^{|\Group_\gridx| \times \NT\NR}$ denotes the matrix
that is constituted by the rows of $\VV \mathbf{C}(\ttau_1,\nnu_1)$
indexed by $\Group_\gridx = \bijG(\Block_\blidx)$. Furthermore, note that due to the block-diagonal structure of $\VV$, we have
$\VV \mathbf{C}(\ttau_1,\nnu_1) = \big( \big( \VV_0 \ist\mathbf{C}_0(\ttau_1,\nnu_1) \big)^{\rmv T}$\linebreak 
$\cdots\, \big(\VV_{D-1} \mathbf{C}_{D-1}(\ttau_1,\nnu_1) \big)^{\rmv T} \big)^{\rmv T}\!$ with
$\mathbf{C}_m(\ttau_1,\nnu_1) \triangleq$\linebreak 
$\Big( \mathbf{c}_m\big(\tau_1^{(\tth_1)} \!,\nu_1^{(\tth_1)}\big)\cdots\, \mathbf{c}_m\big(\tau_1^{(\tth_{\NT\NR})}\!,\nu_1^{(\tth_{\NT\NR})}\big) \Big)$.
It follows that each summand $\big\| \big(\VV
\mathbf{C}(\ttau_1,\nnu_1)\big)[\blidx] \big\|_{\textnormal{F}}$ in
\eqref{recall_norms} involves only the submatrices $\VV_m
\mathbf{C}_m(\ttau_1,\nnu_1)$ of $\VV \mathbf{C}(\ttau_1,\nnu_1)$ for those
$m$ for which there is an $\irange$ such that $\bijG(m,i) \rmv\in\rmv
\Group_\blidx\,$ (or equivalently $(m,i) \rmv\in\rmv \Block_\blidx\,$).
Therefore, for any fixed $\midx \rmv\in\rmv \{1,\dots,D/\Delta m'\}$, the set
of summands $\big\{ \big\| \big(\VV \mathbf{C}(\ttau_1,\nnu_1)\big)[\blidx]
\big\|_{\textnormal{F}} \big\}_{\blidx \in \mathcal{A}_{\midx}}$ involves
exactly the set of sub\-matrices $\big\{ \VV_m \mathbf{C}_m(\ttau_1,\nnu_1)
\big\}_{m \in \mathcal{M}_{\midx}}$. As a consequence, a reordering of the
summands in \eqref{recall_norms} yields 
\begin{align}
\sum_{\blidx=1}^{\blnum} \big\|\big(\VV
\mathbf{C}(\ttau_1,\nnu_1)\big)[\blidx] \big\|_{\textnormal{F}} \ist=
\sum_{\midx=1}^{D/\Delta m'} \!\!\sum_{\blidx \in \mathcal{A}_{\midx}}
\!\big\|\big(\VV \mathbf{C}(\ttau_1,\nnu_1)\big)[\blidx]
\big\|_{\textnormal{F}} \,. \nonumber \\[-3mm]
\label{reord_summ} \\[-8mm]
\nonumber 
\end{align}
Finally, using \eqref{recall_norms} and
\eqref{reord_summ}, the function minimized in \eqref{bas_opt} can be
developed as
\begin{align*}
&\sum_{\rho} {\| \VV \mathbf{C}((\ttau_1,\nnu_1)_{\rho}) \|}_{\textnormal{F}|\Part}\\[-1.5mm]
&\hspace{10mm}=\ist \sum_{\rho} \! \sum_{\midx=1}^{D/\Delta m'} \!\!\sum_{\blidx \in \mathcal{A}_{\midx}} \!\big\|\big( \VV \mathbf{C}((\ttau_1,\nnu_1)_{\rho}) \big)[\blidx] \big\|_{\textnormal{F}}\\[0mm]
&\hspace{10mm}= \sum_{\midx=1}^{D/\Delta m'} \!\!Y_{\midx}\big( {\{ \VV_m \}}_{m \in \mathcal{M}_{\midx}} \big) \,,\\[-9mm]
\end{align*}
with
\begin{align*}
&Y_{\midx}\big( {\{ \VV_m \}}_{m \in \mathcal{M}_{\midx}} \big)\\[1mm]
&\;\triangleq\, \sum_{\rho} \!\sum_{\blidx \in \mathcal{A}_{\midx}} \!\!\big\|\big( \VV \mathbf{C}((\ttau_1,\nnu_1)_{\rho}) \big)[\blidx] \big\|_{\textnormal{F}} \\[-1mm]
&\;=\, \sum_{\rho} \!\sum_{\blidx \in \mathcal{A}_{\midx}} \!\!\Bigg(  \sum_{\xi=1}^{\NT\NR} \!\sum_{(m,i)\in\Block_\blidx}
  \!\big|\big[\VV \mathbf{C}((\ttau_1,\nnu_1)_{\rho})\big]_{\bijG(m,i),\xi} \big|^2\Bigg)^{\!\!1/2}  \\[1mm]
&\;=\, \sum_{\rho} \!\sum_{\blidx \in \mathcal{A}_{\midx}} \!\Bigg( \sum_{\xi=1}^{\NT\NR} \!\sum_{m \in \mathcal{M}_{\midx}} \sum_{i:(m,i)\in\Block_\blidx} \\[-1.5mm]
&\hspace*{27mm}\big|\big[\VV_m \mathbf{C}_m((\ttau_1,\nnu_1)_{\rho})\big]_{i+J/2+1,\xi} \big|^2 \Bigg)^{\!\!1/2} \!. \\[-7mm]
\end{align*}
Because $Y_{\midx}\big( {\{ \VV_m \}}_{m \in \mathcal{M}_{\midx}} \big)$
involves only the
$\VV_m$ for $m \in \mathcal{M}_{\midx}$ and the
sets $\mathcal{M}_{\midx}$ are pairwise disjoint, the minimization problem
\eqref{bas_opt} reduces to the $D/\Delta m'$ separate
\vspace{-1mm}
problems
\be
\label{sep_min}
{\{ \hat{\VV}_m \}}_{m \in \mathcal{M}_{\midx}} =\, \underset{}{\arg \min}\;\ist Y_{\midx}\big( {\{ \VV_m \}}_{m \in \mathcal{M}_{\midx}} \big) \ist,
\vspace{-2mm}
\ee
for $ \midx \rmv\in\rmv \{1,\dots,D/\Delta m'\}$. Here, the
minimization is with respect to ${\{ \VV_m \}}_{m \in \mathcal{M}_{\midx}}\!$
with $\VV_m \in \mathcal{U}$, where $\mathcal{U}$ denotes the
\emph{nonconvex} set of unitary $J \times J$ matrices. Note that each
problem
\eqref{sep_min} is only of dimension $\Delta m' J^2$, since $|\mathcal{M}_{\midx}|=\Delta m'$ and $\VV_m \in
\C^{J \times J}\!$, whereas
problem \eqref{bas_opt} has dimension $D J^2\!$.\linebreak 
%
%
%
Typically, $\Delta m'$ is
small because $\phi^{(\nu^{(\tth)})}(m-\tau^{(\tth)}/\Ts)$ decays fast. The final matrix $\hat{\VV} \in
\mathbb{C}^{\JD \times \JD}$
 minimizing \eqref{bas_opt} is then given
as $\hat{\VV}=\textnormal{diag}\{\hat{\VV}_0,\dots,\hat{\VV}_{D-1}\}$.

To (approximately) solve \eqref{sep_min}, we use the fact that
a unitary matrix $\VV_m \!\in \mathcal{U}$ can be 
approximated as $\VV_m \rmv= e^{\jmath \AA_m} \rmv\approx \mathbf{I}_{J} + \jmath \AA_m$, where $\AA_m$ is a Her\-mitian $J \rmv\times\rmv J$ matrix
and $\mathbf{I}_{J}$ denotes the $J  \times J$ identity matrix \cite{Horn99}.
This approximation is good 
if $\AA_m$ is sufficiently ``small.'' Therefore, following \cite{GT_jstsp10}, 
we construct ${\{ \VV_m \}}_{m \in \mathcal{M}_{\midx}}\!$ iteratively
by performing a sequence of \emph{small} updates. To guarantee that the
iterated $\VV_m$ are 
unitary, we use the approximations $\VV_m \approx \mathbf{I}_{J} + \jmath \AA_m$ in the optimization criterion
but not for actually updating $\VV_m$. The resulting iterative basis
optimization algorithm is a straightforward adaptation of the algorithm
presented in \cite{GT_jstsp10} and will  be stated without a detailed
discussion. 
In iteration $\idxx \ge 1$, 
a standard convex optimization technique \cite{boyd_conv_opt01} is used to solve the \emph{convex}
\vspace{-1mm}
problem
\begin{align*}
&\hspace*{-2mm}{\big\{ \hat{\AA}_m^{(\idxx)} \big\}}_{m \in \mathcal{M}_{\midx}} \nonumber \\[0mm]
&\;\,=\!\!
    \underset{{\{ \AA_m \}}_{m \in \mathcal{M}_{\midx}} \rmv\in\ist \mathcal{A}^{(\idxx)}}{\arg \min} \!\! Y_{\midx}\big( {\big\{
    (\mathbf{I}_{J} + \jmath \AA_m )\VV_m^{(\idxx)} \big\}}_{m \in
    \mathcal{M}_{\midx}} \big)\ist, \!\!
\\[-6mm]
\nonumber
\end{align*}
where $\mathcal{A}^{(\idxx)}$ is
the set of all sets of $|\mathcal{M}_{\midx}| \!=\! \Delta m'$ Hermitian $J \!\times\! J$ matrices $\AA$ satisfying
$\max_{i,j} |{[\AA]}_{i,j}| \!<\rmv \varepsilon^{(\idxx)}$. Then, if
$Y_{\midx}\big( {\big\{ e^{\jmath \hat{\AA}_m^{(\idxx)}} \VV_m^{(\idxx)} \big\}}_{m \in \mathcal{M}_{\midx}} \big) <
Y_{\midx}\big( {\big\{  \VV_m^{(\idxx)} \big\}}_{m \in \mathcal{M}_{\midx}} \big)$, the algorithm sets
    $\VV_m^{(\idxx+1)} \!=\rmv e^{\jmath \hat{\AA}_m^{(\idxx)}} \VV_m^{(\idxx)}$, $m \!\in\! \mathcal{M}_{\midx}$
and $\varepsilon^{(\idxx+1)} \rmv=\varepsilon^{(\idxx)}$; otherwise
$\VV_m^{(\idxx+1)} \!=\rmv \VV_m^{(\idxx)}$, $m \!\in\! \mathcal{M}_{\midx}$  and
$\varepsilon^{(\idxx+1)} \rmv= \varepsilon^{(\idxx)}/2$.
The algorithm stops
either if the threshold $\varepsilon^{(\idxx)}$ falls below a prescribed value or after a prescribed
number of iterations.
It is initialized 
with an initial threshold $\varepsilon^{(1)}$
and initial matrices $\VV_m^{(1)}$ that are chosen as unitary
$J \!\times\! J$ DFT matrices. For this choice, the analysis in Section
\ref{sec:Del_Dop_Sp} shows that the coefficients $\GGmith = \sqrt{JD}
\,\Fmith$ are already jointly group sparse to a certain degree, so that it
can be expected that the algorithm converges to a ``good'' local minimum.

The algorithm
reduces the objective function in \eqref{bas_opt}
in each iteration. Although it only aims at maximizing joint group sparsity and does not take into
account CS-relevant
properties of the resulting measurement matrices $\bPhitra\rmv$
(in particular, $\mu_{\UU}$, cf.\ \eqref{RIP_unitary}),
our simulation results in Section \ref{sec:Sim_Res} demonstrate the excellent performance
of the
optimized basis. Note that the algorithm
has to be performed only once before the start of data transmission because it does not involve the receive signal.


\vspace{-2mm}

\section{Simulation Results}\label{sec:Sim_Res}

We present simulation results demonstrating the performance gains of the proposed MGCS channel estimator
relative to existing compressive channel estimators
\cite{GT_jstsp10,DE_GT_icassp10,DE_GT_spawc10}. We also consider the special case of a SISO system.

\vspace{-.5mm}

\subsection{Simulation Setup} \label{sub-sec:Sim_Setup}

We simulated
CP MIMO-OFDM systems with $\NT \rmv=\rmv \NR \in \{1,2,3,4\}$ transmit and receive antennas,
$K\!=\rmv 512$ subcarriers, 
symbol duration $N \rmv=\rmv 640$, CP length $N \rmv-\rmv
K\!=\rmv 128$, carrier frequency $f_0 \rmv=\rmv 5\,$GHz, transmit bandwidth
$1/\Ts \rmv= 5\,$MHz, and $L\!=\!32$ transmitted OFDM symbols.
The 
filters $f_1(t) \!=\! f_2(t)$ 
were
root-raised-cosine filters with roll-off factor $0.25$. The size of the pilot
sets was $\Pil \rmv=\rmv |\Pils| \rmv=\rmv 1024$.
Thus, the total number of pilot symbols was $\Pil\NT^2 = 1024 \cdot\rmv
\NT^2$, corresponding to a fraction of $6.25\cdot\NT\ist \%$ of all the
$KL\NT \rmv= 16.384\cdot\rmv \NT$ transmitted symbols. The pilot
time-frequency positions $\mu_q^{(\ss)}\rmv$ 
were chosen uniformly at random from a subsampled time-frequency grid $\mathcal{G}$ with spacings $\Delta L \!=\! 1$ and $\Delta K \!=\! 4$, and
partitioned into $\NT$
pairwise disjoint pilot time-frequency position sets $\Pils\!$, $\ss\in\{1,\dots,\NT$\}.
Note that $J=L/\Delta L = 32$ and $D=K/\Delta K=128$.
The pilot matrix
$\PP = \big( \pp^{(1)} \cdots\ist \pp^{(\NT)} \big)$ had a constant diagonal and was zero otherwise; the pilot (QPSK) symbol on the diagonal was
scaled such that its power was equal to the total power of
$\NT$ data (QPSK) symbols.

We used the geometry-based channel simulator
IlmProp \cite{ilmprop} to generate 500 realizations of a doubly selective
MIMO channel during blocks of $L\!=\!32$ OFDM symbols. Transmitter and receiver were separated by
about 1500\,m. 
Seven clusters of ten specular scatterers each were randomly placed
in an area of size 2500\,m $\!\times\!$ 800\,m; additionally, three clusters
of ten specular scatterers each were randomly placed within a circle of
radius 100\,m around the receiver. For each cluster and
the receiver, the speed was uniform
on $[0, 50]$\,m/s, the acceleration was
uniform on $[0,7]$\,m/s$^2$, and the angles
of the velocity and acceleration vectors were uniform on
$[0^{\circ}\rmv,360^{\circ})$.
In the MIMO case, the transmit antennas as
well as the receive antennas were spaced $c/(2 f_0)$ apart. The noise
$\zz[n]$ in \eqref{io_channel_impresp} was independent and identically distributed across time $n$ and the vector entries,
and circularly symmetric complex Gaussian with component variance $\sigma_z$ chosen such that a prescribed
signal-to-noise-ratio (SNR) was achieved. Here, the SNR is defined as
the mean received signal power averaged over one block of length $LN$ and all
receive antennas, divided by $\sigma_z^2$.


The reconstruction method employed by the proposed
channel estimator was G-BPDN or G-OMP in the SISO case and
G-BPDN (operating in MGCS mode, cf.\ Section \ref{sub-sec:MGCS}), DCS-SOMP, or G-DCS-SOMP in the MIMO case.
The performance of G-CoSaMP (not shown to avoid cluttered
figures) was observed to be intermediate between
G-BPDN and
G-DCS-SOMP.
The pdf 
for basis optimization
(see Section \ref{sec:Bas_Opt_frame}) was constructed as
$p(\ttau_1,\nnu_1) \!=\! p\big(\tau_1^{(\tth_1)}\!,\nu_1^{(\tth_1)}\big) \, p\big(\tau_1^{(\tth_2)} \!,$\linebreak 
$\nu_1^{(\tth_2)} \big| \tau_1^{(\tth_1)}\!,\nu_1^{(\tth_1)}\big) \ist\cdots\ist p\big(\tau_1^{(\tth_{\NT\NR})}\!,\nu_1^{(\tth_{\NT\NR})} \big| \tau_1^{(\tth_1)},\nu_1^{(\tth_1)}\big)$.
Here, the first factor was uniform on $[0,\tau_{\textnormal{max}}] \times [-\nu_{\textnormal{max}},\nu_{\textnormal{max}}]$,
where $\tau_{\textnormal{max}} \!=\rmv 25.6 \mu s$ is the CP length and $\nu_{\textnormal{max}} \!\approx\rmv 293\ist$Hz is
$3 \%$ of the subcarrier spacing.
The remaining factors were uniform on $\{0\} \times [-1.4,1.4]\ist$Hz,
i.e., the time delays of the individual component channels were 
equal whereas the Doppler frequency shifts 
differed by at most $\pm1.4$ Hz.
The channel estimation performance was measured by the empirical mean square error (MSE) normalized by the
mean energy of the channel coefficients.

\vspace{-.5mm}

\subsection{Performance Gains Due to Exploiting Group Sparsity}
\label{sub-sec:Sim_SISO}

For the SISO case, we compare the performance of the proposed compressive channel estimator leveraging group sparsity---i.e.,
using G-BPDN or G-OMP as GCS reconstruction method---with that of the conventional compressive channel estimator
using BPDN or OMP \cite{GT_jstsp10}.
\begin{figure}[t]
\vspace*{1mm}
\centering
\subfigure{
\includegraphics[width=5.4cm, height=4.4cm]{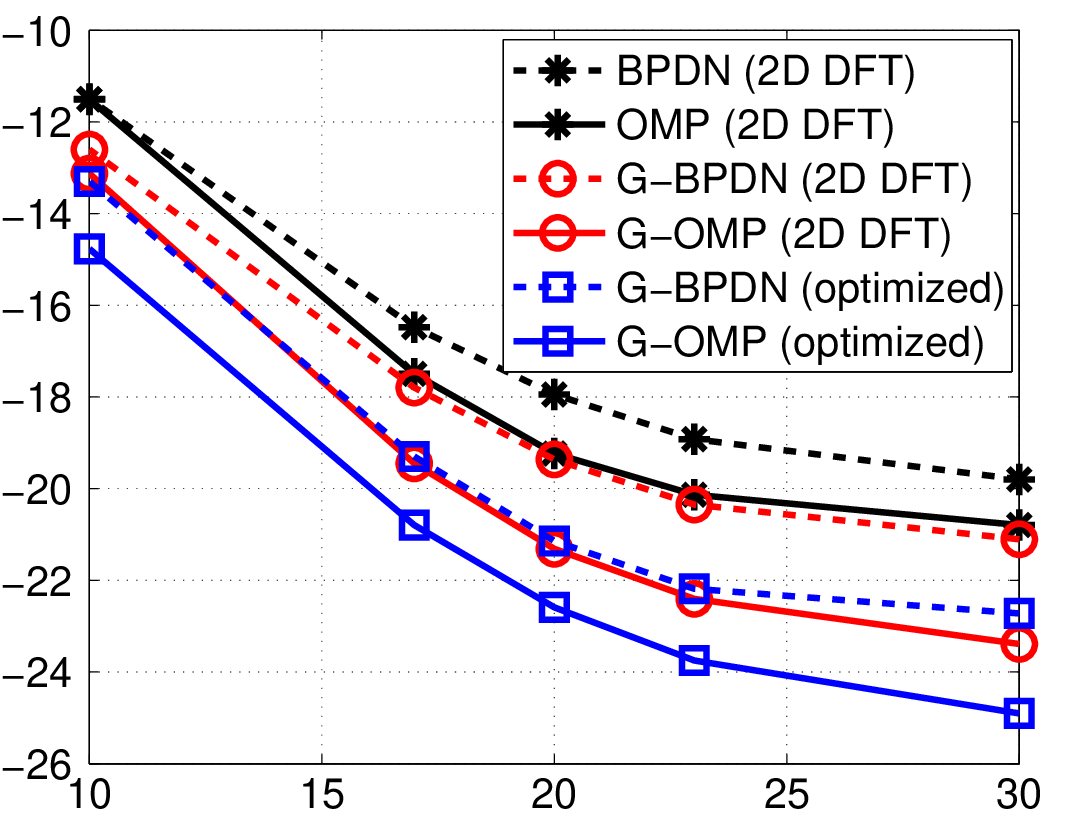} \label{fig:SISO_SNR}
\scriptsize
\put(-91,-7){SNR [dB]}
\put(-167,49){\begin{sideways}{\scriptsize MSE [dB]}\end{sideways}}   
\footnotesize
\put(-81.5,-20){(a)}
\normalsize
}
\qquad
\subfigure{
\includegraphics[width=5.4cm, height=4.4cm]{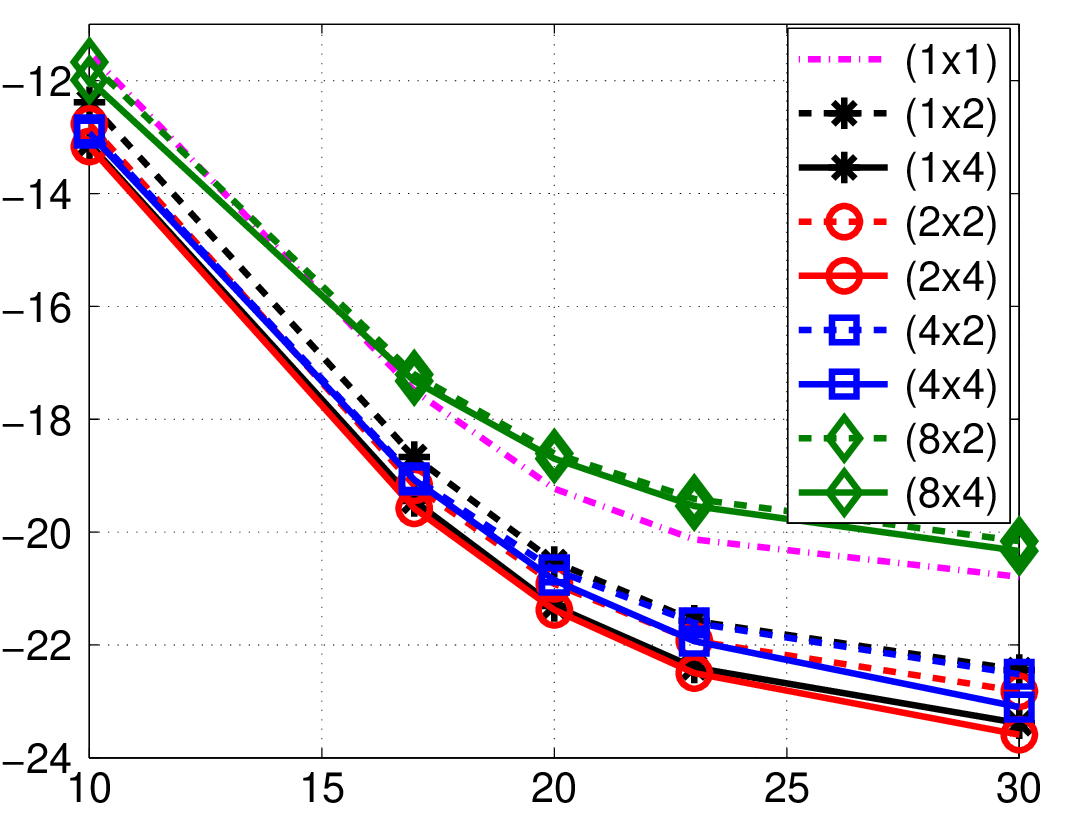} \label{fig:SISO_blocksize}
\scriptsize
\put(-91,-7){SNR [dB]}
\put(-167,45){\begin{sideways}{\scriptsize MSE [dB]}\end{sideways}}   
\footnotesize
\put(-81.5,-20){(b)}
\normalsize
}
\vspace{-.5mm}
\caption{MSE versus SNR for various compressive estimators in the SISO case:
(a) using different CS reconstruction methods, the 2D DFT basis and the optimized basis, and block size $\Delta m' \!\times\rmv \Delta i' = 1 \!\times\rmv 4$;
(b) using G-OMP, the 2D DFT basis, and different block sizes $\Delta m' \!\times\rmv \Delta i'$.}
\label{fig:SISO}
\end{figure}
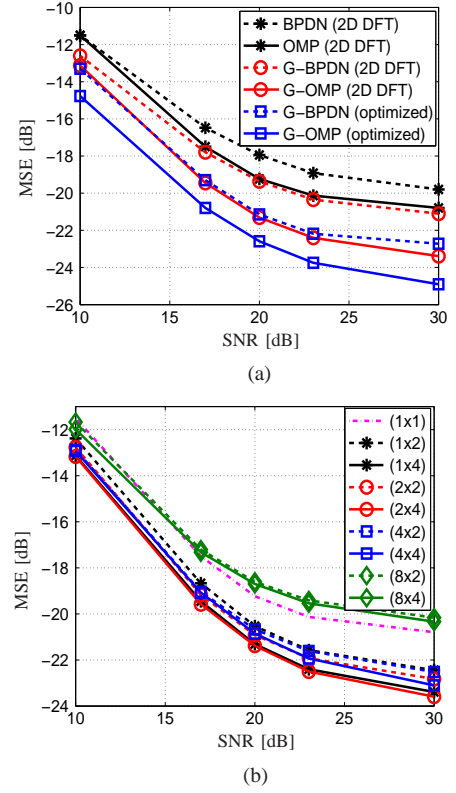
Fig.\ \ref{fig:SISO_SNR} shows the channel estimation MSE versus the SNR.
The blocks $\Block_\gridx$ of the delay-Doppler tiling
used in the definition of group sparsity (see Section \ref{sub-sec:Gr_Sp}) were of size $\Delta m' \!\times\rmv \Delta i' = 1
\!\times\rmv 4$. For the proposed channel estimator, we used both the 2D DFT basis and the optimized basis.
It is seen that exploiting
the inherent group sparsity of the channel yields a substantial reduction of the MSE,
and an additional substantial MSE reduction is obtained by using the optimized basis.

Fig.\ \ref{fig:SISO_blocksize} shows the MSE
versus the SNR for the proposed channel estimator using G-OMP,
the 2D DFT basis, and
different block sizes $\Delta m' \!\times\rmv \Delta i'$ (note that the case $\Delta m' \!\times\rmv \Delta i' = 1 \!\times\rmv 1$ corresponds to the
conventional compressive channel estimator of  \cite{GT_jstsp10}). One can observe a
strong dependence of the performance on the block size. This can be explained by the fact that if the blocks $\Block_\gridx$ are chosen too large in a certain direction,
many entries not belonging to the (effective) support of $\tilde{\xx}$ in \eqref{MIMO ChaEst eq vec} 
will be assigned nonzero values during reconstruction
since they belong to blocks containing some large entries.

\vspace{-1mm}

\subsection{Performance Gains Due to Exploiting
Joint Sparsity}
\label{sub-sec:Sim_MIMO}

Next, we consider the MIMO case.
We first compare our
channel estimator leveraging only joint sparsity
(hereafter referred to as MCS channel estimator, cf.\ also \cite{DE_GT_icassp10})
with the conventional compressive channel estimator.
At this point, the proposed estimator does not exploit
group sparsity; it uses G-BPDN
or DCS-SOMP, where G-BPDN  is based on blocks $\Block_\gridx$ of size $\Delta
m' \!\times\rmv \Delta i' = 1 \!\times\rmv 1$ but runs in MGCS mode
in order to exploit joint sparsity (this will be abbreviated as ``MG-BPDN-$1\!\times\!1$'').
The reason for choosing MG-BPDN-$1\!\times\!1$ instead of M-BPDN is its ability 
\pagebreak 
to handle the different measurement matrices $\bPhi^{(\MCSidx)}$ in
\eqref{meas-eq-MCS} 
(note that our application involves different measurement 
matrices, cf.\ \eqref{MIMO ChaEst eq vec}).
As a performance benchmark, we also consider a conventional compressive channel estimator that uses BPDN or OMP for each component channel individually.

\begin{figure}[t]
\vspace*{1mm}
\centering
\subfigure{
\includegraphics[width=5.4cm, height=4.4 cm]{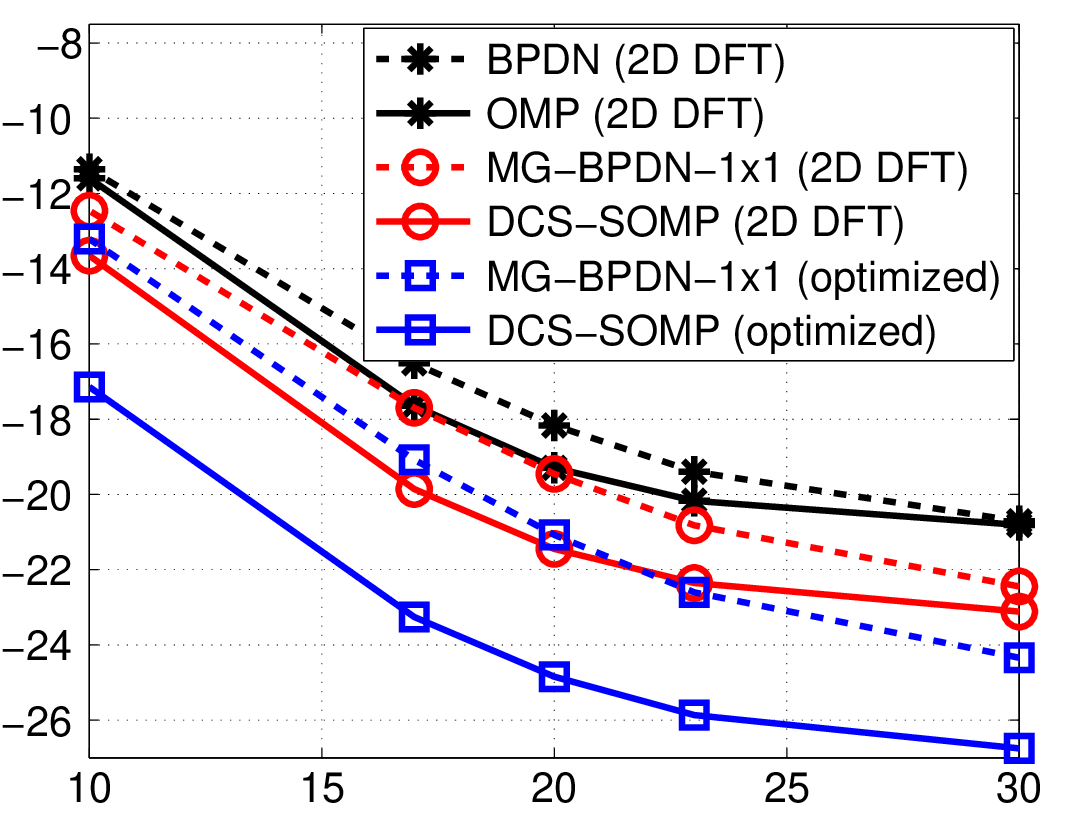} \label{fig:MIMO_MCS_SNR}
\scriptsize
\put(-91,-7){SNR [dB]}
\put(-167,51){\begin{sideways}{\scriptsize MSE [dB]}\end{sideways}}   
\footnotesize
\put(-81.5,-20){(a)}
\normalsize
}
\qquad
\subfigure{
\includegraphics[width=5.4cm, height=4.4 cm]{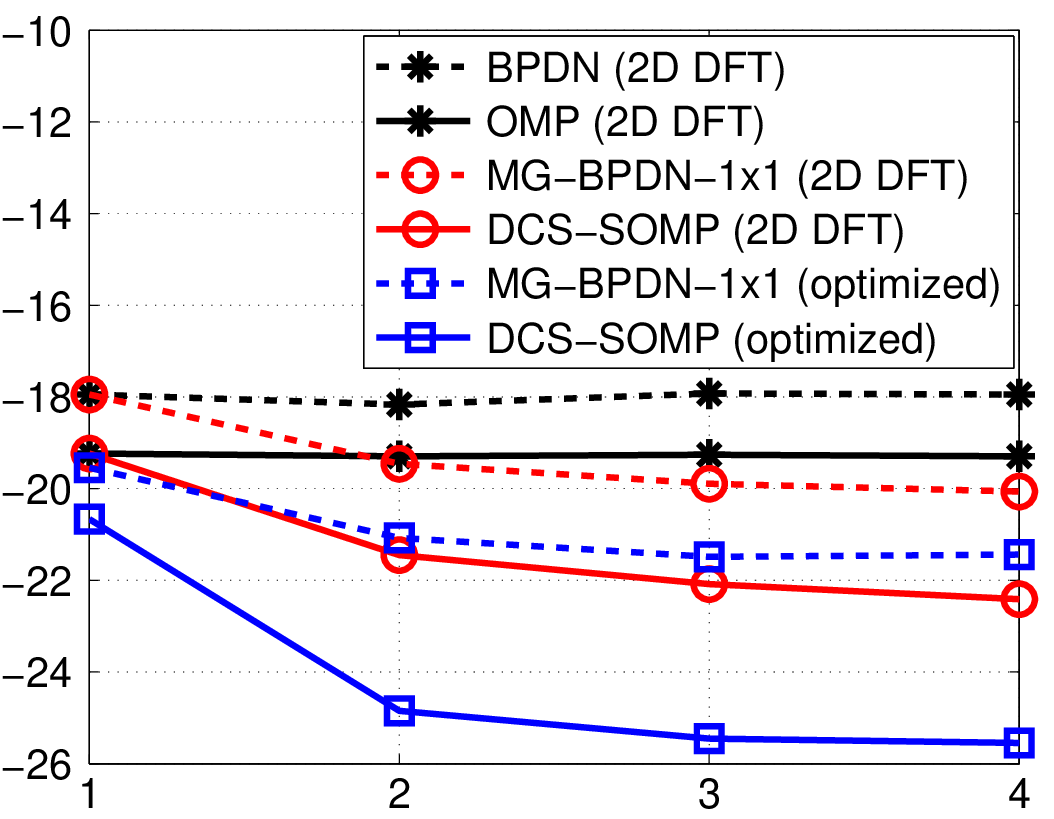} \label{fig:MIMO_MCS_Ant}
\scriptsize
\put(-91,-7){$\NT \!=\! \NR$}
\put(-167,49){\begin{sideways}{\scriptsize MSE [dB]}\end{sideways}}   
\footnotesize
\put(-81.5,-20){(b)}
\normalsize
}
\caption{MSE performance
(a)
versus SNR for a $2 \!\times\! 2$ MIMO system,
(b)
versus the number $\NT=\NR$ of transmit/receive antennas at an SNR of 20\,dB.}
\label{fig:MIMO}
\vspace{-1mm}
\end{figure}

\begin{figure}[t]
\vspace*{2mm}
\centering
\includegraphics[width=8.3cm]{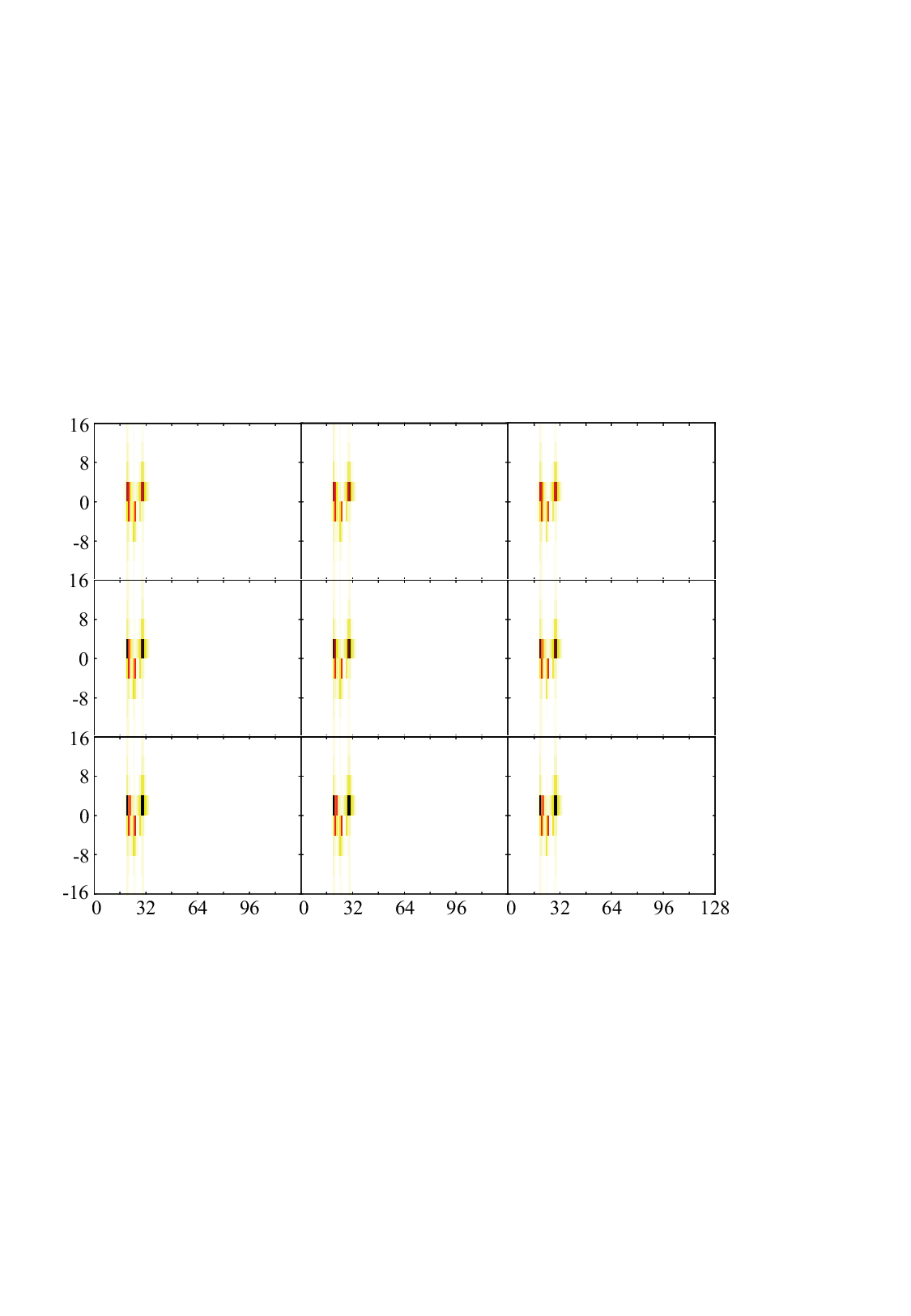}
\small
\put(-242,143.5){ $i$}
\put(-242,88.5){ $i$}
\put(-242,33){ $i$}
\put(-194,-8){ $m$} \put(-121,-8){ $m$} \put(-48,-8){ $m$}
\footnotesize
\put(-193,163){ $\theta_1\!=\!(1,1)$} \put(-120.5,163){ $\theta_2\!=\!(1,2)$} \put(-47,163){ $\theta_3\!=\!(1,3)$}
\put(-193,108){ $\theta_4\!=\!(2,1)$} \put(-120.5,108){ $\theta_5\!=\!(2,2)$} \put(-47,108){ $\theta_6\!=\!(2,3)$}
\put(-193,53){ $\theta_7\!=\!(3,1)$} \put(-120.5,53){ $\theta_8\!=\!(3,2)$} \put(-47,53){ $\theta_9\!=\!(3,3)$}
\normalsize
\vspace{1mm}
\caption{Visualization of the joint group sparsity of the 2D DFT
coefficients $\Fmith$ for a $3 \!\times\! 3$ MIMO system and
block size $\Delta m' \!\times\rmv \Delta i' = 1 \!\times\rmv 4$.}
\label{fig:jg-sparsity}
\vspace{-1mm}
\end{figure}
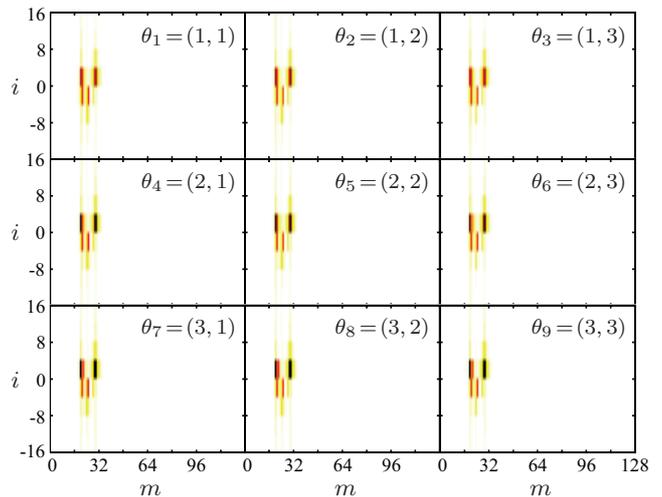

Fig.\ \ref{fig:MIMO_MCS_SNR} shows the
MSE versus the SNR for a MIMO system with $\NT \!=\! \NR \!=\! 2$ transmit/receive
antennas. It is seen that substantial reductions of the MSE are obtained by exploiting the channel's joint sparsity via MCS methods
and, additionally, by using the optimized basis.
Fig.\ \ref{fig:MIMO_MCS_Ant} shows the MSE versus 
$\NT \!=\! \NR \rmv\in \{1,2,3,4\}$ 
at an
SNR of 20\,dB.
The MSE is seen to
decrease for an increasing number of 
antennas.
This is because the estimation of the joint support becomes more accurate when a larger number of jointly sparse signals are available;
this behavior has been studied in \cite{HolgerAverageCaseAnalysis} for
M-BPDN and SOMP. The flattening of the MSE curves is caused by the fact that the component channels,
besides being jointly sparse in the sense of similar effective supports, are also similar with respect to the values of their nonzero entries.
As explained in \cite{HolgerAverageCaseAnalysis}, the case where all jointly sparse signals are equal is
a worst-case scenario for MCS, since no additional support information can be gained from additional signals.
In our case, this effect is alleviated since the jointly sparse signals are observed through
different measurement matrices
$\bPhi^{(\ss)}\!$, $\ss\in\{1,\dots,\NT\}$.
The MSE of the conventional compressive channel estimator is essentially
independent of the number of antennas.

\vspace{-.5mm}

\subsection{Performance Gains Due to
Exploiting Joint Group Sparsity}
\label{sub-sec:Sim_MIMO_Gr}

Next, we consider the case where the proposed channel estimator fully exploits
the available structure, 
i.e., the joint group sparsity of the
$\Fmith$ or $\GGmith$. Fig.\ \ref{fig:jg-sparsity} shows the energy of the 2D DFT
coefficients $\Fmith$ accumulated in
blocks $\Block_\gridx$ of size
$\Delta m' \!\times\rmv \Delta i' = 1 \!\times\rmv 4$ for the nine component channels
$\theta=$\linebreak 
$(1,1),(1,2),\ldots,(3,3)$ of a $3 \!\times\! 3$ MIMO system.
It is seen that the $\Fmith$ are effectively supported on the same blocks for all
$\theta$.
This demonstrates the strong available joint group sparsity,
and thus suggests that significant performance gains can be obtained by using MGCS channel estimation.

\begin{figure}[t]
\vspace*{1mm}
\centering
\subfigure{
\includegraphics[width=5.4cm, height=4.4 cm]{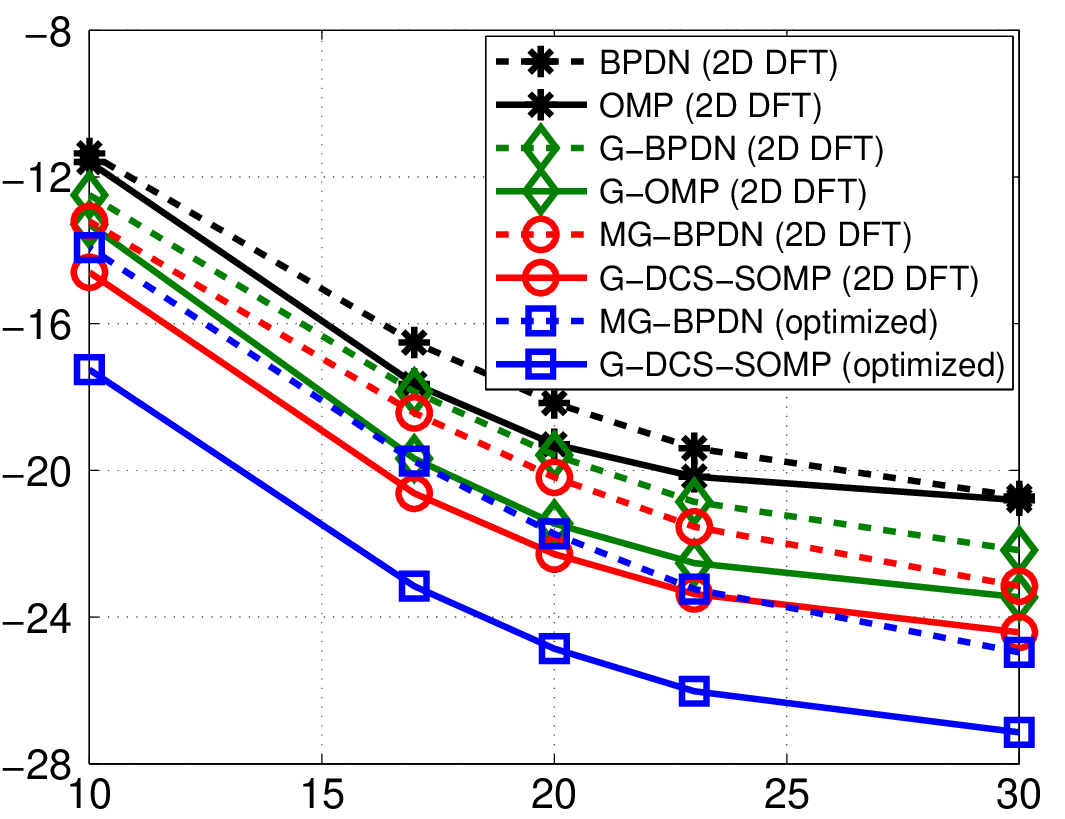} \label{fig:MIMO_MGCS_GCS}
\scriptsize
\put(-91,-7){SNR [dB]}
\put(-167,49){\begin{sideways}{\scriptsize MSE [dB]}\end{sideways}}   
\footnotesize
\put(-81.5,-20){(a)}
\normalsize
}
\qquad
\subfigure{
\includegraphics[width=5.4cm, height=4.4 cm]{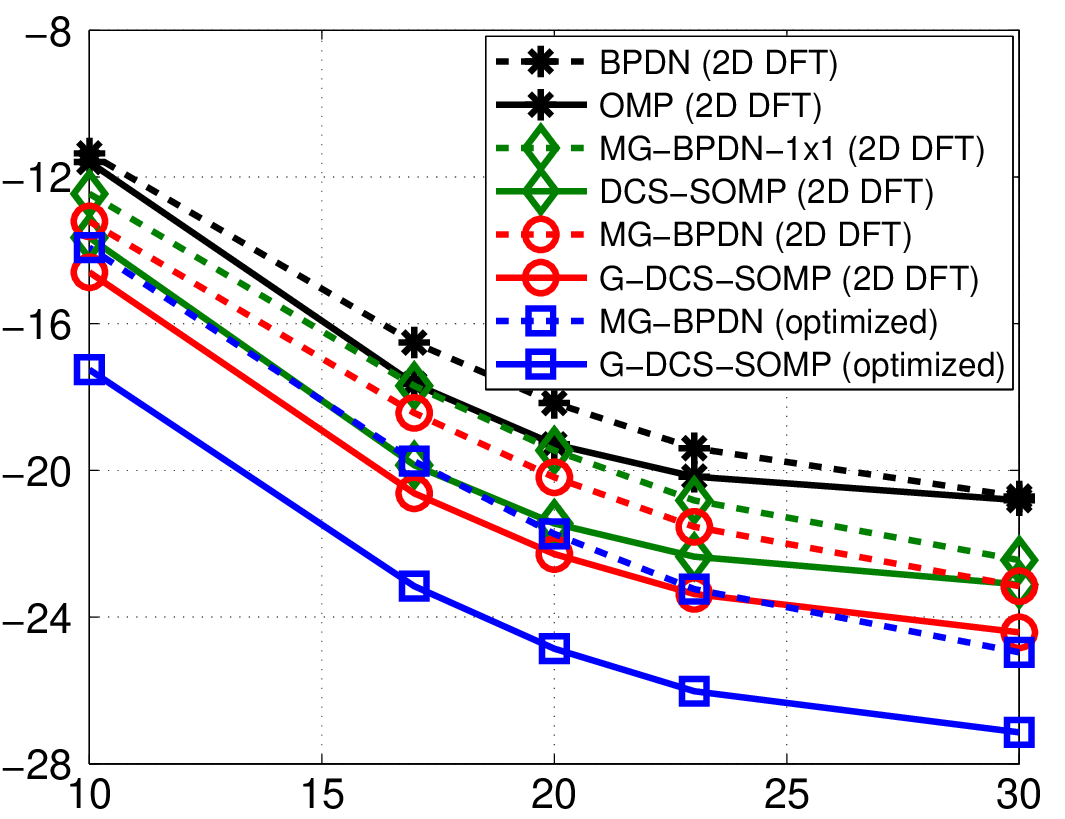} \label{fig:MIMO_MGCS_MCS}
\scriptsize
\put(-91,-7){SNR [dB]}
\put(-167,49){\begin{sideways}{\scriptsize MSE [dB]}\end{sideways}}   
\footnotesize
\put(-81.5,-20){(b)}
\normalsize
}
\caption{MSE versus SNR for
a $2 \!\times\! 2$ MIMO system:
(a) MGCS estimators, GCS estimators, and conventional compressive estimators;
(b) MGCS estimators, MCS estimators, and conventional compressive estimators.}
\label{fig:MIMO_MGCS}
\vspace{-1mm}
\end{figure}
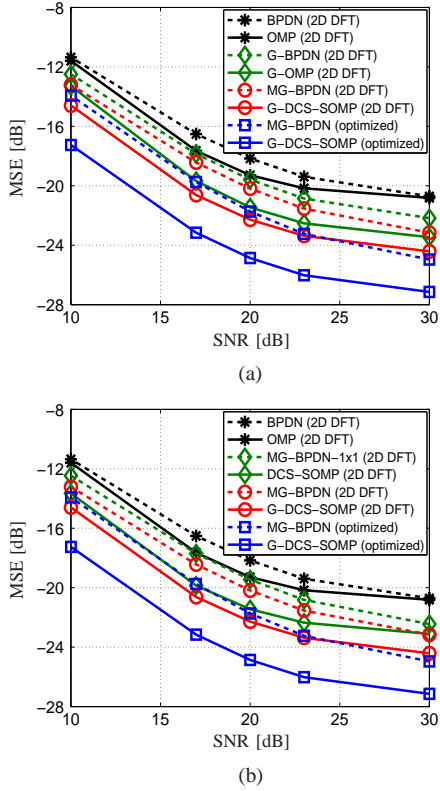

To assess the actual
gains, we simulated 
MGCS estimators,
GCS estimators (exploiting
only group sparsity),
MCS estimators (exploiting
only joint sparsity),
and 
conventional compressive estimators for a $2 \!\times\! 2$ MIMO system.
We used the 2D DFT basis
and, for the MGCS estimator, additionally an optimized basis.
Fig.\ \ref{fig:MIMO_MGCS} shows
the
MSE versus the SNR.
Both parts (a) and (b) show identical MSE curves for the 
MGCS estimators (using MG-BPDN or G-DCS-SOMP)
and the conventional compressive estimators (using BPDN or OMP); however, part (a) compares these curves with
the GCS estimators (using G-BPDN or G-OMP) and
part (b)
with
the MCS estimators (using MG-BPDN-$1\!\times\!1$ or DCS-SOMP). For the MGCS and GCS estimators, the
blocks $\Block_\gridx$
were of size $\Delta m' \!\times\rmv \Delta i' \!=\rmv 1 \!\times\rmv 4$.
It can
be seen\linebreak 
that exploiting group
or joint sparsity separately already outperforms conventional compressive channel estimation.
Moreover, 
substantial additional
 gains are obtained by exploiting the joint group sparsity structure through the proposed MGCS estimator,
and even larger gains are achieved when the MGCS estimator is used with an optimized basis.


\section{Conclusion}\label{sec:Concl}

For
multicarrier MIMO systems transmitting over doubly selective
channels, we demonstrated that leakage effects
induce
an approximate \emph{group} sparsity
structure of the individual component channels in the delay-Doppler domain.
We also
showed that the effective delay-Doppler supports of the
component channels overlap significantly, which implies that these channels
are
approximately \emph{jointly} group sparse. 
Using the methodology of multichannel group sparse compressed sensing (MGCS),
we then
devised a
compressive channel estimator that
leverages the
joint group sparsity structure.
We also presented an upper bound on the
MSE of this
estimator, and we analyzed the estimator's computational complexity. Furthermore,
we proposed an optimization of a basis involved in the estimator
that aims at maximizing
joint group sparsity.
We
presented an iterative approximate optimization algorithm
consisting of a sequence of convex programming problems.
Statistical information about the channel can be incorporated in this algorithm if available.

Simulations using a geometry-based channel simulator demonstrated substantial performance gains over
conventional compressive channel estimation.
Large gains can already be obtained by exploiting only group sparsity or joint sparsity,
and the combined MGCS approach yields an even larger
gain. An additional gain
results from the proposed basis optimization.


\section*{Acknowledgment}

The authors would like to thank 
H.\ Rauhut for helpful 
\vspace*{-1mm}
comments.

\section*{Appendix A:\, Proof of Inequality \eqref{norm_diffs_ineq} }\label{sec_appB}

Supposing, without loss of generality, that $\norm{\mathbf{a}}_2 \geq \norm{\mathbf{b}}_2$, the inequality \eqref{norm_diffs_ineq}
is equivalent 
\pagebreak 
to
\[
\norm{\frac{\mathbf{a}}{\norm{\mathbf{a}}_2}-\frac{\mathbf{b}}{\norm{\mathbf{b}}_2}}_2^2 \ist\leq\, \frac{\norm{\mathbf{a} \rmv-\rmv \mathbf{b}}_2^2}{\norm{\mathbf{b}}_2^2} \;,
\vspace{-.5mm}
\]
and, expanding the squared norms,
\vspace{-.5mm}
to
\[
2 - 2 \ist \frac{\mathbf{a}^T\mathbf{b}}{\norm{\mathbf{a}}_2 \rmv \norm{\mathbf{b}}_2}
    \,\leq\, \frac{\norm{\mathbf{a}}_2^2 + \norm{\mathbf{b}}_2^2 - 2\ist \mathbf{a}^T\mathbf{b}}{\norm{\mathbf{b}}_2^2} \;.
\]
Rearranging terms, this is furthermore equivalent to
\be
\label{norm_diffs_ineq_equiv}
\frac{\norm{\mathbf{a}}_2^2}{\norm{\mathbf{b}}_2^2} \ist\ist+\ist 2 \ist \frac{\mathbf{a}^T \mathbf{b}}{\norm{\mathbf{b}}_2}
  \rmv\left( \frac{1}{\norm{\mathbf{a}}_2}-\frac{1}{\norm{\mathbf{b}}_2}\right) \ist\geq\ist 1 \,.
\ee
To prove \eqref{norm_diffs_ineq_equiv}, we use the Cauchy-Schwarz inequality, noting that $\frac{1}{ \norm{\mathbf{a}}_2} - \frac{1}{\norm{\mathbf{b}}_2} \leq 0$:
\begin{align*}
&\frac{\norm{\mathbf{a}}_2^2}{\norm{\mathbf{b}}_2^2} \ist\ist+\ist 2 \ist \frac{\mathbf{a}^T \mathbf{b}}{\norm{\mathbf{b}}_2}
  \rmv\left( \frac{1}{\norm{\mathbf{a}}_2}-\frac{1}{\norm{\mathbf{b}}_2}\right)\\[1mm]
&\hspace{7mm}\geq\, \frac{\norm{\mathbf{a}}_2^2}{\norm{\mathbf{b}}_2^2} \ist\ist+\ist 2 \ist \frac{\norm{\mathbf{a}}_2 \rmv\norm{\mathbf{b}}_2}{\norm{\mathbf{b}}_2}
  \rmv\left( \frac{1}{\norm{\mathbf{a}}_2}-\frac{1}{\norm{\mathbf{b}}_2}\right)\\[1mm]
&\hspace{7mm}=\, \frac{\norm{\mathbf{a}}_2^2}{\norm{\mathbf{b}}_2^2} \ist\ist+\ist 2 \rmv\left( \rmv 1- \frac{\norm{\mathbf{a}}_2}{\norm{\mathbf{b}}_2}\right)\\[1mm]
&\hspace{7mm}=\, \frac{\norm{\mathbf{a}}_2^2}{\norm{\mathbf{b}}_2^2} \ist-\ist 2 \frac{\norm{\mathbf{a}}_2}{\norm{\mathbf{b}}_2} \ist+\ist 2\\[1mm]
&\hspace{7mm}=\ist \left( \frac{\norm{\mathbf{a}}_2}{\norm{\mathbf{b}}_2} -\rmv1 \right)^{\!\rmv 2} \rmv+ 1\\[1mm]
&\hspace{7mm}\geq\, 1 \ist.
\end{align*}

\section*{Appendix B:\, Proof of Theorem \ref{Theorem_MIMO} }\label{sec_appA}

\vspace{1mm}



Let $\mathbf{h}^{(\tth)} \!\triangleq \vecr_{l,k}\big\{H^{(\tth)}_{l,k}
\big\} \rmv\in\rmv \C^{K \rmv L}\!$, i.e., ${[\mathbf{h}^{(\tth)}]}_{k+lK+1} \!=\rmv
H^{(\tth)}_{l,k}$;\linebreak 
let $\mathbf{f}^{(\tth)} \rmv\triangleq \vecr_{m,i}\big\{\Fmith \big\} \in \C^{K \rmv L}$, i.e.,
${[\mathbf{f}^{(\tth)}]}_{mL+i+L/2+1} = \Fmith$; and let
$\mathbf{U}_{\textnormal{F}} \in \C^{K \rmv L\times K \rmv L}$ be the unitary
matrix with entries ${[\mathbf{U}_{\textnormal{F}}]}_{k+lK+1,\ist mL+i+L/2+1}
= \frac{1}{\sqrt{K\rmv L}} \, e^{-\jmath 2 \pi (\frac{km}{K}-\frac{li}{L})}$,
where $l\in\{0,\dots,L-1\}$, $k\in\{0,\dots,K-1\}$, $m\in\{0,\dots,K-1\}$,
and $i\in\{-L/2,\dots, L/2-1\}$. Then, \eqref{System_Channel_Det} can be
written as $\mathbf{h}^{(\tth)} = \sqrt{K\rmv L} \,
\mathbf{U}_{\textnormal{F}} \ist \mathbf{f}^{(\tth)}\rmv$,
which
\vspace{-1mm}
implies
\be
{\| \mathbf{h}^{(\tth)} \|}_2 \ist=\ist \sqrt{K\rmv L} \, {\| \mathbf{f}^{(\tth)} \|}_2 \,.
\label{eq_appA_h}
\vspace{-1mm}
\ee
Furthermore, let $\mathbf{h}^{(\tth)}_{\Delta} \!\triangleq \vecr_{\lambda,\kappa}\big\{H^{(\tth)}_{\lambda \Delta L,\kappa \Delta K} \big\} \!\in\! \C^{J \rmv D}\rmv$, i.e., 
${[\mathbf{h}^{(\tth)}_{\Delta}]}_{\kappa +\lambda D+ 1} = H^{(\tth)}_{\lambda \Delta L,\kappa \Delta K}$; let
$\tilde{\mathbf{f}}^{(\tth)} \in \C^{J \rmv D}$ be defined by
${[\tilde{\mathbf{f}}^{(\tth)}]}_{m J +i+J/2+1} = \Fmith$ (this is the
subvector of $\mathbf{f}^{(\tth)}$ that corresponds to the restriction of
$\Fmith$ to $\{0,\dots,D-1\} \times \{-J/2,\dots,J/2-1\}$); and let
$\tilde{\mathbf{U}}_{\textnormal{F}} \in \C^{J\rmv D \times J\rmv D}$ be the
unitary
matrix with entries 
${[\tilde{\mathbf{U}}_{\textnormal{F}}]}_{\kappa +\lambda D+ 1,\ist m J +i+J/2+1} = \frac{1}{\sqrt{J\rmv D}} \, e^{-\jmath 2 \pi
(\frac{\kappa m}{D}-\frac{\lambda i}{J})}$, where $\kappa \rmv\in\rmv \{0,\dots,D \rmv-\! 1\}$,
$\lambda \rmv\in\rmv \{0,\dots,$\linebreak 
$J \rmv-\! 1\}$, $m \rmv\in\rmv \{0,\dots,D \rmv-\! 1\}$, and $i \rmv\in\rmv \{-J/2,\dots, J/2 \rmv-\! 1\}$. Then, we can rewrite
\eqref{System_Channel_Det_subs}
as $\mathbf{h}^{(\tth)}_{\Delta} = \sqrt{J\rmv D} \, \tilde{\mathbf{U}}_{\textnormal{F}} \ist \tilde{\mathbf{f}}^{(\tth)}\rmv$.
\pagebreak 
We thus
\vspace{-2mm}
obtain
\be
{\| \mathbf{h}^{(\tth)}_{\Delta} \|}_2 \ist=\ist \sqrt{J\rmv D} \, {\| \tilde{\mathbf{f}}^{(\tth)} \|}_2 \ist=\ist \sqrt{J\rmv D} \, {\|\mathbf{f}^{(\tth)}\|}_2 \,,
\label{eq_appA_htheta}
\ee
since $\mathbf{f}^{(\tth)}$ differs from its subvector
$\tilde{\mathbf{f}}^{(\tth)}$ only by additional zero entries. Finally, let
$\mathbf{g}^{(\tth)}\triangleq\vecr_{m,i}\big\{\GGmith\big\} \in \C^{J
\rmv D}$, i.e.,
${[\mathbf{g}^{(\tth)}]}_{m J +i+J/2+1} = \GGmith$, and
let $\mathbf{U} \in \C^{J\rmv D \times J\rmv D}$ be the unitary matrix with
entries ${[\mathbf{U}]}_{\kappa +\lambda D+ 1,\ist m J +i+J/2+1} =
u_{m,i}[\lambda,\kappa]$. Then \eqref{Gen_Basis_Exp} can be written as
$\mathbf{h}^{(\tth)}_{\Delta} = \mathbf{U} \mathbf{g}^{(\tth)}\rmv$, which
\vspace{0mm} 
implies
\be
{\| \mathbf{h}^{(\tth)}_{\Delta} \|}_2 \ist=\ist {\| \mathbf{g}^{(\tth)} \|}_2 \,.
\label{eq_appA_hk}
\vspace{.5mm}
\ee
Combining \eqref{eq_appA_h}--\eqref{eq_appA_hk}, we obtain ${\|\mathbf{h}^{(\tth)}\|}_2 = \sqrt{\frac{K\rmv L}{J\rmv D}} \, {\|\mathbf{g}^{(\tth)}\|}_2$
and, furthermore,
\begin{align}
\hspace{-2mm}\Bigg( \sum_{l=0}^{L-1} \sum_{k=0}^{K-1} \big| \hat{H}^{(\tth)}_{l,k} \rmv\rmv-\rmv H^{(\tth)}_{l,k} \big|^2 \Bigg)^{\!\!1/2}\! &\rmv=\ist \big\| \hat{\mathbf{h}}^{(\tth)} \rmv\rmv-\rmv \mathbf{h}^{(\tth)} \big\|_2 \nonumber\\[-1mm]
\hspace{-2mm}&\rmv=\ist \sqrt{\frac{K\rmv L}{J\rmv D}} \, \big\| \hat{\mathbf{g}}^{(\tth)} \rmv\rmv-\rmv \mathbf{g}^{(\tth)} \big\|_2 \,,
\label{eq_h-k-diffnorm} 
\end{align}
where $\hat{\mathbf{h}}^{(\tth)} \triangleq \vecr_{l,k}\big\{\hat{H}^{(\tth)}_{l,k} \big\}$ and
$\hat{\mathbf{g}}^{(\tth)}\triangleq\vecr_{m,i}\big\{\hat{G}^{(\tth)}_{m,i}\big\}$
denote the estimates of $\mathbf{h}^{(\tth)}$ and $\mathbf{g}^{(\tth)}$,
respectively.

Next, let $\GGmiti$ be the $\NR \!\times\! \NT$ matrix with entries
$\big[\GGmiti\big]_{\rec,\tra} \!\triangleq \tilde{G}_{m,i}^{(\rec,\tra)}$ for $\rr\in\{1,\dots,\NR\}$ and $\ss\in\{1,\dots,\NT\}$,
and let
\vspace{-1.5mm}
(cf.\ \eqref{eq_D-x-G})
\be
\ggantti \ist\triangleq\, \vecr_{m,i}\big\{\Gmiantti\big\} \ist=\ist \sqrt{\frac{J \rmv D}{\Pil}} \, \xxant .
\label{eq_tildeg_G_x}
\vspace{-1mm}
\ee
By the definition of $\Gmiantti$
in \eqref{eq_G-k}, we have $\GGmiti = \GGmi \ist \PP$ and, in turn, $\GGmi =
\GGmiti \ist \PP^{-1}$. Similarly, we have $\hat{\mathbf{G}}_{m,i} =
\GGmitihat \ist \mathbf{P}^{-1}\rmv$, where $\GGmitihat$ denotes the estimate
of $\GGmiti$ (cf. step 2 in Section \ref{sub-sec:Estimator}). We then obtain
\begin{align}
\sum_{\tth \in \Tth} \big\| \hat{\mathbf{g}}^{(\tth)} \rmv\rmv-\rmv \mathbf{g}^{(\tth)} \big\|_2^2\,
  &\overset{(a)}{\eq} \,\sum_{m=0}^{D-1} \sum_{i=-J/2}^{J/2-1} \rmv \big\| \hat{\mathbf{G}}_{m,i} \rmv-\rmv \GGmi \big\|_{\text{F}}^2 \nonumber\\
  &\eq\,\sum_{m=0}^{D-1} \sum_{i=-J/2}^{J/2-1} \rmv \big\| \big( \GGmitihat \rmv-\rmv \GGmiti \big) \mathbf{P}^{-1} \big\|_{\text{F}}^2 \nonumber\\
  &\,\overset{(b)}{\leq}\, \sum_{m=0}^{D-1} \sum_{i=-J/2}^{J/2-1} \rmv \big\| \GGmitihat \rmv-\rmv \GGmiti \big\|_{\text{F}}^2 \,\| \mathbf{P}^{-1} \|^2 \nonumber\\
  &\overset{(c)}{\eq} \| \mathbf{P}^{-1} \|^2 \sum_{\tth \in \Tth} \big\| \hat{\tilde{\mathbf{g}}}^{(\tth)} \rmv\rmv-\rmv \tilde{\mathbf{g}}^{(\tth)} \big\|_2^2 \nonumber\\[1mm]
  &\eq \| \mathbf{P}^{-1} \|^2 \, \frac{\JD}{\Pil} \sum_{\ant \in \Ant} \big\| \xxanthat \rmv\rmv-\rmv \xxant \big\|_{2}^2  \nonumber\\
  &\eq \frac{\JD}{\Pil} \, \| \mathbf{P}^{-1} \|^2 \, {\| \xxhat \rmv-\rmv \xx \|}_{2}^2
  \label{pars_upper1}\;, \\[-6mm]
  \nonumber
\end{align}
where $\norm{\cdot}$ denotes the spectral norm 
\cite{Horn99}, $\xx \triangleq \big(\xx^{(\MCSidx_1)\ist T} \cdots$\linebreak 
$\xx^{(\MCSidx_{\NT\NR})\ist T}\big)^T\rmv$,
and $\xxhat \triangleq \big( \hat{\xx}^{(\MCSidx_1)\ist T} \cdots\,
\hat{\xx}^{(\MCSidx_{\NT\NR})\ist T}\big)^T\rmv$. Here, $(a)$ and $(c)$ are obtained by reordering the sums, and $(b)$ follows by
the general inequality ${\|\mathbf{A}\mathbf{B}\|}_{\text{F}} \leq
{\|\mathbf{A}\|}_{\text{F}} \ist \|\mathbf{B}\|$ \cite[problem 20 in 
ch.\ 5.6]{Horn99}.
A combination of \eqref{eq_h-k-diffnorm} and \eqref{pars_upper1} then 
\vspace{-1mm}
yields
\begin{align}
E &\,=\ist \Bigg( \sum_{\tth \in \Tth} \sum_{l=0}^{L-1} \sum_{k=0}^{K-1} \big| \hat{H}^{(\tth)}_{l,k} \rmv\rmv-\rmv H^{(\tth)}_{l,k} \big|^2 \Bigg)^{\!\!1/2} \nonumber\\[0mm]
&\eq \bigg( \sum_{\ant\in\Ant}\frac{\KL}{\JD} \, \big\| \hat{\mathbf{g}}^{(\tth)} \rmv\rmv-\rmv \mathbf{g}^{(\tth)} \big\|_{2}^2 \bigg)^{\rmv\!1/2} \nonumber\\[0mm]
&\,\leq\, \sqrt{\frac{\KL}{\Pil}}\, \| \mathbf{P}^{-1} \| \, {\| \xxhat \rmv-\rmv \xx \|}_{2} \,.
\label{th MCE proof}\\[-7mm]
\nonumber
\end{align}

We now consider the first part of Theorem \ref{Theorem_MIMO}, concerning the use of G-BPDN for MGCS reconstruction. 
We will use the following result on the performance of G-BPDN.
Let $\xxhat$ denote the solution of \eqref{G-BPDN}, and let $\xx^{S|\Part}\rmv$ denote the best group $S$-sparse approximation of $\xx$ with respect to $\Part$, 
i.e., the $\xx' \!\in \Spset{S|\Part}$
minimizing ${\|\xx' \!-\rmv \xx\|}_{2|\Part}$.
Then, as shown in \cite{Eldar_GBPDN}, if
$\yy$ satisfies ${\|\bPhi \xx \rmv-\rmv \yy\|}_2 = {\|\zz\|}_2 \leq \epsilon$ and
$\bPhi$ satisfies the group restricted isometry property of order $2S$ with G-RIC
$\delta_{2S|\Part} \leq \sqrt{2}-\rmv 1$,
then\footnote{We 
note that this result was formulated in \cite{Eldar_GBPDN} for the special case of block
sparsity; however, it extends to the general group sparse setting in a straightforward
way.} 
\begin{equation}
\label{GBPDN_res}
{\|\xxhat \rmv-\rmv \xx\|}_2 \,\leq\, \frac{c_0}{\sqrt{S}} \,{\|\xx \rmv-\rmv \xx^{S|\Part}\|}_{2|\Part} \ist+\ist c_1 \epsilon \,,
\vspace{-1mm}
\end{equation}
with $c_{0}$ and $c_{1}$ as given in Theorem \ref{Theorem_MIMO}.
This result bounds the reconstruction error ${\|\xxhat \rmv-\rmv \xx\|}_2$ in terms of ${\|\xx \rmv-\rmv \xx^{S|\Part}\|}_{2|\Part}$, which
characterizes the deviation of $\xx$ from being group $S$-sparse with respect to $\Part$,
and in terms of $\epsilon$.

Recall \eqref{eq-delta-max}, i.e., the fact that
the G-RIC of the stacked measurement matrix $\bPhi$ in \eqref{Phi_MCS_GCS} with respect to the associated partition $\tilde{\Part}$ 
(cf.\ \eqref{Group_joint}) satisfies $\delta_{S|\tilde{\Part}} = \max_{\tra} \delta_{S|\Part}^{(\tra)}$.
Our assumption on the $\bPhitra\rmv$, i.e., $\delta_{2S|\Part}^{(\tra)} \rmv\leq\rmv \sqrt{2}-1$
for all $\tra\in\{1,\dots,\NT\}$, then implies that
$\delta_{2S|\tilde{\Part}}\leq\sqrt{2}-1$. Thus, with our additional assumption that 
$\big( \sum_{\tth \in \Tth} {\|\zzant\|}_2^2\big)^{1/2} \rmv\leq\rmv \epsilon$,
we have (cf.\ \eqref{GBPDN_res})
${\|\xxhat \rmv-\rmv \xx\|}_2 \leq \frac{c_0}{\sqrt{S}} \ist {\|\xx \rmv-\rmv \xx^{S|\tilde{\Part}}\|}_{2|\tilde{\Part}} + c_1 \epsilon$. Inserting into \eqref{th MCE proof} yields the bound
\be
E \,\leq\, \sqrt{\frac{\KL}{\Pil}}\, \| \mathbf{P}^{-1} \| \, \bigg( \frac{c_0}{\sqrt{S}} \ist {\|\xx \rmv-\rmv \xx^{S|\tilde{\Part}}\|}_{2|\tilde{\Part}} + c_1 \epsilon \bigg) \,.
\label{th MCE proof_1}
\ee
Now recall that $\xx^{S|\tilde{\Part}}$ is the group $S$-sparse vector $\xx' \!\in\rmv \Sigma_{S|\tilde{\Part}}$ 
minimizing $\|\xx \rmv-\rmv \xx'\|_{2|\tilde{\Part}}$,
and note that the subvectors
$\xx^{S|\tilde{\Part}}[\gridx]$ coincide with the subvectors $\xx[\gridx]$
for $\gridx \in \mathcal{T}$,
where $\mathcal{T} \subseteq \{1,\dots,\grnum\}$ denotes the set of those $S$ group indices that yield the largest
norms ${\|\xx[b]\|}_2\ist$,
and
$\xx^{S|\tilde{\Part}}[\gridx] = \mathbf{0}$ for
$\gridx \notin \mathcal{T}$.
\vspace{-.5mm}
Therefore,
\begin{align} 
{\big\|\xx \rmv-\rmv \xx^{S|\Partti}\big\|}_{2|\tilde{\Part}} = \sum_{\gridx \ist=1}^{\grnum} \big\|\xx [\gridx]-\rmv \xx^{S|\Partti}[\gridx]\big\|_{2} = \sum_{\gridx \ist\notin \mathcal{T}} \|\xx [\gridx]\|_{2}\,. \nonumber\\[-3mm]
\label{bestappr}\\[-6mm]
\nonumber
\end{align}
Moreover, by the definition of $\mathcal{T}$, we have
${\|\xx [\gridx]\|}_{2} \geq {\|\xx [\gridx']\|}_{2}$ for all $\gridx \!\in\! \mathcal{T}$ and $\gridx' \!\notin\! \mathcal{T}$,
which yields  
$\sum_{\gridx \ist\in \mathcal{T}} {\|\xx [\gridx]\|}_{2} \geq \sum_{\gridx \ist\in \mathcal{T}'} {\|\xx [\gridx]\|}_{2}$
for any set $\mathcal{T}' \!\subseteq\rmv \{1,\dots,\grnum\}$ of
cardinality $|\mathcal{T}'| \rmv=\rmv |\mathcal{T}| \rmv=\rmv S$, and in turn
$\sum_{\gridx \ist\notin \mathcal{T}} {\|\xx [\gridx]\|}_{2} \ist\leq \sum_{\gridx \ist\notin \mathcal{T}'} {\|\xx [\gridx]\|}_{2}$.
Inserting into \eqref{bestappr} gives
\be\label{apprsets}
{\big\|\xx \rmv-\rmv \xx^{S|\Partti}\big\|}_{2|\tilde{\Part}} \ist\leq \sum_{\gridx \ist\notin \mathcal{T}'} \!{\|\xx [\gridx]\|}_{2} \,,
\vspace{-1mm}
\ee
for any such set $\mathcal{T}'$ of cardinality $S$.
Then, with $\bcalSi$ defined as in Section \ref{sub-sec:Perf_Ana}, we 
obtain
\begin{align}
{\big\|\xx \rmv-\rmv \xx^{S|\Partti}\big\|}_{2|\tilde{\Part}} 
&\overset{(a)}{\,\leq\,}  \sum_{\gridx \ist\notin \bcalSi} {\|\xx [\gridx]\|}_{2} \nonumber\\[-1mm]
&\!\overset{\eqref{x_b_theta_vector}}{\,\eq\,} \sum_{\gridx \ist\notin \bcalSi} \bigg( \sum_{\tth \in \Tth} \big\| \xxant [\gridx] \big\|_{2}^2 \bigg)^{\rmv\!1/2}\nonumber\\[.5mm]
&\!\overset{\eqref{eq_tildeg_G_x}}{\,\eq\,} \sqrt{\frac{\Pil}{\JD}} \sum_{\gridx \ist\notin \bcalSi} \bigg(\sum_{\tth \in \Tth} \big\| \ggantti  [\gridx] \big\|_{2}^2 \bigg)^{\rmv\!1/2} \;,
\label{th MGCE proof end_0}\\[-6.5mm]
\nonumber
\end{align}
where $(a)$ follows from 
\eqref{apprsets} and the fact that $|\bcalSi|=S$. Now
for each group $\Group_\blidx$ of $\Part$ we have (with $\Block_\blidx = \bijG^{-1}(\Group_\blidx)$, cf.\
\eqref{bij} and the discussion following \eqref{bij})
\begin{align}
\sum_{\tth \in \Tth} \big\| \ggantti[\gridx] \big\|_2^2
&\ist\overset{(a)}{\eq}\ist \sum_{\tth \in \Tth} \sum_{(m,i) \in\Block_\blidx} \!\! |\Gmiantti|^2\nonumber\\[0mm]
&\ist\overset{(b)}{\eq}\ist\!\!\! \sum_{(m,i) \in \Block_\blidx} \!\! \big\| \GGmiti \big\|_{\textnormal{F}}^2\nonumber\\[0mm]
&\ist\overset{(c)}{\leq}\!\sum_{(m,i) \in \Block_\blidx} \!\! {\|\GGmi\|}_{\textnormal{F}}^2 \ist \|\PP\|^2\nonumber\\[0mm]
&\overset{(d)}{\eq}\ist \|\PP\|^2 \sum_{\tth \in \Tth} \big\| \mathbf{g}^{(\tth)}[\gridx] \big\|_2^2 \,.
\label{norm_g_gtilde}\\[-7mm]
\nonumber
\end{align}
Here, $(a)$ follows from $\ggantti \rmv= \vecr_{m,i}\big\{\Gmiantti\big\}$,
$(b)$ follows from $\tilde{G}_{m,i}^{(\rec,\tra)} = \big[\GGmiti\big]_{\rec,\tra}$,
$(c)$ follows from $\GGmiti \rmv=\rmv \GGmi \ist \PP$ and ${\|\mathbf{AB}\|}_{\textnormal{F}} \leq
{\|\mathbf{A}\|}_{\textnormal{F}} \|\mathbf{B}\|$, and $(d)$ follows from
$\mathbf{g}^{(\tth)} \rmv= \vecr_{m,i}\big\{\GGmith\big\}$.
Inserting \eqref{norm_g_gtilde} into \eqref{th MGCE proof end_0} yields
\begin{align}
{\big\|\xx \rmv-\rmv \xx^{S|\Partti}\big\|}_{2|\tilde{\Part}} 
&\,\leq\, \sqrt{\frac{\Pil}{\JD}} \, \|\PP\| \sum_{\gridx \ist\notin \bcalSi} \bigg( \ist\sum_{\tth \in \Tth} \big\| \mathbf{g}^{(\tth)} [\gridx] \big\|_{2}^2 \bigg)^{\rmv\!1/2}\nonumber\\[0mm]
&\overset{(a)}{\eq}\ist \sqrt{\frac{\Pil}{\JD}} \, \|\PP\| \, C_{G,\bcalSi,\Part} \,,
  \label{th MGCE proof end}\\[-6.5mm]
\nonumber
\end{align}
where $(a)$ follows from $\mathbf{g}^{(\tth)} \rmv=\vecr_{m,i}\big\{\GGmith\big\}$ and \eqref{C_GSJ}.
Inserting this bound into \eqref{th MCE proof_1} finally yields
\[
E \,\leq\, c_{0} \ist \sqrt{\frac{\LK}{\JD}} \ist \| \PP^{-1} \| \rmv \norm{\PP} \frac{C_{G,\bcalSi,\Part}}{\sqrt{S}} +\ist c_{1} \ist \sqrt{\frac{\LK}{\Pil}} \ist \| \PP^{-1} \| \ist \epsilon \,,
\vspace{-2mm}
\]
which is \eqref{th MGCE  BPDN}.


Next, we consider 
the second part of Theorem \ref{Theorem_MIMO}, concerning the use of G-CoSaMP for MGCS reconstruction. We will use the following result on
the performance of G-CoSaMP, obtained by specializing results from \cite{Bar_model08}.
Consider a partition $\Part$ with groups of equal size.
If $\yy$ satisfies ${\|\bPhi\xx \rmv-\rmv \yy\|}_2 \leq \epsilon$ and $\bPhi$ satisfies the group restricted isometry property of order $4S$ with respect to $\Part$ with G-RIC
$\delta_{4S|\Part} \leq 0.1$, the result
$\hat\xx$ of G-CoSaMP after $\idx$ iteration steps
satisfies\footnote{Here, 
we have used the inequality $\norm{\xx}_{2}\leq\norm{\xx}_{2|\Part}$, which can be shown as follows:
\vspace{-3.5mm}
\begin{align*}
\norm{\xx}_{2} &\eq\!
\Bigg( \sum_{\gridx=1}^{\grnum} \sum_{\vecidx\in \Group_\gridx} \rmv \big| {[\xx]}_{\vecidx} \big|^2 \Bigg)^{\!\!1/2} \\[0mm]
&\,\leq\, \sum_{\gridx=1}^{\grnum}\Bigg( \sum_{\vecidx\in\Group_\gridx} \rmv \big| {[\xx]}_{\vecidx} \big|^2\Bigg)^{\!\!1/2} \\[0mm]
&\eq \sum_{\gridx=1}^{\grnum} \rmv \norm{\xx[\gridx]}_{2} \\[1mm]
&\eq \norm{\xx}_{2|\Part} .\\[-11mm]
\end{align*}
} 
\be
\label{MB-CoSaMP_res}
{\|\hat\xx \rmv-\rmv \xx\|}_2 \,\leq\, \frac{1}{2^{\idx}}
\ist {\|\xx\|}_2 \ist+\ist 20 \ist \bigg(\rmv 1+\frac{1}{\sqrt{S}}\bigg) {\|\xx \rmv-\rmv \xx^{S|\Part}\|}_{2|\Part} \ist+\ist 20 \ist \epsilon \,.
\ee

Under our assumption on the $\bPhitra\rmv$, i.e., $\delta_{4S|\Part}^{(\tra)} \rmv\leq\rmv 0.1$, the
G-RIC of the stacked measurement matrix $\bPhi$ with respect to the associated partition $\tilde{\Part}$
satisfies $\delta_{4S|\tilde{\Part}}\rmv\leq\rmv 0.1$ (cf. \eqref{eq-delta-max}).
With our additional assumption that $\big( \sum_{\tth \in \Tth} {\|\zzant\|}_2^2\big)^{1/2} \rmv\leq\rmv \epsilon$, we obtain (cf.\ \eqref{MB-CoSaMP_res})
${\|\xxhat \rmv-\rmv \xx\|}_2 \leq \frac{1}{2^{\idx}} \ist {\|\xx\|}_2 + 20 \ist \big(\rmv 1+\frac{1}{\sqrt{S}}\big) {\|\xx \rmv-\rmv \xx^{S|\Partti}\|}_{2|\Partti} + 20 \ist \epsilon$.
Inserting into \eqref{th MCE proof} yields the bound
\begin{align}
&E \,\leq\, \sqrt{\frac{\KL}{\Pil}}\, \| \mathbf{P}^{-1} \| \, \bigg[ \frac{1}{2^{\idx}} \ist {\|\xx\|}_2
  \ist+\ist 20 \ist \bigg(\rmv 1\rmv+\rmv \frac{1}{\sqrt{S}}\bigg) \nonumber\\[0mm] 
  & \hspace{37mm} \times\! \big\|\xx \rmv-\rmv \xx^{S|\Partti} \big\|_{2|\Partti} +\ist 20 \ist \epsilon \bigg] \,.
\label{th MCE proof_2}\\[-9mm]
\nonumber
\end{align}
We have
\be
\label{norm_ineq_0}
{\|\xx\|}_2^2 \ist\overset{\eqref{x_b_theta_vector}}{\,\eq\,} \sum_{\ant \in \Ant} \big\|\xxant\big\|_{2}^2
  \ist\overset{\eqref{eq_tildeg_G_x}}{\,\eq\,} \frac{\Pil}{\JD}\sum_{\ant \in \Ant} \big\|\ggantti\big\|_{2}^2 \,.
\vspace{.5mm}
\ee
Following a similar reasoning as in \eqref{norm_g_gtilde}, we obtain
\begin{align*}
\sum_{\tth \in \Tth} \big\| \ggantti \big\|_2^2 \rmv
&\eq\rmv \sum_{\tth \in \Tth} \sum_{m=0}^{D-1} \sum_{i=-J/2}^{J/2-1} \rmv\! \big|\Gmiantti\big|^2\\[0mm]
&\eq\rmv\rmv \sum_{m=0}^{D-1} \sum_{i=-J/2}^{J/2-1} \!\! \big\| \GGmiti \big\|_{\textnormal{F}}^2\\[0mm]
&\,\leq\ist \sum_{m=0}^{D-1} \sum_{i=-J/2}^{J/2-1} \!\! {\|\GGmi\|}_{\textnormal{F}}^2 \ist \|\PP\|^2\\[0mm]
&\eq\rmv \|\PP\|^2 \sum_{\tth \in \Tth} \big\| \mathbf{g}^{(\tth)} \big\|_2^2 \,.\\[-6.5mm]
\end{align*}
Thus, \eqref{norm_ineq_0} becomes
\vspace{-1mm}
further
\begin{align}
\label{norm_ineq}
{\|\xx\|}_2^2 &\,\leq\, \frac{\Pil}{\JD} \, {\|\PP\|}^2 \sum_{\ant \in \Ant} \big\|\mathbf{g}^{(\tth)}\big\|_{2}^2 \nonumber\\[0mm]
&\eq \frac{\Pil}{\JD}\big\|\PP\big\|^2 \sum_{\ant \in \Ant} \sum_{m=0}^{D-1} \sum_{i=-J/2}^{J/2-1} \!\rmv \absb{\Gmiant}^2.
\end{align}
Inserting \eqref{norm_ineq} and \eqref{th MGCE proof end} into \eqref{th MCE proof_2}, we finally obtain
\begin{align*}
E &\,\leq\, \frac{1}{2^{\idx}} \ist \sqrt{\frac{\LK}{\JD}} \ist \| \mathbf{P}^{-1} \| \| \mathbf{P} \| \ist
\Bigg( \sum_{\ant \in \Ant} \sum_{m=0}^{D-1} \sum_{i=-J/2}^{J/2-1} \!\rmv \absb{\Gmiant}^2 \Bigg)^{\!\!1/2}\\[-1mm]
  &\hspace*{15mm}  +\, 20 \ist \sqrt{\frac{\LK}{\JD}} \ist \| \mathbf{P}^{-1} \| \| \mathbf{P} \| \, C_{G,\bcalSi,\Part} \bigg(\rmv 1\rmv+\rmv \frac{1}{\sqrt{S}}\bigg)\\[1mm]
  &\hspace*{15mm}+\ist \sqrt{\frac{\KL}{\Pil}} \ist \| \mathbf{P}^{-1} \| \, 20 \ist \epsilon \ist\,, \\[-8mm]
\end{align*}
which is \eqref{theorem_est_error}.

\section*{Appendix C:\, Proof of Equation \eqref{BasOpt_Calc} }\label{sec_appC}

We will calculate the entries $\GGmith$ of $\mathbf{G}$ for elementary single-scatterer
channels $h^{(\tth)}(t,\tau)=\delta(\tau-\tau_1^{(\tth)}) \, e^{\jmath 2 \pi \nu_1^{(\tth)} t}$,\linebreak 
$\MCSsett$. Combining \eqref{System_Channel_Det_subs}
and \eqref{Gen_Basis_Exp} and using \eqref{eq:u_v}, we 
have
\begin{align*}
&\sum_{m=0}^{D-1}\sum_{i=-J/2}^{J/2-1} \!\!\GGmi \, \frac{1}{\sqrt{D}} \, v_{m,i}[\lambda] \, e^{-\jmath 2 \pi \frac{\kappa m}{D}} \\[0mm]
&\quad\;=\ist \sum_{m=0}^{D-1}\sum_{i=-J/2}^{J/2-1} \!\!\Fmi \, e^{-\jmath 2 \pi ( \frac{\kappa m}{D}-\frac{\lambda i}{J} )} \ist , \\[-7.5mm]
\end{align*}
or, equivalently,
\[
\frac{1}{\sqrt{D}} \rmv \sum_{i=-J/2}^{J/2-1} \!\!\GGmi \, v_{m,i}[\lambda] \,=\rmv \sum_{i=-J/2}^{J/2-1} \!\!\Fmi \, e^{\jmath 2 \pi \frac{\lambda i}{J} } \ist.
\]
Expressing the $\Fmi$ by \eqref{F_S_A}, the previous relation written entrywise becomes
\begin{align*}
&\frac{1}{\sqrt{D}} \rmv \sum_{i=-J/2}^{J/2-1} \!\!\GGmith \, v_{m,i}[\lambda]\\[.5mm]
& \,=\rmv \sum_{i=-J/2}^{J/2-1} \sum_{q=0}^{N-1} \rmv S_h^{(\theta)}[m,i+qL] \, A^{*}_{\gamma,g} \bigg( \rmv m \ist , \frac{i+qL}{\Lr} \bigg) \ist
    e^{\jmath 2 \pi \frac{\lambda i}{J} } \rmv, 
\end{align*}
for $\MCSsett$.
Inserting \eqref{Spread_fct_mod} (specialized to $P \!=\! 1$, i.e., a single-scatterer channel with $\eta_1^{(\tth)}=1$) and \eqref{leakage_kernel} yields
\[
\frac{1}{\sqrt{D}} \rmv \sum_{i=-J/2}^{J/2-1} \!\!\GGmith \, v_{m,i}[\lambda] \,=\, \phi^{(\nu_1^{(\tth)})} \bigg(m-\frac{\tau_1^{(\tth)}}{\Ts}\bigg) \ist C^{(\nu_1^{(\tth)})}[m,\lambda] \,,
\vspace{-.7mm}
\]
with $C^{(\nu)}[m,\lambda]$ as defined in \eqref{BasOpt_C}. Since ${\{ v_{m,i}[\lambda] \}}_{i=-J/2}^{J/2\rmv-1}$
is an orthonormal basis, the last relation is equivalent to the following expression
\vspace{-.5mm}
of $\GGmith$:
\[
\GGmith \,=\, \sqrt{D} \,\ist \phi^{(\nu_1^{(\tth)})} \bigg(m-\frac{\tau_1^{(\tth)}}{\Ts}\bigg)
  \sum_{\lambda=0}^{J-1} v^*_{m,i}[\lambda] \, C^{(\nu_1^{(\tth)})}[m,\lambda] \,.
\]
For $\tth \rmv=\rmv \tth_\xi$ with $\xi \rmv\in\rmv \{1,\ldots,\NT\NR\}$, this can be rewritten 
\vspace{-.5mm}
as
\begin{align*}
G_{m,i}^{(\tth_\xi)} &\,=\, \big[\VV_m \ist \mathbf{c}_m \big(\tau_1^{(\tth_\xi)} \rmv,\nu_1^{(\tth_\xi)}\big) \big]_{i+J/2+1} \\[0mm]
  &\,=\, \big[ \VV \mathbf{c}\big(\tau_1^{(\tth_\xi)} \rmv,\nu_1^{(\tth_\xi)}\big) \big]_{\bijG(m,i)}\\[0mm]
  &\,=\, \big[\VV \mathbf{C}(\ttau_1,\nnu_1) \big]_{\bijG(m,i),\xi} \,,\\[-6.5mm]
\end{align*}
and finally, because $G_{m,i}^{(\tth_\xi)} =
{[\mathbf{G}]}_{\bijG(m,i),\xi}\ist$ (see \eqref{Gmith}),
\vspace{-.5mm}
as
\[\mathbf{G} =
\VV \mathbf{C}(\ttau_1,\nnu_1).
\]



\bibliography{tf-zentral1,bib_MGSCS}
\bibliographystyle{ieeetr}

\end{document}